\documentclass[rmp,aps,nofootinbib]{revtex4}
\usepackage{amsmath}
\usepackage{amssymb}
\usepackage{graphics}
\usepackage{epsfig}

\begin{document}

\title{Landau-Zener-St\"{u}ckelberg interferometry}
\author{S. N. Shevchenko}
\email{sshevchenko@ilt.kharkov.ua}
\affiliation{B.Verkin Instiute for Low Temperature Physics and Engineering, Kharkov,
Ukraine}
\affiliation{RIKEN Advanced Science Institute, Wako-shi, Saitama, Japan}
\author{S. Ashhab}
\affiliation{RIKEN Advanced Science Institute, Wako-shi, Saitama, Japan}
\affiliation{Department of Physics, The University of Michigan, Ann Arbor, Michigan, USA}
\author{Franco Nori}
\affiliation{RIKEN Advanced Science Institute, Wako-shi, Saitama, Japan}
\affiliation{Department of Physics, The University of Michigan, Ann Arbor, Michigan, USA}

\begin{abstract}
A transition between energy levels at an avoided crossing is known as a
Landau-Zener transition. When a two-level system (TLS) is subject to
periodic driving with sufficiently large amplitude, a sequence of
transitions occurs. The phase accumulated between transitions (commonly
known as the St\"{u}ckelberg phase) may result in constructive or
destructive interference. Accordingly, the physical observables of the
system exhibit periodic dependence on the various system parameters. This
phenomenon is often referred to as Landau-Zener-St\"{u}ckelberg (LZS)
interferometry. Phenomena related to LZS interferometry occur in a variety
of physical systems. In particular, recent experiments on LZS interferometry
in superconducting TLSs (qubits) have demonstrated the potential for using
this kind of interferometry as an effective tool for obtaining the
parameters characterizing the TLS as well as its interaction with the
control fields and with the environment. Furthermore, strong driving could
allow for fast and reliable control of the quantum system. Here we review
recent experimental results on LZS interferometry, and we present related
theory.
\end{abstract}

\date{\today }
\maketitle
\tableofcontents

%03.67.Lx Quantum computation architectures and implementations \\
%32.80.Xx Level crossing and optical pumping \\
%42.50.Hz Strong-field excitation of optical transitions in quantum systems;
%multiphoton processes; dynamic Stark shift \\
%85.25.Am Superconducting device characterization, design, and modeling \\
%85.25.Cp Josephson devices \\
%85.25.Hv Superconducting logic elements and memory devices;
%microelectronic circuits

\section{Introduction}

\subsection{Strongly driven TLSs: a brief historical review}

The model of a quantum two-level system (TLS), with only two relevant
quantum states, is used to describe a variety of physical systems. At first,
it was used in relation to spins and atomic collisions, which describe
natural microscopic systems. It has been used to describe various other
naturally occurring settings ever since. Furthermore, some artificial
mesoscopic systems that were realized recently in semiconductor quantum dots
and superconducting circuits can operate as effective TLSs. These
solid-state TLSs have received increasing attention recently for two main
reasons: they exhibit fundamental quantum phenomena on macroscopic scales,
and they are considered possible candidates to operate as quantum bits
(qubits) in future quantum information processing devices.

It is quite common that the two energy levels of a quantum TLS exhibit an
\textit{avoided level crossing} or anticrossing as some external parameter
is varied, as shown in Fig.~\ref{E(e)}. The physical properties of the two
energy eigenstates are typically exchanged when going from one side of the
avoided crossing to the other side. If the external control parameter is
varied in time such that the system traverses the avoided crossing region, a
non-adiabatic transition between the two energy levels can occur. The
transition probability is usually named after Landau and Zener (LZ) in
recognition of their pioneering work on this subject, \cite{Landau:1932b}
and \cite{Zener:1932}.

If a LZ transition is one step in a sequence of coherent processes occurring
during the dynamics, not only the change in the occupation probabilities but
also the change in the relative \textit{phase} between the quantum states is
relevant. In particular, in this review we shall focus on the situation
where the control parameter is varied periodically and strongly, such that
the system keeps going back and forth across the avoided crossing region. In
this case, the physical observables exhibit periodic dependence on the phase
acquired between the transitions \cite{Stueckelberg:1932, Landau:1932a}.
This periodicity, which is the basis of Landau-Zener-St\"{u}ckelberg (LZS)
interferometry, provides a useful tool that allows the characterization of
the parameters defining the quantum TLS and its interaction with the control
fields and the environment. We note here that the formula for the
single-passage transition probability can also be found in the paper \cite%
{Majorana:1932}, and it can alternatively be named the Landau-Zener-St\"{u}%
ckelberg-Majorana formula, as proposed in \cite{Giacomo:2005}.

\subsubsection{Atomic physics}

The problem of strongly driven TLSs appears in a number of contexts in
atomic physics. In atomic-collision problems, the electronic state of the
atom pair can experience a LZ transition from the ground states of the
incoming atoms into a hybridized intermediate state and then back to the
electronic ground states of the outgoing atoms. In this case the avoided
crossing region is passed twice \cite{Child:1974, Nikitin:1984}, and the St%
\"{u}ckelberg phase typically has large variations and hence its
interferometric consequences are washed out \cite{Stueckelberg:1932,
Landau:1932a}.

An atom in an intense laser field can also experience LZ transitions, but in
contrast to atomic-collision problems, the atom is driven periodically in
time and multiple LZ transitions occur \cite{Delone:1985}. In the case of an
atom in a laser field, (regardless of the field intensity) if the atom is
driven at a frequency that matches its energy-level splitting, it can be
resonantly excited to an upper level. Then, at low driving amplitudes, the
level occupation probabilities experience Rabi oscillations \cite{Rabi:1937}%
. When the driving amplitude is increased, multiphoton processes become
relevant. These processes occur when the energy-level splitting matches the
energy of an integer number of photons \cite{Shirley:1965}. These
multiphoton processes have traditionally been described by making use of
Floquet theory [\cite{Autler:1955}, \cite{Ritus:1967}, \cite{Sambe:1973},
\cite{Zel'dovich:1973}, \cite{Barone:1977}, \cite{Aravind:1984}; for a
review see \cite{Chu:2004}]. In addition to controlling atomic states,
resonant driving of atoms has also been studied as a means for finding
atomic parameters, and the parameters of quantum systems in general \cite%
{Coffey:1969, Henry:1977}.

\subsubsection{Other systems}

In addition to atomic systems, LZS interference has been studied in a number
of other systems. For example, periodic dependence and time-domain
oscillations related to LZS interference have been predicted in the
electronic transport in semiconductor superlattices \cite%
{Rotvig:1995,Rotvig:1996}. St\"{u}ckelberg oscillations were also studied in
superconducting quantum point contacts biased with a dc voltage and exposed
to periodic driving \cite{Gorelik:1998}. This effect was proposed as a
sensitive tool for microwave-photon detection. Another example of
interference in microscopic systems arises in the study of interacting
magnetic molecules [\cite{Wernsdorfer:2000}, \cite{Garanin:2004}], molecular
nanomagnets \cite{Calero:2005, Foldi:2007, Foldi:2008}, and single nitrogen
vacancy centers in diamond \cite{Fuchs:2009}. Interference effects have also
been predicted to take place in the interlayer tunneling in
quasi-one-dimensional layered materials, such as organic conductors [\cite%
{Cooper:2006}, \cite{Banerjee:2008}], and in the tunneling of a periodically
driven two-mode Bose-Einstein condensate \cite{Zhang:2008a,Zhang:2008b}.
Recently, LZ tunneling and St\"{u}ckelberg oscillations were measured in a
gas of ultracold molecules \cite{Mark:2007a, Mark:2007b} and in the
dipole-dipole interaction between Rydberg atoms \cite{van Ditzhuijzen:2009}.

Avoided level crossings can also be used in the study of second-order phase
transition problems \cite{Damski:2006, Dziarmaga:2009}. Periodic driving
around a phase transition can then result in interference fringes \cite%
{Mukherjee:2009}. The LZS model has also been used to describe the
singlet-triplet transition for the state of two electrons in a semiconductor
double quantum dot [\cite{Ribeiro:2009}, \cite{Petta:2010}, \cite%
{Burkard:2010}, \cite{Ribeiro:2010}] as well as the transport of photons
between two halves of an optical cavity divided by a dielectric membrane
\cite{Heinrich:2010}. Recent technological advances have also allowed the
realization of the strong-driving regime in superconducting qubits, where
such interference has been observed in several experiments [\cite%
{Oliver:2005}, \cite{Sillanpaa:2006}, \cite{Wilson:2007}, \cite%
{Izmalkov:2008}, \cite{Sun:2009}, \cite{LaHaye:2009}]. Such superconducting
qubit systems will be discussed in more detail below.

\subsubsection{Theory of LZS interferometry}

The theory of driven quantum TLSs has been studied in numerous papers [for
reviews see \cite{Leggett:1987}, \cite{Grifoni:1998}, \cite{Thorwart:2001},
\cite{Vitanov:2001}, \cite{Zhu:2001}, \cite{Frasca:2003}]. Some of the
recent work has focused on the impact of the environment, e.g. [\cite%
{Ao:1991}, \cite{Thorwart:2000}, \cite{Wubs:2006}, \cite{Zueco:2008}, \cite%
{Nalbach:2009}]. In particular, some papers pointed out that if the
relaxation in the system is small enough, interference between multiple LZ
transitions becomes relevant for the system's dynamics, e.g. \cite%
{Shimshoni:1991}, \cite{Saito:2002, Saito:2007}; when the avoided crossing
is passed several times, interference can take place and the dynamics is
sensitive to the phase acquired between successive crossings, e.g. \cite%
{Kayanuma:1994}.

LZS interference has also been recently proposed as a possible approach to
achieve perfect population transfers and to implement quantum gates for
quantum-control and quantum-computing purposes [\cite{Teranishi:1998}, \cite%
{Zhu:2001}, \cite{Gaitan:2003}, \cite{Zwanziger:2003}, \cite{Nagaya:2007},
\cite{Li:2009}] [related work on combining adiabatic LZ crossings with
coherent dynamics can be found in \cite{Goswami:2003}, \cite{Vitanov:2003},
\cite{Kral:2007}, \cite{Oh:2008}, \cite{Wei:2008}].

The purpose of the present review is to present the theory of strongly
driven TLSs and LZS interferometry and recent experimental results on the
subject, particularly in relation to superconducting qubits.

\subsection{Superconducting qubits}

Superconducting qubits are Josephson-junction-based circuits that can behave
as effective two-level systems. There are several types of superconducting
qubits, differing mostly by their topology and by their physical parameters.
They are generally called phase, charge, or flux qubits. For reviews see [%
\cite{Makhlin:2001}, \cite{Devoret:2004}, \cite{You:2005}, \cite{Wendin:2007}%
, \cite{Zagoskin:2007}, and \cite{Clarke:2008}].

Superconducting qubits have several properties that make them unique for
studying quantum effects. The qubits have mesoscopic size, which means that
they allow the observation of quantum effects on a macroscopic scale. They
can be integrated in electrical circuits, which provides different ways of
controlling and probing their states. Their parameters are adjustable: they
can be tuned by changing the applied bias current, gate voltage, or magnetic
flux. The qubit's state can be probed by measuring the charge or current
induced in it. These unique properties of superconducting qubits allow the
realization of the strong-driving regime, and they have allowed the
realization of several experiments on LZS interferometry in the past few
years.

\subsection{Fano and Fabry-Perot interferometry using superconducting qubits}

Superconducting qubits can be used for LZS interferometry, but also to study
other types of interferometry, including Fano and Fabry-Perot interferometry
[\cite{Zhou:2008a,Zhou:2008b}, \cite{Liao:2010}]. These latter devices can
be implemented using quasi-one-dimensional open systems where photons are
injected from the left and move towards the right side of the device. Along
the way, the photons interact with either one or two qubits acting as
tunable mirrors, controlled by changing the applied electric and/or magnetic
fields on the qubits. These qubits, working as tunable mirrors, can change
the reflection and transmission coefficients of the photons confined in
waveguides.

Let us first consider the case of a single superconducting qubit interacting
with an incoming photon \cite{Zhou:2008a}. When the energy of the incoming
photon matches the energy spacing of the qubit, the photon is reflected,
otherwise it is transmitted. This type of single-photon switch exhibits
Breit-Wigner scattering: now in one dimension instead of the standard three
dimensional case for natural atoms. This Breit-Wigner scattering produces a
symmetric Lorentzian peak in the reflection coefficient, versus frequency,
of the photon. This situation occurs when the dispersion relation of the
incoming photon is linear, as in a transmission line resonator, acting as a
``rail" guiding the motion of the photons. When the photon dispersion
relation is nonlinear, for long-wavelength photons propagating in a
quasi-one-dimensional array of coupled cavities, the reflection coefficient
exhibits an asymmetric Fano line shape, due to the interference between the
continuous mode of the incoming photon and the discrete energy levels of the
qubit. Thus, for single-photon transport in a one-dimensional waveguide, the
photons can be partially or totally reflected by a controllable two-level
system which can act as a tunable mirror.

It is known that the Fabry-Perot interferometer, which consists of two
highly reflecting planar mirrors, provides the simplest cavity. It is then
natural to ask the question: ``is it possible to construct a quantum version
of a Fabry-Perot interferometer?" Namely, to build a resonator, in a
one-dimensional waveguide, with two tunable-mirrors made of quantum
scatterers. Reference \cite{Zhou:2008b} focused on this question and studied
quantum analogs of the Fabry-Perot interferometer. They found that two
separate two-level systems interacting with photons in a waveguide can also
create, between them, single-photon quasi-bound states. A recent work \cite%
{Liao:2010} studied a more complex form of interferometry, with the goal of
controlling the transport of single photons by tuning the frequency of
either one or two cavities in an array of coupled cavities guiding the 1D
motion of photons. In some regimes, the photons can be localized around the
scatterers, which act as impurities, producing isolated states in the gap of
the energy spectrum [\cite{Zhou:2008b}, \cite{Liao:2010}].

\subsection{Hamiltonian and bases}

The main subject of this review is a strongly and periodically driven TLS. A
quantum TLS with energy bias $\varepsilon $ and tunnelling amplitude $\Delta
$\ is described by the Hamiltonian

\begin{equation}
H(t)=-\frac{\Delta }{2}\sigma _{x}-\frac{\varepsilon (t)}{2}\sigma _{z}
\label{Hamiltonian}
\end{equation}%
in terms of the Pauli matrices $\sigma _{x,z}$ ($\hbar =1$ is assumed
throughout). One generally thinks of $\Delta $ as being fixed by the system
properties, and $\varepsilon $ as being a tunable control parameter; hence $%
\varepsilon $ is a function of time $t$ and $\Delta $ is not. We assume the
monochromatic time-dependent bias%
\begin{equation}
\varepsilon (t)=\varepsilon _{0}+A\sin \omega t,  \label{eps(t)}
\end{equation}%
with amplitude $A$, frequency $\omega $ and offset $\varepsilon _{0}$.

Note that this problem with the time-dependent component along the $z$-axis
can be related to the one with the time-dependent component along the $x$%
-axis by a $\pi /2$-rotation around the $y$-axis:%
\begin{equation}
H^{\prime }(t)=e^{-i\frac{\pi }{4}\sigma _{y}}H(t)e^{i\frac{\pi }{4}\sigma
_{y}}=-\frac{\Delta }{2}\sigma _{z}+\frac{\varepsilon (t)}{2}\sigma _{x}%
\text{.}
\end{equation}%
The latter Hamiltonian is typical for problems such as an atom in a laser
field. It should also be noted that having different signs in the
Hamiltonian than the ones given above does not cause any nontrivial change
in the results.

The instantaneous eigenvalues of $H(t)$ depend on the bias $\varepsilon (t)$
as follows:%
\begin{eqnarray}
E_{\pm }(t) &=&\pm \frac{1}{2}\Omega (t),  \label{Energy(t)} \\
\Omega (t) &=&\sqrt{\Delta ^{2}+\varepsilon (t)^{2}}.
\end{eqnarray}%
The $\varepsilon $-dependence of the energy levels is shown in Fig.~\ref%
{E(e)}. In particular, it shows an avoided level crossing at $\varepsilon =0$%
.\newline

\begin{figure}[h]
\includegraphics[width=8 cm]{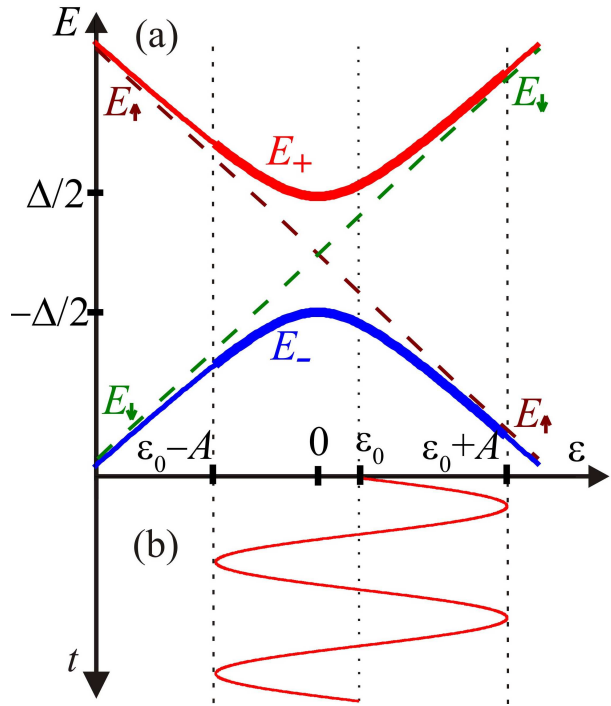}
\caption{(Color online) (a) Energy levels $E$ versus the bias $\protect%
\varepsilon$. The two solid curves (red and blue) represent the \textit{%
adiabatic energy levels}, $E_{\pm }$, which display avoided crossing with
energy splitting $\Delta $. The dashed lines show the crossing \textit{%
diabatic energy levels} $E_{\uparrow ,\downarrow }$, corresponding to the
diabatic states $\protect\varphi _{\uparrow }$ and $\protect\varphi %
_{\downarrow }$. (b) The bias $\protect\varepsilon$ represents the driving
signal, and it oscillates between $\protect\varepsilon _{\min }=\protect%
\varepsilon _{0}-A$ and $\protect\varepsilon _{\max }=\protect\varepsilon %
_{0}+A$ with a sinusoidal time dependence: $\protect\varepsilon (t)=\protect%
\varepsilon _{0}+A\sin \protect\omega t$.}
\label{E(e)}
\end{figure}

Although the TLS is a rather simple-looking model, concrete calculations on
it can present significant difficulties. Specifically, the Schr\"{o}dinger
equation can be written as a second-order differential equation with
periodic coefficients, the Hill equation, which is not solvable in analytic
closed form \cite{Grifoni:1998}. Nevertheless, different theoretical
approaches can be used to obtain approximate analytical results in various
parameter regions.

As we shall explain in detail below, one key relation for purposes of
deciding the suitability of a given theoretical approach is that between the
driving frequency $\omega $ and the minimal energy splitting $\Delta $. In
superconducting qubits, both limits, $\omega \ll \Delta $ \cite%
{Sillanpaa:2006} and $\omega \gg \Delta $ \cite{Oliver:2005}, have been
realized in recent experiments. Another key relation concerns the velocity
of passing the avoided level region, which is of the order of $A\omega $ (we
also introduce the adiabaticity parameter $\delta $ later in Eqs.~(\ref%
{P=P_LZ}-\ref{delta})). We will distinguish between the slow driving regime,
$A\omega \lesssim \Delta ^{2}$, and the fast driving regime, $A\omega \gg
\Delta ^{2}$. The natural representations for the two regimes are formulated
in the adiabatic and diabatic bases, respectively. The \textit{diabatic basis%
} $\{\varphi _{\uparrow },\varphi _{\downarrow }\}$ is formed with the
eigenstates of $\sigma _{z}$: $\sigma _{z}\varphi _{\uparrow }=\varphi
_{\uparrow }$, $\sigma _{z}\varphi _{\downarrow }=-\varphi _{\downarrow }$
(note that these states would be the eigenstates of the Hamiltonian if $%
\Delta $ vanished). The \textit{adiabatic basis} consists of the
instantaneous eigenstates of the time-dependent Hamiltonian: $H(t)\varphi
_{\pm }(t)=E_{\pm }(t)\varphi _{\pm }(t)$. The two bases are related to each
other by the time-dependent coefficients $\beta _{\pm }(t)$ (see Appendix A
for details):%
\begin{eqnarray}
\varphi _{\pm }(t) &=&\beta _{\mp }(t)\varphi _{\uparrow }\mp \beta _{\pm
}(t)\varphi _{\downarrow }, \\
\beta _{\pm }(t) &=&\sqrt{\frac{\Omega (t)\pm \varepsilon (t)}{2\Omega (t)}}.
\label{beta}
\end{eqnarray}%
Any given wave function of the quantum TLS can be decomposed in either one
of these two bases; the coefficients in the quantum superposition are the
probability amplitudes for the respective states.

\section{Theory: adiabatic-impulse model}

There are a number of theoretical approaches that can be used to study the
problem of a strongly driven TLS. Among these approaches, the
adiabatic-impulse model, which we shall explain in detail below (and in
Appendices A and B), is perhaps the most intuitive whenever the TLS is
repeatedly driven through the avoided crossing. We shall therefore put
special emphasis on this theoretical approach in this review. Three
alternative theoretical approaches, which in certain parameter regimes can
be well suited for the study of this problem, will be discussed in
Appendices C (rotating-wave approximation), D (Floquet theory) and E
(dressed-state picture).

\begin{figure}[h]
{\Huge \includegraphics[width=11cm]{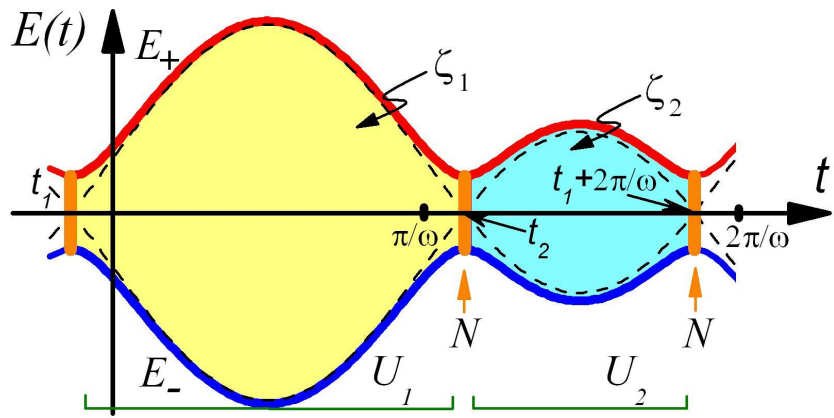} }
\caption{(Color online) Time evolution of the energy levels during one
period. The time-dependent adiabatic energy levels define a two-stage
evolution: transitions at the non-adiabatic regions, described by the
evolution matrix $N$, and the adiabatic evolution, described by the matrices
$U_{1,2}=\exp (-i\protect\zeta _{1,2}\protect\sigma _{z})$. The acquired
phases $\protect\zeta _{1,2}$ have a geometrical interpretation: they are
equal to the area under the curves, shown by the yellow and blue regions.
The diabatic energy levels, $\pm \protect\varepsilon (t)/2$, are shown by
the dashed lines.}
\label{Epm(t)}
\end{figure}

The adiabatic energy levels $E_{\pm }$ introduced in Sec.~I.C have a minimum
distance of $\Delta $, and this minimum distance is realized at times $%
t_{1,2}+2\pi n/\omega $, where $\omega t_{1}=\arcsin (-\varepsilon _{0}/A)$
and $\omega t_{2}=\pi -\omega t_{1}$, see Fig.~\ref{Epm(t)}; here $n$ is an
integer. Because the energy eigenstates, i.e.~the states of the adiabatic
basis, change rather rapidly around the avoided-crossing region and are
approximately constant far to the right or far to the left of that region,
one would intuitively expect that the system will evolve almost
adiabatically far from the points of avoided crossing and the evolution
becomes nonadiabatic in the vicinity of these points. One could therefore
make the approximation that beyond a certain boundary the evolution is
completely adiabatic. However, it is convenient not to introduce an
arbitrary boundary between the adiabatic and nonadiabatic regions and
instead consider the evolution to be adiabatic everywhere except at the
points of minimum energy separation, where the system experiences a sudden
mixing in the populations of the two energy levels. Naturally, these sudden
transitions do not occur in reality. However, provided that the system
follows a linear ramp traversing the avoided crossing region, one can still
obtain an accurate description of the net result of the adiabatic and
nonadiabatic parts of the evolution. This discretized picture only
simplifies the algebra that arises in the relevant calculations. The
calculation presented in Appendix A can be used to obtain the appropriate
assignment for the adiabatic phases and the mixing matrices at the crossing
points in order to obtain the correct expressions describing the system's
dynamics. Following \cite{Damski:2006} (see also \cite{Zurek:1996} and \cite%
{Damski:2005}) we call this picture the adiabatic-impulse approximation to
emphasize the two-stage character of the model. In particular, this name
emphasizes that the non-adiabatic transitions are described as
instantaneous, which is just a convenient description of the continuous
dynamics; see also original articles, where the adiabatic-impulse theory was
developed \cite{Delone:1985}, \cite{Averbukh:1985}, \cite{Vitanov:1996},
\cite{Garraway:1997}, \cite{Damski:2006}].

\subsection{Adiabatic evolution}

The wave function $\psi (t)$ that describes the quantum state of the TLS as
a function of time can be decomposed in the adiabatic basis (see Appendix A):%
\begin{eqnarray}
\psi (t) &=&\sum\limits_{\pm }b_{\pm }(t)\varphi _{\pm }(t)=\sum\limits_{\pm
}c_{\pm }(t)\varphi _{\pm }(t)\exp \left\{ \mp i\left( \zeta +\frac{\pi }{4}%
\right) \right\} ,  \label{decomposition} \\
\zeta &=&\frac{1}{2}\int \Omega (t)dt.
\end{eqnarray}%
Since in the adiabatic approximation $c_{\pm }$ are time-independent
coefficients between crossings, the adiabatic evolution from $t=t_{\mathrm{i}%
}$ to $t=t_{\mathrm{f}}$ (assuming that no LZ transitions are encountered)
can be described by the evolution matrix $U$, defined as follows%
\begin{eqnarray}
\mathbf{b}(t_{\mathrm{f}}) &=&U(t_{\mathrm{f}},t_{\mathrm{i}})\mathbf{b}(t_{%
\mathrm{i}}), \\
\mathbf{b}(t) &\equiv &\left(
\begin{array}{c}
b_{+} \\
b_{-}%
\end{array}%
\right) , \\
U(t_{\mathrm{f}},t_{\mathrm{i}}) &=&\left(
\begin{array}{cc}
e^{-i\zeta (t_{\mathrm{f}},t_{\mathrm{i}})} & 0 \\
0 & e^{i\zeta (t_{\mathrm{f}},t_{\mathrm{i}})}%
\end{array}%
\right) =e^{-i\zeta (t_{\mathrm{f}},t_{\mathrm{i}})\sigma _{z}}, \\
\zeta (t_{\mathrm{f}},t_{\mathrm{i}}) &=&\frac{1}{2}\int\limits_{t_{\mathrm{i%
}}}^{t_{\mathrm{f}}}\Omega (t)dt.
\end{eqnarray}%
In particular, the phases acquired during the adiabatic stages (see Fig.~\ref%
{Epm(t)}) are given by:%
\begin{equation}
\zeta _{1}=\frac{1}{2}\int\limits_{t_{1}}^{t_{2}}\Omega (t)dt,\ \ \ \ \
\zeta _{2}=\frac{1}{2}\int\limits_{t_{2}}^{t_{1}+2\pi /\omega }\Omega (t)dt.
\label{zeta_1_2}
\end{equation}

\subsection{Single passage: Landau-Zener transition}

Consider the nonadiabatic region in the vicinity of $t_{1,2}$: $%
t=t_{1,2}+t^{\prime }$, $\omega \left\vert t^{\prime }\right\vert \ll 1$.
Then the bias can be linearized:%
\begin{equation}
\varepsilon (t_{1,2}+t^{\prime })\approx \pm v t^{\prime },  \label{e_linrzd}
\end{equation}%
where%
\begin{equation}
v=A \omega \left\vert \cos \omega t_{1,2}\right\vert =A \omega \sqrt{%
1-\left( \frac{\varepsilon _{0}}{A}\right) ^{2}}.
\end{equation}

The linearized Hamiltonian
\begin{equation}
H(t^{\prime })=-\frac{\Delta }{2}\sigma _{x}\mp \frac{vt^{\prime }}{2}\sigma
_{z}  \label{H_linearzd}
\end{equation}%
is exactly that of the LZ problem. The solution of the LZ problem in terms
of the parabolic cylinder function is presented in detail in Appendix A. For
discussions of different theoretical approaches to the LZ problem, see e.g.
\cite{Delos:1972}, \cite{Benderskii:2003}, \cite{Wittig:2005}. Here, in the
main text, we also present the shortest solution, based on the fact that the
transition under the adiabatic perturbation has quasiclassical character,
where the change of the action (given by the integral $\int E(t)dt$) is
large \cite{Landau:1977}. Then the problem of the transition under the
adiabatic perturbation is formally fully analogous to the problem of the
quasiclassical over-barrier reflection.

\begin{figure}[h]
\includegraphics[width=9cm]{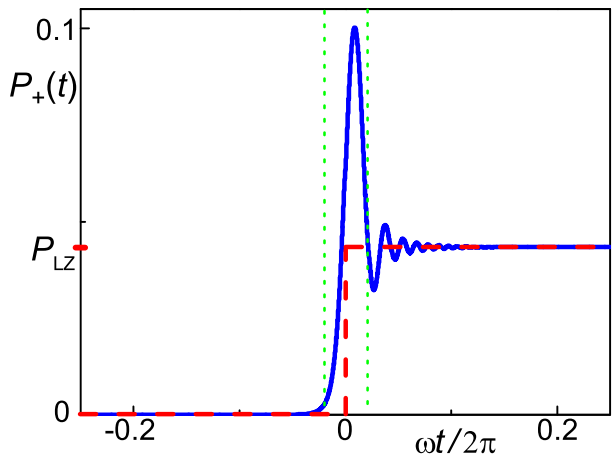}
\caption{(Color online) One-passage LZ transition. The time dependence of
the upper level occupation probability $P_{+}(t)$ is plotted for the
following parameters: $A/\Delta =10$, $\protect\omega /\Delta =0.05$, $%
\protect\varepsilon _{0}=0$. The blue solid curve presents the exact
numerical solution, while red dashed curve corresponds to the analytical
solution. The time interval between the vertical green dotted lines
corresponds to the crossover region, which is characterized by $t_{\mathrm{LZ%
}}$.}
\label{LZ_vs_t}
\end{figure}

Assuming that the system is initially in the lower energy level, the
transition probability from the lower to the upper level during the
single-sweep process is described within the quasiclassical approximation as
follows \cite{Landau:1977}. The energy levels coincide at two points in the
complex plane: at $t^{\prime }=\pm t_{0}=\pm i\Delta /v$ one finds that $%
E_{+}(t_{0})=E_{-}(t_{0})$. Accordingly, the probability $P_{+}$ that the
system ends up in the upper level is determined by the contour integral in
the plane of complex time:%
\begin{equation}
P_{+}=\exp \left( -2\text{Im}\int\limits_{0}^{t_{0}}[E_{+}(t^{\prime
})-E_{-}(t^{\prime })]dt^{\prime }\right) .
\end{equation}%
Calculating this integral for $2E_{\pm }=\pm \sqrt{\Delta ^{2}+(vt^{\prime
})^{2}}$ and $t_{0}=i\Delta /v$, we obtain the LZ transition probability:
\begin{eqnarray}
P_{+} &=&P_{\mathrm{LZ}}=\exp \left( -2\pi \delta \right) ,  \label{P=P_LZ}
\\
\delta &=&\frac{\Delta ^{2}}{4v}.  \label{delta}
\end{eqnarray}%
Equation (\ref{P=P_LZ}) describes the transition probability for an
arbitrary value of the exponent \cite{Nikitin:1996}. As the driving velocity
$v$ is changed from $0$ (adiabatic limit) to $\infty $ (sudden-change
limit), the transition probability $P_{+}$ varies from zero to unity\textit{%
. }In Fig.~\ref{LZ_vs_t} the red dashed curve shows the instantaneous
transition of the adiabatic-impulse model, with the mixing probability given
by Eq.~(\ref{P=P_LZ}). The blue solid curve is calculated numerically by
integrating the Schr\"{o}dinger equation \cite{Shevchenko:2005}. Note that
the LZ formula (\ref{P=P_LZ}) accurately describes the final upper level
occupation probability $P_{+}$, but it does not describe the transient
dynamics in the vicinity of the energy-level avoided crossing. This issue
was studied in detail in Refs. \cite{Mullen:1989}, \cite{Vitanov:1999}, and
\cite{Zenesini:2009}. In particular, it was shown that the duration of the
transition (the region between the green dotted lines in Fig.~\ref{LZ_vs_t})
can be estimated as follows:
\begin{equation}
t_{\mathrm{LZ}}\sim \frac{2}{\Delta }\sqrt{\delta }\max (1,\sqrt{\delta }).
\label{tLZ}
\end{equation}%
Here we would like to note that, as argued in \cite{Garraway:1997}, Eq.~(\ref%
{tLZ}) gives only the upper limit of the transition time, and one can think
of the transition time is being shorter than the one given in Eq.~(\ref{tLZ}%
). The conditions of the applicability of the adiabatic-impulse model, which
describes the LZS interferometry, require that the relations $\Delta /\omega
$ and/or $A/\omega $ are large \cite{Garraway:1997}:
\begin{equation}
\Delta ^{2}+A^{2}\gg \omega ^{2}.  \label{cond_of_validity}
\end{equation}%
In particular, the multiphoton fringes were shown to be described by this
model at $A\lesssim \omega \lesssim \Delta $ in \cite{Krainov:1980}.

The LZ formula (\ref{P=P_LZ}) describes the upper level occupation
probability $P_{+}$, which is the square of the absolute value of the
probability amplitude. However, when interference is relevant (as will be
discussed in detail below), one has also to know the change in the relative
phase between the two components of the wave function as the crossing region
is traversed. One can keep track of this phase by describing the LZ
transition using the (non-adiabatic) unitary evolution matrix $N$, which is
derived in Appendix A:%
\begin{equation}
\mathbf{b}(t_{1,2}+0)=N\text{ }\mathbf{b}(t_{1,2}-0),
\end{equation}%
\begin{equation}
N=\left(
\begin{array}{cc}
\sqrt{1-P_{\mathrm{LZ}}}\text{ }e^{-i\widetilde{\varphi }_{\mathrm{S}}} & -%
\sqrt{P_{\mathrm{LZ}}} \\
\sqrt{P_{\mathrm{LZ}}} & \sqrt{1-P_{\mathrm{LZ}}}\text{ }e^{i\widetilde{%
\varphi }_{\mathrm{S}}}%
\end{array}%
\right) ,  \label{N}
\end{equation}%
\begin{equation}
\widetilde{\varphi }_{\mathrm{S}}=\varphi _{\mathrm{S}}-\frac{\pi }{2},
\end{equation}%
\begin{equation}
\varphi _{\mathrm{S}}=\frac{\pi }{4}+\delta (\ln \delta -1)+\text{arg}\Gamma
(1-i\delta )\text{,}  \label{phiS}
\end{equation}%
where $\varphi _{\mathrm{S}}$\ is the so-called Stokes phase (see e.g. \cite%
{Child:1974}, \cite{Kayanuma:1997}, \cite{Wubs:2005}) and $\Gamma $ is the
gamma function.

\subsection{Double passage: St\"{u}ckelberg phase}

Consider now the double-passage process, where the avoided crossing region
is passed twice at the same speed. This double-passage problem corresponds
to one full driving cycle of the periodic driving that we are considering.
The probability for the upper level $P_{+}$ after one period is given by
Eq.~(\ref{P_I(m)}) with $n=1$:

\begin{eqnarray}
P_{+} &=&4P_{\mathrm{LZ}}(1-P_{\mathrm{LZ}})\sin ^{2}\Phi _{\mathrm{St}},
\label{with_Fi_St} \\
\Phi _{\mathrm{St}} &=&\zeta _{2}+\widetilde{\varphi }_{\mathrm{S}}\text{.}
\end{eqnarray}%
This result, that the probability of excitation $P_{+}$ is an oscillating
function of a certain phase $\Phi _{\mathrm{St}}$, was first obtained in
\cite{Landau:1932a, Stueckelberg:1932}. The St\"{u}ckelberg phase $\Phi _{%
\mathrm{St}}$ consists of two components: the first $\zeta _{2}$ is acquired
during the adiabatic evolution and the second $\widetilde{\varphi }_{\mathrm{%
S}}$ during the non-adiabatic transitions.

\begin{figure}[h]
\includegraphics[width=7cm]{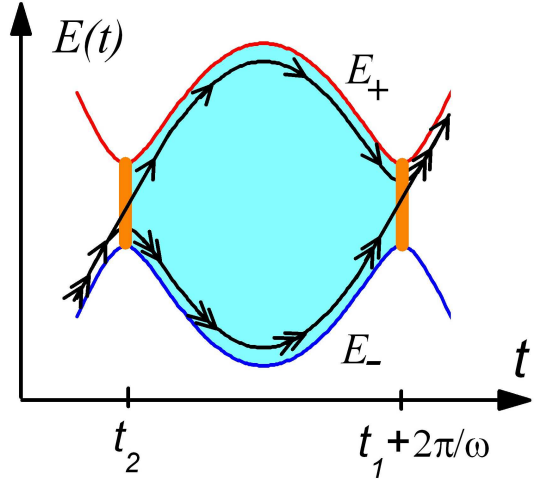}
\caption{(Color online) Double-passage transition. Adiabatic energy levels
as in Fig.~2 are plotted. The lines with one and two arrows show the two
trajectories where the transition to the upper level happens during the
first and the second passages of the avoided level crossing. Their
respective transition probabilities are given by $P_{\mathrm{LZ}}(1-P_{%
\mathrm{LZ}})$ and $(1-P_{\mathrm{LZ}})P_{\mathrm{LZ}}$, while the
interference is described by Eq.~(\protect\ref{with_Fi_St}).}
\label{with_paths}
\end{figure}

St\"{u}ckelberg oscillations were observed in atomic inelastic scattering
cross-sections \cite{Nikitin:1972} and in the microwave excitation of
Rydberg atoms \cite{Baruch:1992, Yoakum:1992}. However, in many cases these
St\"{u}ckelberg oscillations average out and can be neglected, and one only
needs to consider the averaged probability \cite{Nikitin:1996}%
\begin{equation}
\overline{P_{+}}=2P_{\mathrm{LZ}}(1-P_{\mathrm{LZ}}).  \label{without_interf}
\end{equation}%
The above expresion is the sum of two probabilities, which correspond to the
system being excited at the first or at the second passage through the
avoided-crossing region. These two components are schematically illustrated
in Fig.~\ref{with_paths} with trajectories marked by single and double
arrows, respectively.

The quantum-mechanical interference between the different LZ transitions has
the result that the net excitation probability after the double passage can
range from $0$ (destructive interference) to $4P_{\mathrm{LZ}}(1-P_{\mathrm{%
LZ}})$ (constructive interference) according to Eq.~(\ref{with_Fi_St}). The
latter probability is twice as large as the one without interference, Eq.~(%
\ref{without_interf}). This situation is analogous to the one encountered in
the Mach-Zehnder interferometer \cite{Ji:2003, Oliver:2005, Pezze:2007}.

\subsection{Multiple passage}

Let us now study a multiple-passage process, where the system passes the
avoided-crossing region periodically. Consider first the relevant time
scales. The characteristic time for an LZ transition $t_{\mathrm{LZ}}$\ can
be estimated from Eq.~(\ref{tLZ}). The time between subsequent tunnelling
events is of the order of half driving period $T/2=\pi /\omega $. LZS
interference takes place when (\textit{i}) successive LZ transitions events
do not overlap and (\textit{ii}) the phase coherence (characterized by the
dephasing time $T_{2}$) is preserved:%
\begin{equation}
t_{\mathrm{LZ}}<T/2<T_{2}.  \label{condition4LZSI}
\end{equation}%
In this subsection we assume $T/2\ll T_{2}$ and ignore decoherence. For
definiteness, as the initial condition we consider the system to be in the
ground state at time $t_{1}+0$. Then the state at time $t$, after $n$ full
periods, is described by the following evolution matrices (see Fig.~\ref%
{Epm(t)}):%
\begin{equation}
U\left( t,t_{1}+\frac{2\pi n}{\omega }\right) \left( NU_{2}NU_{1}\right) ^{n}%
\text{ for }t-\frac{2\pi n}{\omega }\in \left( t_{1},t_{2}\right) ,
\label{(I)}
\end{equation}%
\begin{equation}
U\left( t,t_{2}+\frac{2\pi n}{\omega }\right) NU_{1}\left(
NU_{2}NU_{1}\right) ^{n}\text{ for }t-\frac{2\pi n}{\omega }\in \left(
t_{2},t_{1}+\frac{2\pi }{\omega }\right) ,  \label{(II)}
\end{equation}%
\begin{equation}
U_{1,2}=\exp (-i\zeta _{1,2}\sigma _{z}).
\end{equation}

\begin{figure}[h]
\includegraphics[width=8cm]{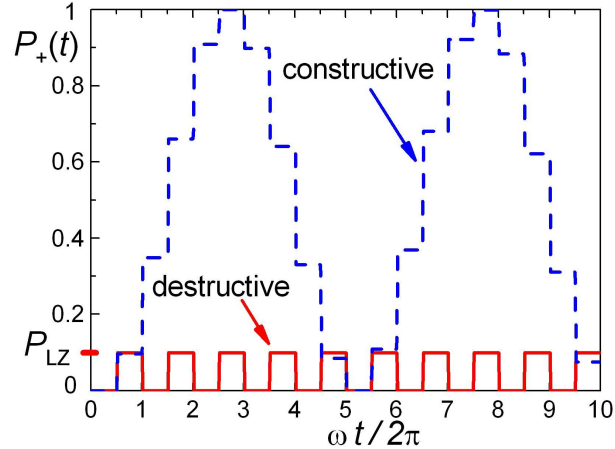}
\caption{(Color online) Constructive (blue dashed line) versus destructive
(red continuous line) interference. Time dependence of the upper level
occupation probability $P_{+}$ with constructive ($\protect\zeta _{1}=%
\protect\pi /2+k\protect\pi $) and destructive ($\protect\zeta _{1}=k\protect%
\pi $) interference at $\protect\varepsilon _{0}=0$ and $P_{\mathrm{LZ}}=0.1$%
. The TLS is taken to be initially in the lower level.}
\label{interference}
\end{figure}

One can now derive expressions for the evolution matrices in terms of the
parameters in the Hamiltonian (\ref{Hamiltonian}), as explained in Appendix
B. Using these matrices one can analyze the dynamics for any given initial
state. Figure \ref{interference} illustrates two different possibilities
that correspond to constructive and destructive interference between
consecutive crossings.\footnote{%
These cases are also clearly illustrated in the two computer animations
available online at http://dml.riken.jp/animations.php.} In addition to the
time dependence, one can also derive expressions for the average populations
of the two quantum states as functions of the system parameters, since it is
common to experimentally measure the steady-state populations of these
states. In general, one needs to make approximations in order to obtain
clear analytical results. Below we consider two opposite limits, depending
on whether the individual crossings are slow (i.e., almost adiabatic) or
fast (i.e., almost sudden). As explained in Appendix B (see also \cite%
{Ashhab:2007}), in both cases a resonance condition can be derived that
determines whether the system will exhibit oscillations between the two
basis states. This condition is given by
\begin{equation}
(1-P_{\mathrm{LZ}})\sin \zeta _{+}-P_{\mathrm{LZ}}\sin \zeta _{-}=0,
\end{equation}%
where
\begin{eqnarray}
\zeta _{+} &=&\zeta _{1}+\zeta _{2}+2\widetilde{\varphi }_{\mathrm{S}}, \\
\zeta _{-} &=&\zeta _{1}-\zeta _{2}.
\end{eqnarray}%
As we shall see shortly, this resonance condition results in drastically
different patterns in the slow- and fast-passage limits.

\subsubsection{Slow-passage limit}

\begin{figure}[h]
\includegraphics[width=8cm]{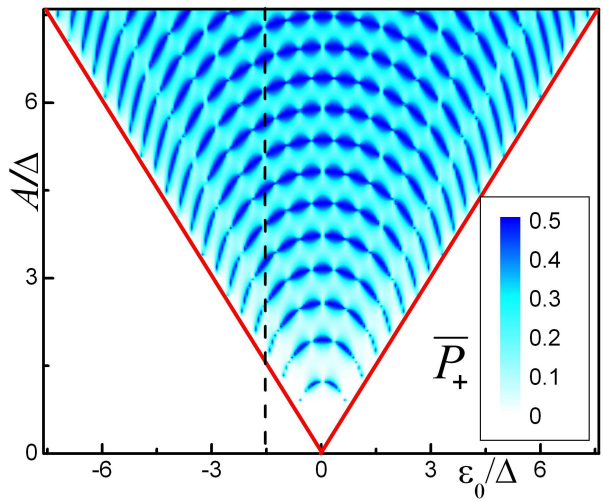}
\caption{(Color online) Slow-driving LZS interferometry for $A\protect\omega %
\lesssim \Delta ^{2}$. The time averaged upper level occupation probability $%
\overline{P_{+}}$\ as a function of the energy bias $\protect\varepsilon %
_{0} $ and the driving amplitude $A$. The graph is calculated with Eq.~(%
\protect\ref{Pp_with_offset}) for $\protect\omega /\Delta =0.32<1$. The
inclined red lines mark the region of the validity of the theory: $\protect%
\varepsilon _{0}<A$, which means that the system experiences avoided level
crossings. Outside of this region the excitation probability is negligibly
small. The vertical dashed line shows the alteration of the excitation
maxima and minima.}
\label{Fig:LZSI_for_HUT}
\end{figure}

In the limit $\delta \gg 1$, $P_{\mathrm{LZ}}\ll 1$ and the resonance
condition reduces to
\begin{equation}
\zeta _{+}=\zeta _{1}+\zeta _{2}+2\widetilde{\varphi }_{\mathrm{S}}=k\pi ,
\label{res_cond}
\end{equation}%
for any integer $k$, and $\widetilde{\varphi }_{\mathrm{S}}\approx -\pi /2$.
Going further and deriving an analytic expression in terms of $\Delta $, $%
\varepsilon _{0}$ and $A$ is complicated by the fact that the integrals that
determine $\zeta _{1}$ and $\zeta _{2}$ cannot be evaluated in closed
analytic form. Numerical calculations, however, are straightforward. The
resonance condition (\ref{res_cond}) describes arcs around the point $%
A=\varepsilon _{0}=0$ (see Fig.~\ref{Fig:LZSI_for_HUT}), where we treat $%
\Delta $ as a fixed parameter. Taking the system to be initially in the
lower energy level, the average occupation probability of the upper level is
given by (see Eq.~(\ref{ShIF}) in Appendix B):
\begin{equation}
\overline{P_{+}}=\frac{P_{\mathrm{LZ}}(1+\cos \zeta _{+}\cos \zeta _{-})}{%
\sin ^{2}\zeta _{+}+2P_{\mathrm{LZ}}(1+\cos \zeta _{+}\cos \zeta _{-})}.
\label{Pp_with_offset}
\end{equation}%
From this expression one can see that on resonance (i.e., when $\sin \zeta
_{+}=0$) the upper level occupation probability is maximal, $\overline{P_{+}}%
=1/2$. One can also see that the widths of the resonance lines are modulated
by the numerator in Eq.~(\ref{Pp_with_offset}); it tends to zero at points
where
\begin{equation}
\cos \zeta _{+}\cos \zeta _{-}=-1.  \label{antiresonance}
\end{equation}

Figure \ref{Fig:LZSI_for_HUT} illustrates the interferometric pattern
obtained from Eq.~(\ref{Pp_with_offset}). In this figure we use the
parameters of \cite{Sillanpaa:2006} (see also Table \ref{table1}). Note that
for any value of the energy bias offset $\varepsilon _{0}$, the maxima and
minima alternate with increasing driving amplitude $A$ (see the colour
variation along the vertical dashed line in Fig.~\ref{Fig:LZSI_for_HUT}), as
was studied by \cite{Shytov:2003} and \cite{Shevchenko:2005}.

In the case of zero offset $\varepsilon _{0}=0$, one can go a little bit
further with the analytic derivation. Taking $\varepsilon _{0}=0$ we have $%
\zeta _{1}+\zeta _{2}\approx 2A/\omega $ and $\zeta _{1}-\zeta _{2}=0$ and
the resonance condition (Eq.~(\ref{res_cond})) is satisfied when
\begin{equation}
2A/\omega =k\pi .  \label{res_cond_slow}
\end{equation}%
One can alternatively analyze the dynamics in the time domain. When $%
\varepsilon _{0}=0$, $t_{1}=0$ and $t_{2}=\pi /\omega $, and from Eq.~(\ref%
{P_I(m)}) in the first approximation in $P_{\mathrm{LZ}}$ it follows that
\begin{equation}
P_{+}(t)=4\sin ^{2}\zeta
_{1}P_{\mathrm{LZ}}(1-P_{\mathrm{LZ}})\frac{\sin ^{2}n\phi }{\sin
^{2}\phi }= \frac{P_{\mathrm{LZ}}}{\cos ^{2}\zeta
_{1}+P_{\mathrm{LZ}}\sin ^{2}\zeta _{1}}\sin ^{2}n\phi ,
\label{with_N}
\end{equation}%
\begin{equation}
\zeta _{1}=\frac{1}{2}\int\limits_{0}^{\pi /\omega }\!\Omega (t)dt=\frac{%
\sqrt{\Delta ^{2}+A^{2}}}{\omega }\;\widetilde{E}\!\left( \frac{A}{\sqrt{%
\Delta ^{2}+A^{2}}}\right) ,  \label{zeta1}
\end{equation}%
where $\widetilde{E}(x)$ is the full elliptic integral of the second kind.

Let us now analyze the expression for $P_{+}(t)$, Eq.~(\ref{with_N}). The
prefactor of $\sin ^{2}n\phi $ has a minimum at $\zeta _{1}=k\pi $ (where
this prefactor is equal to $P_{\mathrm{LZ}}$) and a maximum at $\zeta _{1}=%
\frac{\pi }{2}+k\pi $ (where it is equal to $1$). Thus, the former case ($%
\zeta _{1}=k\pi $) corresponds to destructive interference and the latter
case ($\zeta _{1}=\frac{\pi }{2}+k\pi $) corresponds to constructive
interference. These two cases are illustrated in Fig.~\ref{interference}.
Note that the latter case corresponds to the coherent destruction of
tunnelling, which has been studied extensively in the literature [\cite%
{Grossmann:1991a, Grossmann:1991b, Grossmann:1992}, \cite{Kayanuma:2008},
\cite{Eckel:2009}, \cite{Ho:2009}, \cite{Wubs:2010}]. From Eq.~(\ref{with_N}%
) for the cases of constructive and destructive interference we obtain:%
\begin{equation}
\text{constructive: }P_{+}(t)=\sin ^{2}\left( \sqrt{P_{\mathrm{LZ}}}%
\,n\right) \text{, \ }\zeta _{1}=\frac{\pi }{2}+k\pi ,  \label{construct}
\end{equation}%
\begin{equation}
\text{destructive: }P_{+}(t)=P_{\mathrm{LZ}}\sin ^{2}\left( \frac{\pi }{2}%
n\right) \text{, \ }\zeta _{1}=k\pi .  \label{destruct}
\end{equation}%
The dynamics (i.e., the time dependence of $P_{+}$) in the cases of
constructive and destructive interference is illustrated in Fig.~\ref%
{interference}. The resonance condition $\zeta _{1}=\frac{\pi }{2}+k\pi $
with Eq.~(\ref{zeta1}) in the limit of strong driving (i.e.,~$A\gg \Delta $)
takes the form \cite{Shevchenko:2006}:%
\begin{equation}
2A/\omega =\pi (2k+1).  \label{res_cond_slow_large_A_zero_epsilon}
\end{equation}%
The positions of the resonances depend on the ratio $A/\omega $ but not on $%
\Delta $, as was studied in detail by \cite{Shytov:2003}. Note that Eq.~(\ref%
{res_cond_slow}) predicts more resonance peaks than Eq.~(\ref%
{res_cond_slow_large_A_zero_epsilon}). The explanation of this discrepancy
is that at $\varepsilon _{0}=0$ and for even values of $k$ the resonance
peaks of Eq.~(\ref{res_cond_slow}) disappear according to Eq.~(\ref%
{antiresonance}), and we recover Eq.~(\ref%
{res_cond_slow_large_A_zero_epsilon}). Interestingly, if we take Eq.~(\ref%
{zeta1}) in the limit of weak driving (i.e.,~$A\ll \Delta $), we obtain:%
\begin{equation*}
\Delta =(2k+1)\omega .
\end{equation*}%
This relation describes the odd multiphoton resonances, as studied by \cite%
{Shirley:1965} and \cite{Krainov:1980}, where the energy level separation $%
\Delta $ is an odd multiple of a photon energy $\omega $. We have therefore
obtained the correct resonance condition, even though the weak-driving limit
is outside the regime of validity of the above derivation, which is based on
the interference of multiple LZ crossings.

\subsubsection{Fast-passage limit}

In the limit $\delta \ll 1$, $(1-P_{\mathrm{LZ}})\ll 1$ and the resonance
condition reduces to
\begin{equation}
\zeta _{-}=\zeta _{1}-\zeta _{2}=k\pi .  \label{res_cond_fast_zetas}
\end{equation}%
Unlike the slow-passage limit, one can now derive an approximate analytic
expression for $\zeta _{1}-\zeta _{2}$ in the large-amplitude limit (the
reason is that the parts of the yellow and blue areas in Fig.~2 that are
difficult to calculate cancel when taking the difference between the two
areas):
\begin{equation}
\zeta _{-}=\zeta _{1}-\zeta _{2}\approx \frac{1}{2}\int_{t_{1}}^{t_{1}+2\pi
/\omega }\left( \varepsilon _{0}+A\sin \omega t\right) dt=\frac{\pi
\varepsilon _{0}}{\omega }.
\end{equation}%
The resonance condition is therefore given simply by
\begin{equation}
\varepsilon _{0}=k\omega .  \label{res_cond_fast_basic_parameters}
\end{equation}

If the system is driven on resonance (with a given value $k$), and taking
the TLS to be initially in the lower diabatic state (i.e.,~$\varphi
_{\uparrow }$ or $\varphi _{\downarrow }$ depending on the sign of $%
\varepsilon _{0}$), the system exhibits full oscillations between $\varphi
_{\uparrow }$ and $\varphi _{\downarrow }$:
\begin{equation}
P_{\mathrm{up}}^{(k)}(t)=\frac{1}{2}\left( 1-\cos \Omega _{\mathrm{R}%
}^{(k)}t\right) .
\end{equation}%
The oscillation frequency can be calculated from the effective rotation
angle in the full-cycle evolution matrix $(NU_{2}NU_{1})$, as was done in
\cite{Ashhab:2007}. In the limit $A\gg \varepsilon _{0}$ we find that
\begin{equation}
\Omega _{\mathrm{R}}^{(k)}\approx \Delta \sqrt{\frac{2\omega }{\pi A}}%
\left\vert \cos \left( \frac{A}{\omega }-k\frac{\pi }{2}-\frac{\pi }{4}%
\right) \right\vert .  \label{Rabi_frequency_AI_fast}
\end{equation}%
As in the slow-passage limit, the Rabi frequency exhibits periodic behavior
in the fast-passage limit as well. In particular, it vanishes whenever the
cosine function in Eq.~(\ref{Rabi_frequency_AI_fast}) vanishes. The
interferometric pattern that results in this case is shown in Fig.~7.

Above we have used the adiabatic-impulse approach to analyze the response of
the TLS to the driving field. To demonstrate another approach, next we
consider the results of the rotating-wave approximation (RWA); for more
details see Appendix C.

If decoherence processes are not taken into account, the TLS is described by
the Schr\"{o}dinger equation. Its solution gives the probability of the
upper diabatic state $P_{\mathrm{up}}$, which exhibits $k$-photon Rabi
oscillations:%
\begin{eqnarray}
P_{\mathrm{up}}^{(k)}(t) &=&\frac{\Delta _{k}^{2}}{2\Omega _{\mathrm{R}%
}^{(k)2}}\left( 1-\cos \Omega _{\mathrm{R}}^{(k)}t\right) ,\text{\ } \\
\Delta _{k} &=&\Delta J_{k}\left( \frac{A}{\omega }\right) ,  \label{Delta_k}
\\
\Omega _{\mathrm{R}}^{(k)} &=&\sqrt{(k\omega -\varepsilon _{0})^{2}+\Delta
_{k}^{2}},
\end{eqnarray}%
where $J_{k}$\ is the Bessel function. Note that the $k$-photon Rabi
frequency is described by a power law at small driving, $A/\omega \ll 1$;
namely, at $\varepsilon _{0}=k\omega $:
\begin{equation}
\Omega _{\mathrm{R}}^{(k)}\ =\ \Delta _{k}\ \approx \ \frac{\Delta }{k!}%
\left( \frac{A}{2\omega }\right) ^{k}.
\end{equation}%
Changing the system parameters ($\varepsilon _{0}$ or $\omega $), we pass
through different $k$-photon resonances. Thus, the time-averaged probability
$\overline{P}_{\mathrm{up}}$ can be described by the sum of the
Lorentzian-shape $k$-photon resonances:%
\begin{equation}
\overline{P}_{\mathrm{up}}\ =\ \sum_{k}\overline{P}_{\mathrm{up}}^{(k)}=\
\frac{1}{2}\sum\limits_{k}\frac{\Delta _{k}^{2}}{(k\omega -\varepsilon
_{0})^{2}+\Delta _{k}^{2}}.  \label{P}
\end{equation}%
The position of the resonances is (quasi-) periodic in the parameters $%
\varepsilon _{0},$ $\omega ^{-1},$ $A$; this corresponds to the multiphoton
relation $\varepsilon _{0}=k\omega $ and to the quasi-periodic character of
the Bessel functions $J_{k}(A/\omega )$ for large value of the argument:
\begin{equation}
J_{k}\left( \frac{A}{\omega }\right) \approx \sqrt{\frac{2\omega }{\pi A}}%
\cos \!\left( \frac{A}{\omega }-\frac{\pi }{4}(2k+1)\right) .  \label{bessel}
\end{equation}%
Note that the oscillations described by Eq.~(\ref{bessel}) represent the
counterpart to the St\"{u}ckelberg oscillations of the double-passage
problem \cite{Nikitin:1996}.

%\begin{widetext}

\begin{figure}[h]
\includegraphics[width=8cm]{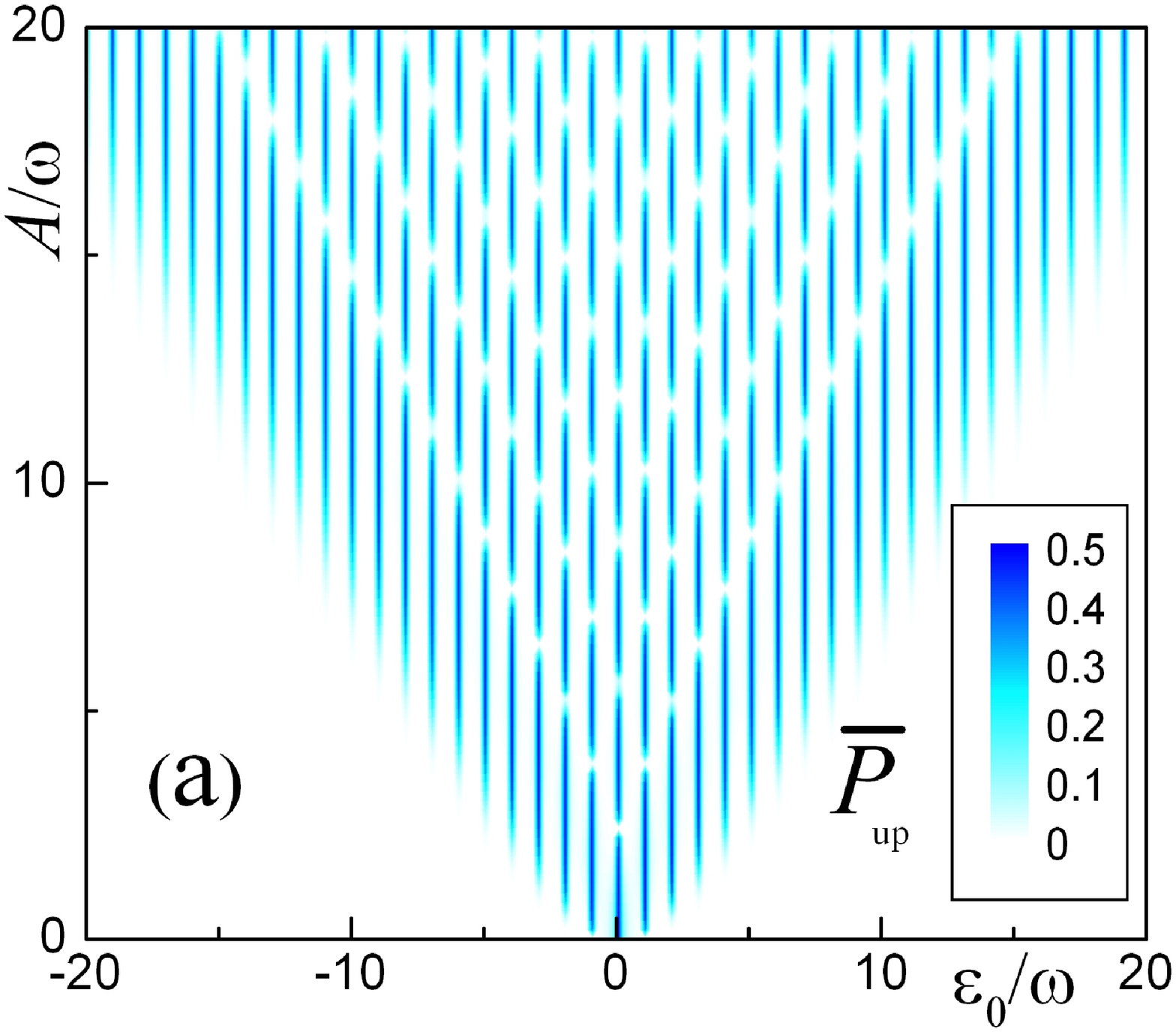} %
\includegraphics[width=8cm]{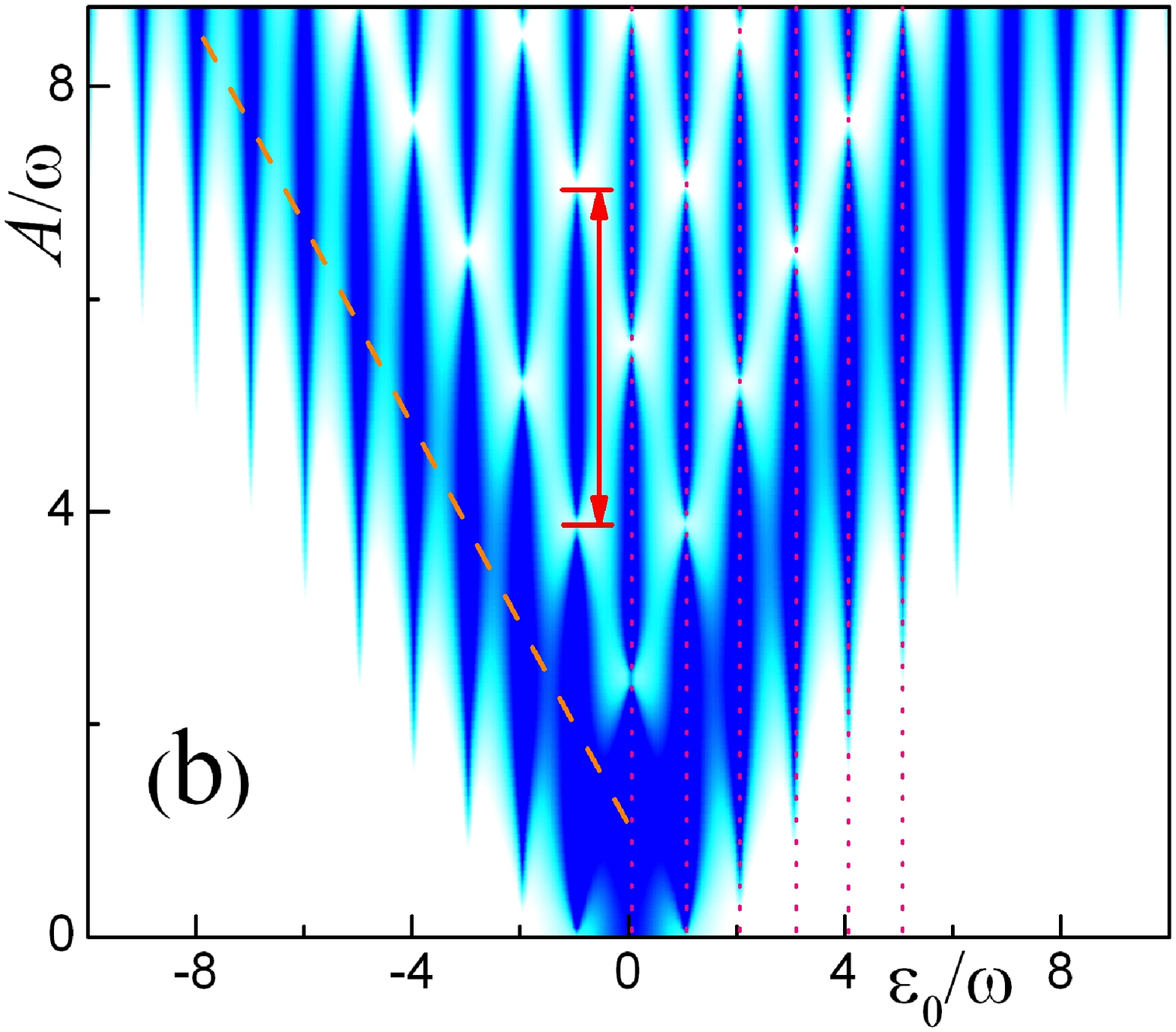}
\caption{(Color online) Fast-driving LZS interferometry for $A\protect\omega %
\gg \Delta ^{2}$: dependence of the time-averaged upper diabatic state
occupation probability $\overline{P}_{\mathrm{up}}$ on $\protect\varepsilon %
_{0}/\protect\omega $ and $A/\protect\omega $. The graphs are plotted using
Eq.~(\protect\ref{P_stat_@T=0}) for $\protect\omega /\Delta =300\gg 1$, $%
\protect\omega T_{1}/(2\protect\pi )=2.4\times 10^{4}$ and $\protect\omega %
T_{2}/(2\protect\pi )=24$ (a) and $\protect\omega /\Delta =1.14>1$, $\protect%
\omega T_{1}/(2\protect\pi )=\protect\omega T_{2}/(2\protect\pi )=6$ (b).
Several multiphoton resonances are shown by the vertical pink dotted lines
at $\protect\varepsilon _{0}=k\protect\omega $ (for $k=0,1,2,...,5$, only)
modulated by Bessel functions. The vertical red double-arrow in (b) shows
the distance between two consecutive zeros of the Bessel function.}
\label{Fig:LZSI_for_MIT_and_IPHT}
\end{figure}

%\end{widetext}

Note here that the results derived in the adiabatic-impulse approach and
those of the RWA are essentially the same. These results can also be derived
using Floquet theory \cite{Son:2009} (see also Appendix D) or using the
dressed-state picture, which represent the hybridization of the qubit and
photon degrees of freedom \cite{Liu:2006, Wilson:2007} (see also Appendix E).

Relaxation can be described using the Bloch equations for the reduced
density matrix; the influence of the environment is taken into account using
the phenomenological energy and phase relaxation times, $T_{1}$ and $T_{2}$.
In Appendix C we obtain the stationary solution of the Bloch equations,
which at zero temperature results in the excitation probability:%
\begin{equation}
\overline{P}_{\mathrm{up}}=\frac{1}{2}\sum_{k}\frac{\Delta _{k}^{2}}{\frac{1%
}{T_{1}T_{2}}+\frac{T_{2}}{T_{1}}(k\omega -\varepsilon _{0})^{2}+\Delta
_{k}^{2}}.  \label{P_stat_@T=0}
\end{equation}%
In the limit where the dephasing and relaxation times are large and equal to
each other, i.e.,~$T_{1}=T_{2}\rightarrow \infty $, the above expression
(somewhat accidentally) coincides with the time averaged solution of the Schr%
\"{o}dinger equation, Eq.~(\ref{P}). In Fig.~\ref{Fig:LZSI_for_MIT_and_IPHT}
we plot the probability $\overline{P}_{\mathrm{up}}$ with Eq.~(\ref%
{P_stat_@T=0}) for the parameters of Refs. \cite{Oliver:2005} and \cite%
{Izmalkov:2008} (see Table \ref{table1}).

We emphasize here that the distinction between the interference patterns in
Figs.~\ref{Fig:LZSI_for_HUT} and \ref{Fig:LZSI_for_MIT_and_IPHT} is
determined not by the ratio $\omega/\Delta$, but rather by the ratio $\omega%
\sqrt{A^2-\varepsilon_0^2}/\Delta^2$. This point is illustrated in Fig.~\ref%
{Fig:From_slow_to_fast_passage}, where one can see the crossover from the
slow-passage to the fast-passage limit as the driving amplitude $A$ is
increased. Note that in order to see this crossover for a fixed frequency,
one needs to take a small value of $\omega/\Delta$.

%\begin{widetext}
\begin{figure}[h]
\includegraphics[width=14cm]{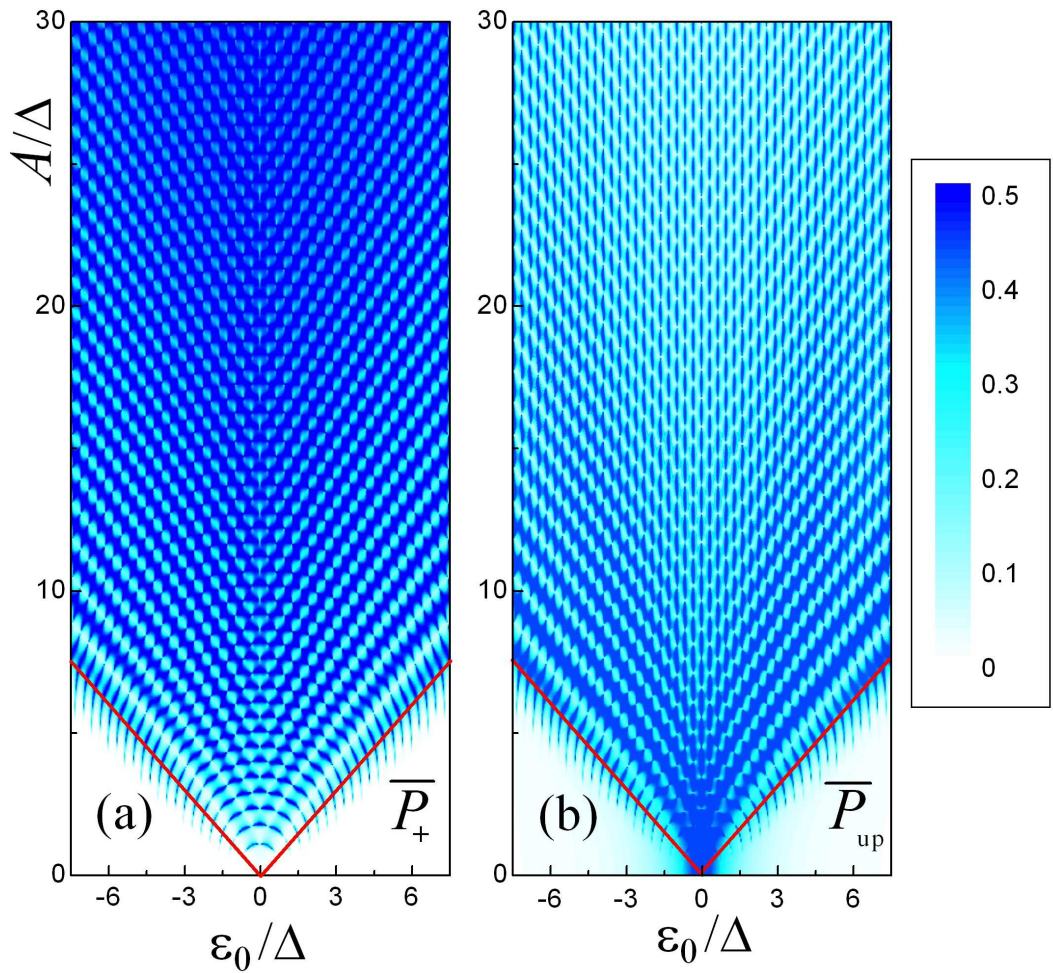}
\caption{(Color online) Crossover from the slow-passage limit
(bottom part of the figure) to the fast-passage limit (top part of
the figure) as the driving amplitude $A$ is increased. On the left
the steady-state probability $\overline{P}_{+}$ of the
\textit{adiabatic} excited state is plotted as a
function of bias offset $\protect\varepsilon _{0}$ and driving amplitude $A$%
. On the right the probability $\overline{P}_{\mathrm{up}}$ of the upper
\textit{diabatic} state is plotted. One can see that the resonance features
are clearest in the adiabatic basis for slow passage and in the diabatic
basis for fast passage. The ratio $\protect\omega /\Delta $ is equal to
0.32, and there is no decoherence. Note that panel (a) differs from Fig.~%
\protect\ref{Fig:LZSI_for_HUT} because that figure was generated using Eq.~(%
\protect\ref{Pp_with_offset}), whereas in this figure we
numerically solve the Bloch equations.}
\label{Fig:From_slow_to_fast_passage}
\end{figure}
%\end{widetext}

The analysis of the positions of the multiphoton resonances allows to do
spectroscopy and to obtain the parameters of the TLS. To calibrate the power
of the generator one can use the distance between the minima or maxima in
the St\"{u}ckelberg oscillations (one of them is shown in Fig.~\ref%
{Fig:LZSI_for_MIT_and_IPHT} with the vertical red double-arrow), or
alternatively, the slope of the interference fringes (shown by the orange
dashed line in Fig.~\ref{Fig:LZSI_for_MIT_and_IPHT}). The analysis of the
interference fringes also allows to obtain the parameters that characterize
the relaxation of the system due to the interaction with the environment.
This was theoretically discussed in \cite{Shytov:2003} and experimentally
realized in [\cite{Oliver:2005}, \cite{Sillanpaa:2006}, \cite{Wilson:2007},
\cite{Izmalkov:2008}]. The longitudinal $T_{1}$ and transverse $T_{2}$
relaxation times can be extracted from the shapes of the resonances in Eq.~(%
\ref{P_stat_@T=0}). Moreover the effective temperature can be estimated
using Eqs.~(\ref{P_stat}-\ref{Z_stat}) from the shape of the first
resonance, where $\Delta E\approx \left\vert \varepsilon _{0}\right\vert
\approx \omega $ which is typically of the order of $1$ GHz and smaller for
superconducting TLSs.

In practice it can be useful to perform more detailed calculations beyond
the stationary solution in Eq.~(\ref{P_stat_@T=0}). We note that the
relaxation times, $T_{1}$ and $T_{2}$, can be introduced phenomenologically
as two fitting parameters (as in \cite{Oliver:2005}) or calculated within
certain model for dissipating environment (as in \cite{Sillanpaa:2006}). For
superconducting qubits a convenient model is the spin-boson model, which
models the environment as a bath of harmonic oscillators \cite{Makhlin:2001}%
, \cite{Wendin:2007}, \cite{McDermott:2009}, \cite{Wilson:2010}; in this
model both $T_{1}$ and $T_{2}$ are parameter-dependent and determined by the
strength of dissipation. However in many cases the relaxation can be
described by effective parameter-independent times $T_{1}$ and $T_{2}$; in
this paper for plotting Fig.~\ref{Fig:LZSI_for_MIT_and_IPHT} we have taken
the relaxation times, $T_{1}$ and $T_{2}$, for simplicity being constant.
Also, the Bloch equations (\ref{Bloch}) can be solved exactly as, e.g., in
\cite{Shevchenko:2008}, which allows one to relax the assumption $\Delta \ll
\omega $; the positions of the multiphoton resonances are then defined by
the relation: $\Delta E\approx \sqrt{\Delta ^{2}+\varepsilon _{0}^{2}}%
\approx k\omega $. Note that the inhomogeneity of the field due to
low-frequency noise cannot be directly described within the Bloch equations
\cite{Abragam:1961}; it results in the broadening of the Lorentzian-shaped
resonances and can be described as the additional term in the expression for
the width of the resonances, as in Ref. \cite{Oliver:2005}.

\subsection{Decoherence}

When studying the effect of decoherence on the dynamics of a quantum system,
it is perhaps most common to start from the case of zero decoherence and
slowly increase the decoherence rates. In the problem of LZS interferometry,
we find it simpler to think about the effect of \textit{coherence} rather
than the effect of decoherence. We study the effect of coherence by
considering the steady-state populations of the two quantum states of the
TLS (as was done in \cite{Berns:2006}). We note here that in the
slow-passage limit we analyze the populations of the adiabatic-basis states
whereas in the fast-passage limit we analyze the populations of the
diabatic-basis states.

\begin{figure}[h]
\includegraphics[width=8cm]{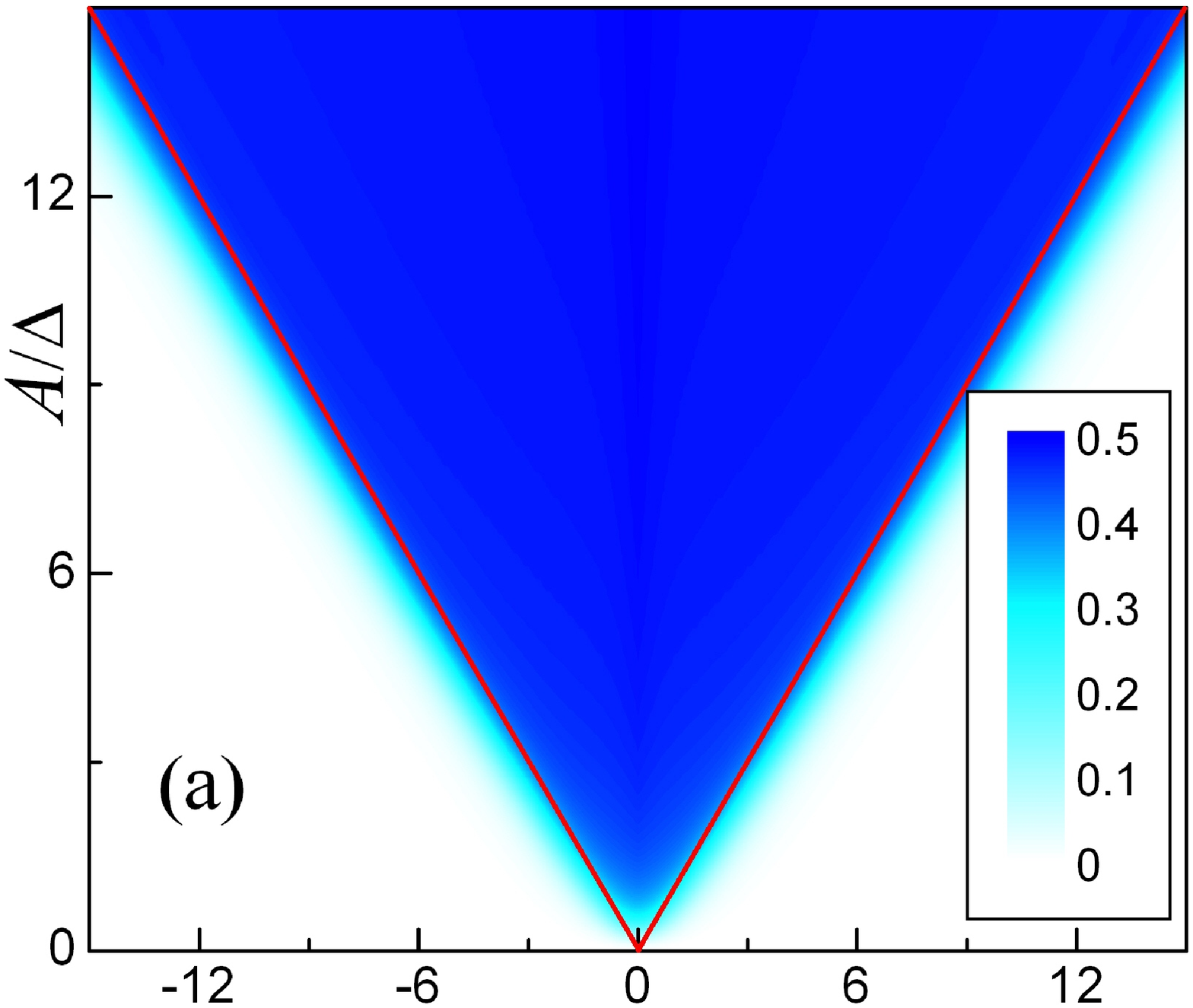} \includegraphics[width=8cm]{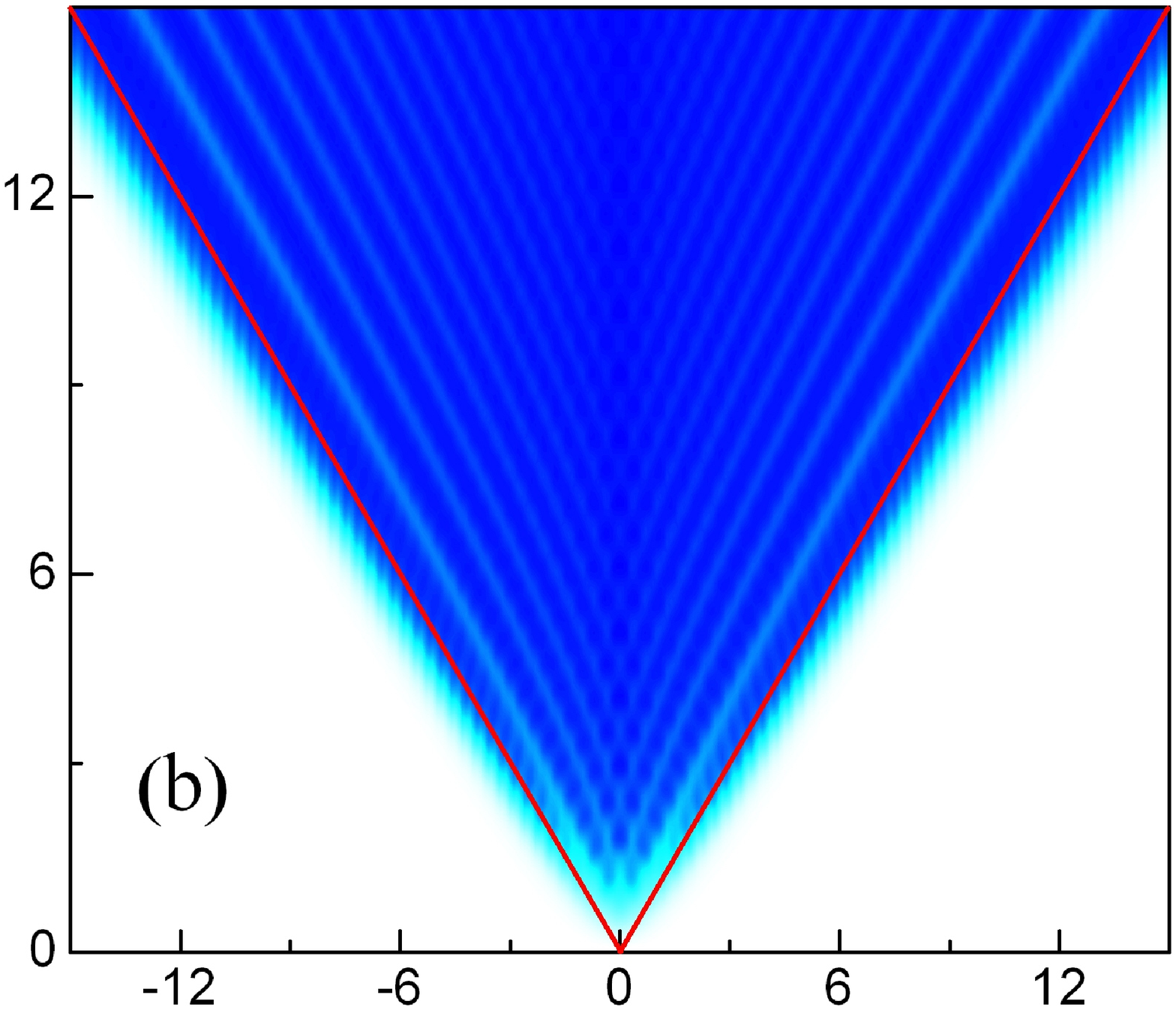}
\includegraphics[width=8cm]{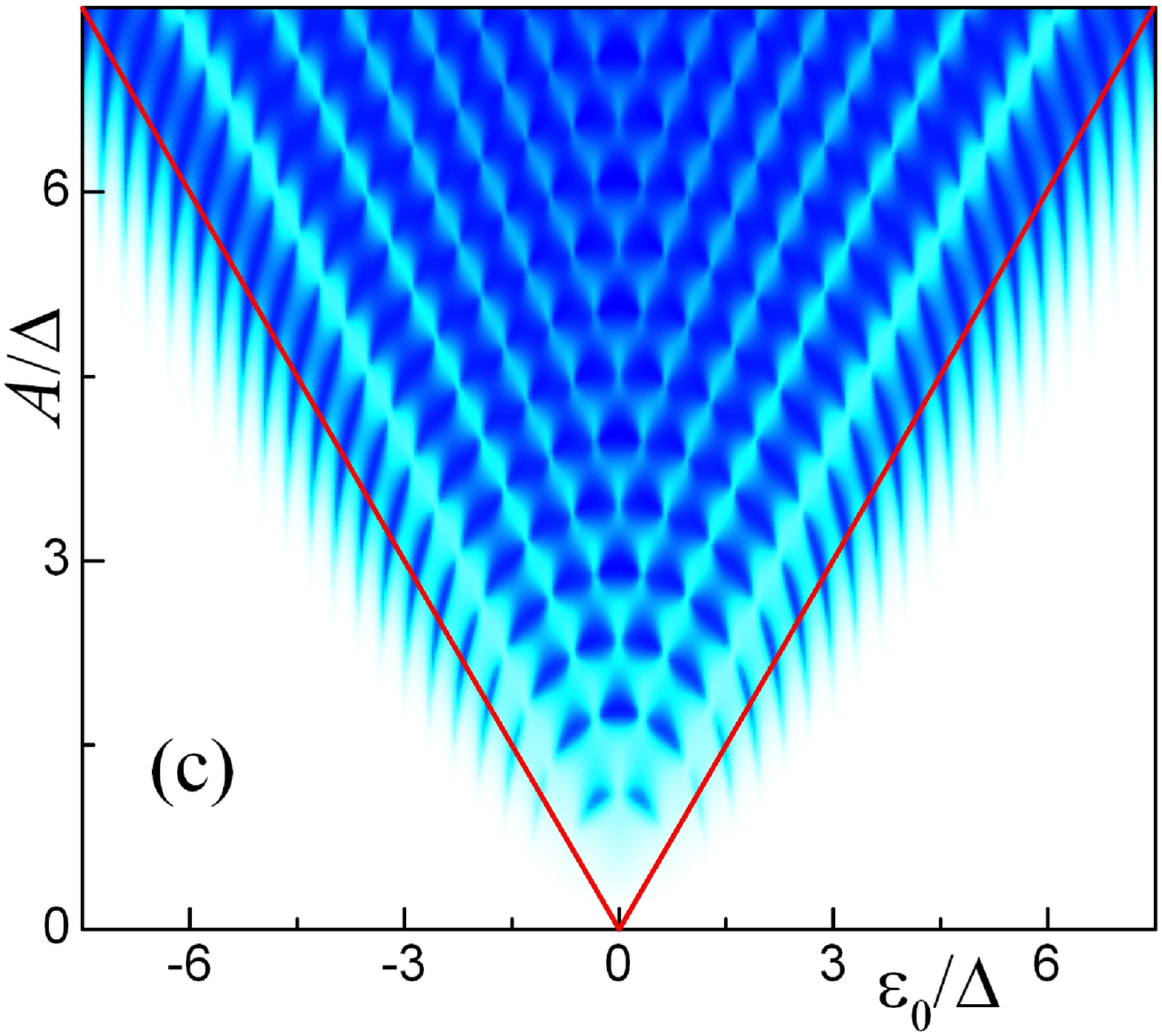} \includegraphics[width=8cm]{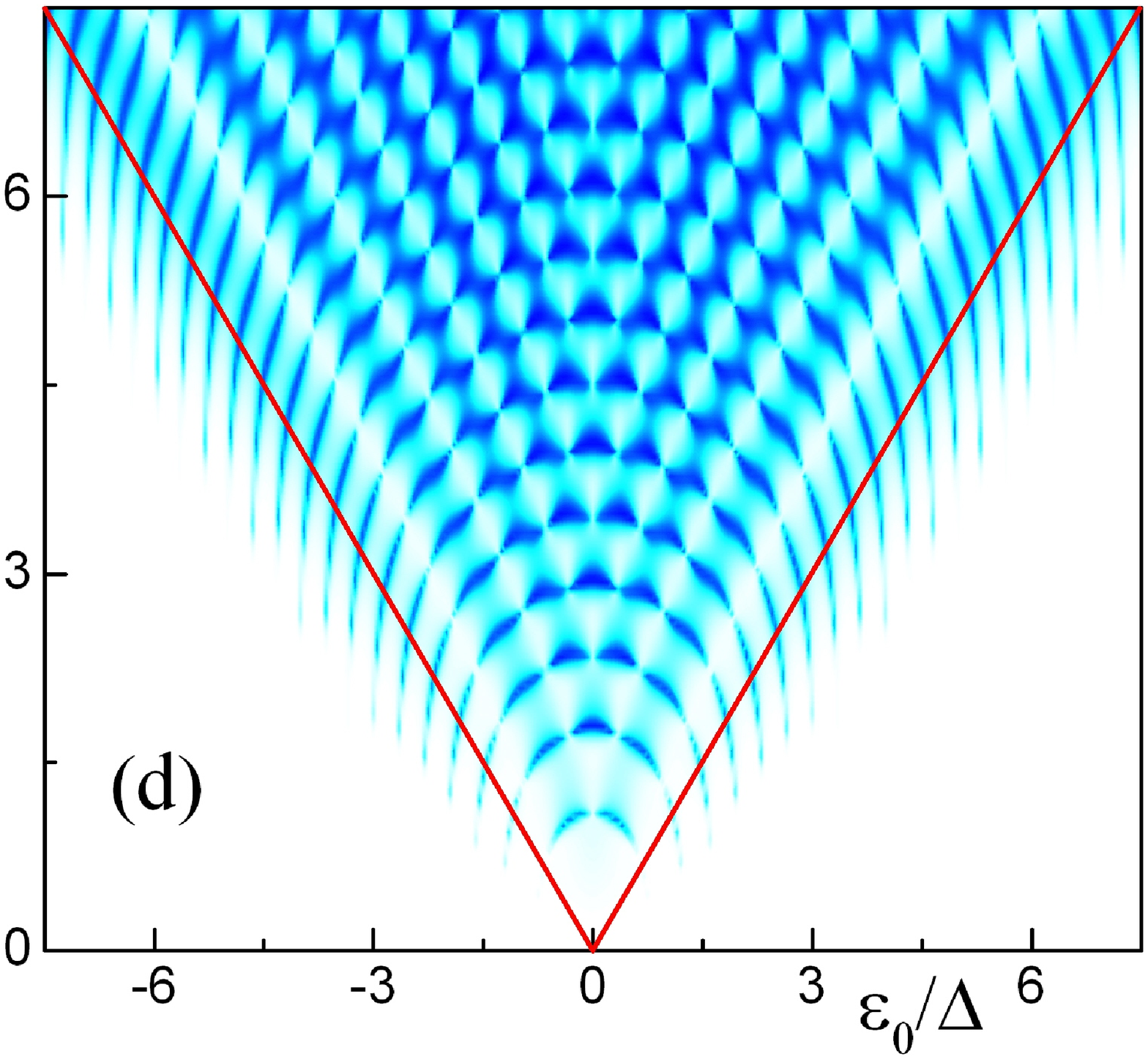}
\caption{(Color online) Same as in Fig.~\protect\ref{Fig:LZSI_for_HUT}
(i.e.~LZS interferometry with low-frequency driving), but including the
effects of decoherence. The time averaged upper level occupation probability
$\overline{P_{+}}$\ was obtained numerically from the Bloch equations with
the Hamiltonian (\protect\ref{Hamiltonian}). The dephasing time $T_{2}$ is
given by $\protect\omega T_{2}/(2\protect\pi )=0.1$ in (a), $1$ in (b), $5$
in (c) and $T_{2}=2T_{1}$ in (d). The relaxation time is given by $\protect%
\omega T_{1}/(2\protect\pi )=10$.}
\label{Fig:low_freq_with_decoherence}
\end{figure}

Starting from the limit of strong decoherence, no interference between the
different LZ transitions occurs. The result is then simple. If the driving
amplitude is large enough to take the TLS through the avoided crossing, each
crossing will act as a small \textquotedblleft kick\textquotedblright\ to
the populations of the two states, and these kicks will eventually add up
such that the TLS reaches a steady state with equal population of the two
energy levels (the ground state can have a higher population if relaxation
is substantial). If the driving amplitude $A$ is smaller than the bias
offset $\left\vert \varepsilon _{0}\right\vert $, no LZ transitions will
occur, and the system will remain in the lower level. This situation is
depicted in Figs.~\ref{Fig:low_freq_with_decoherence}(a) and \ref%
{Fig:high_freq_with_decoherence}(a).

\begin{figure}[h]
\includegraphics[width=8cm]{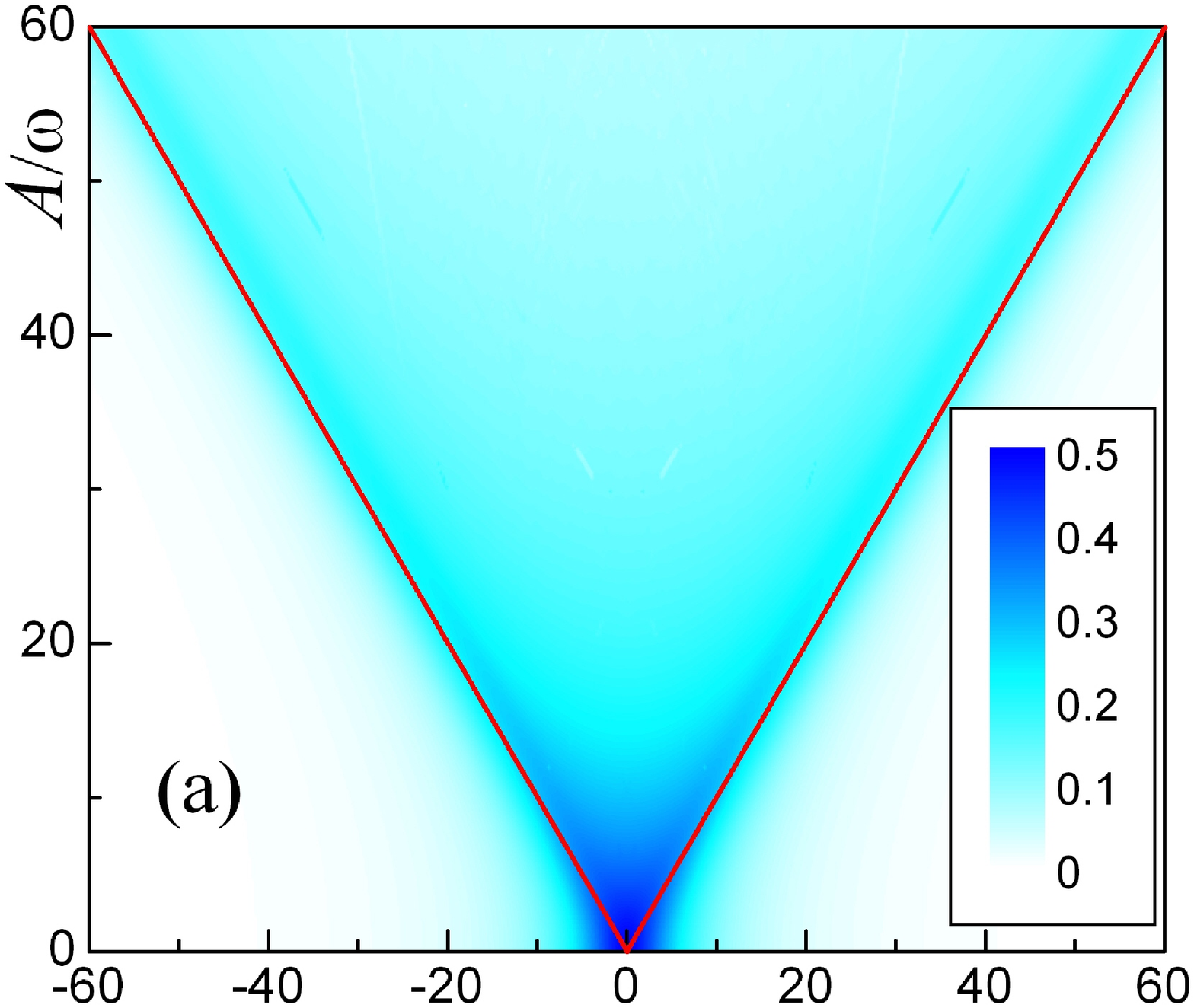} %
\includegraphics[width=8cm]{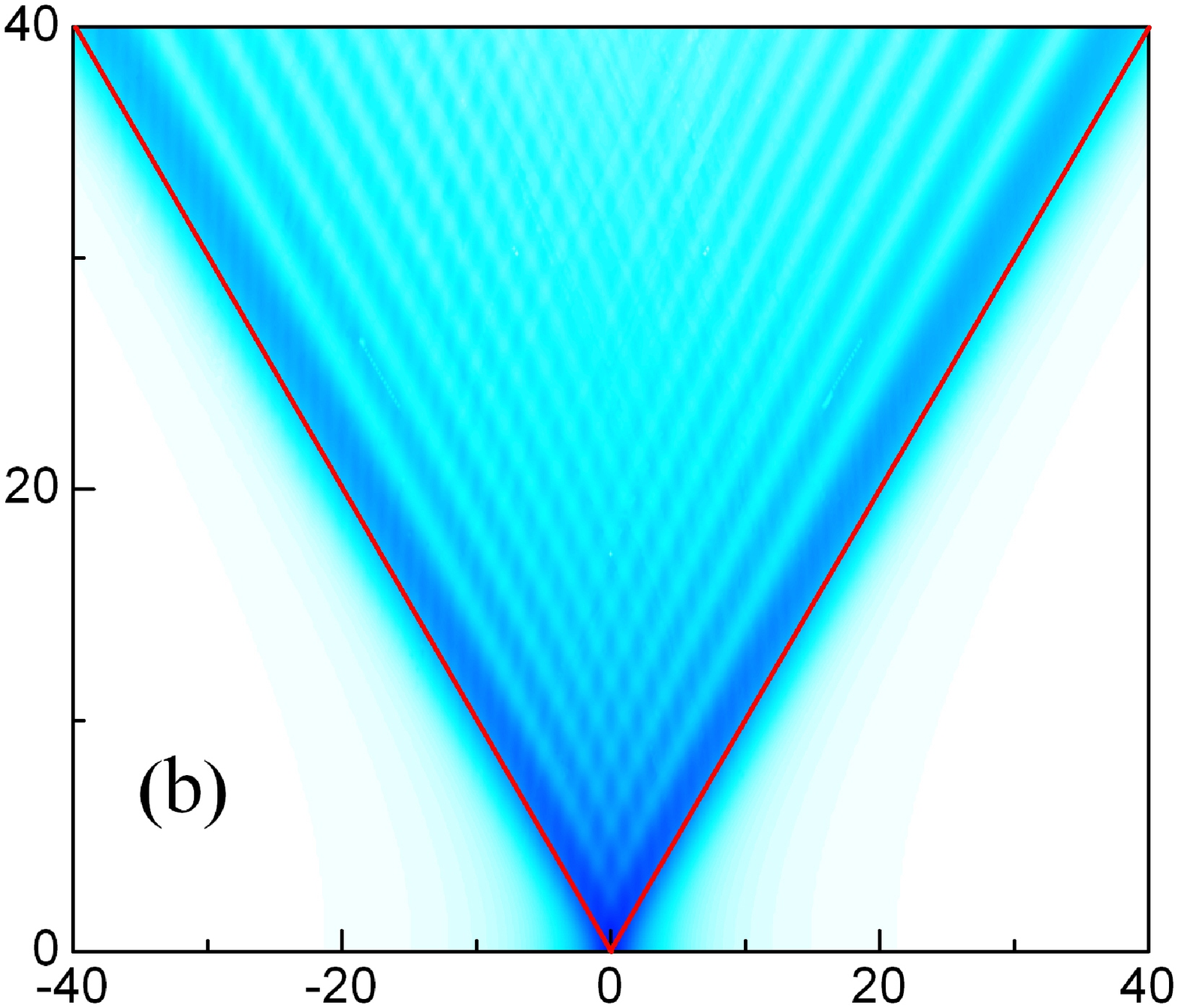} %
\includegraphics[width=8cm]{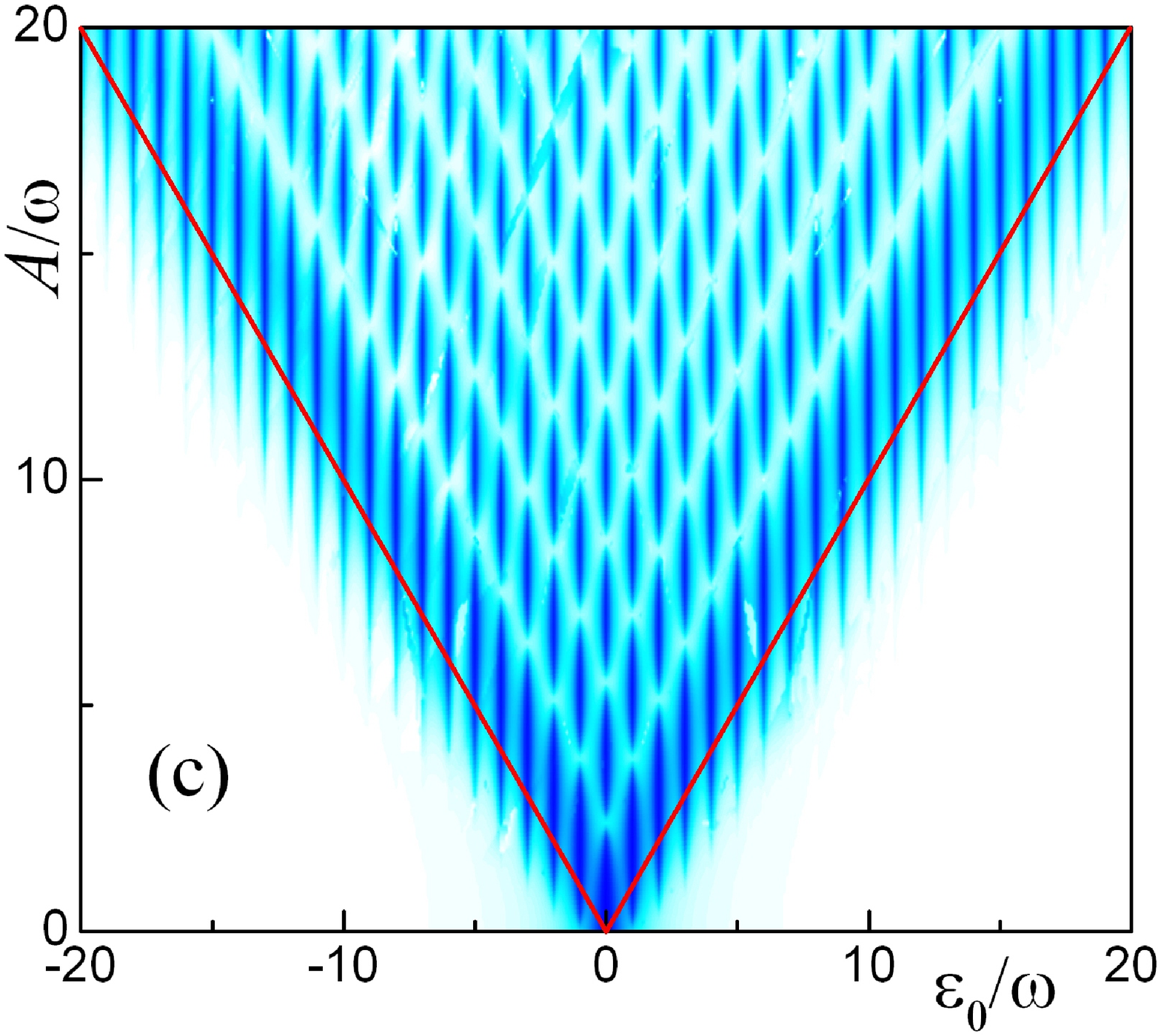} %
\includegraphics[width=8cm]{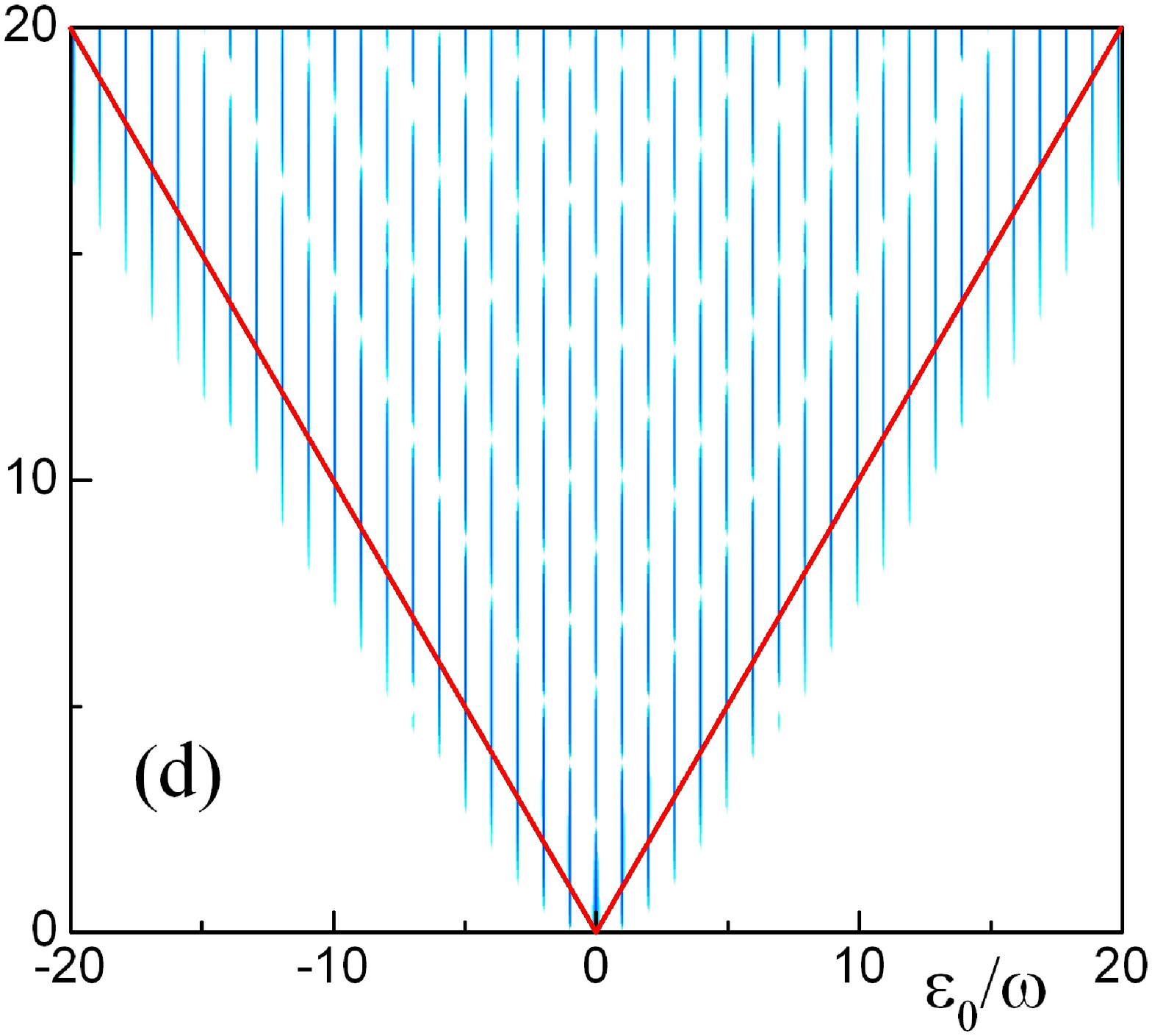}
\caption{(Color online) Same as in Fig.~\protect\ref%
{Fig:LZSI_for_MIT_and_IPHT} (i.e.~LZS interferometry with high-frequency
driving), but including the effects of decoherence. The time-averaged upper
diabatic state occupation probability $\overline{P}_{\mathrm{up}}$ is
obtained numerically by solving the Bloch equations with the Hamiltonian (%
\protect\ref{Hamiltonian}). The dephasing time $T_{2}$ is given by $\protect%
\omega T_{2}/(2\protect\pi )=0.1$ in (a), $0.5$ in (b), $1$ in (c) and $%
T_{2}=2T_{1}$ in (d). The relaxation time is given by $\protect\omega %
T_{1}/(2\protect\pi )=10^{3}$.}
\label{Fig:high_freq_with_decoherence}
\end{figure}

We now increase the coherence time such that the two transitions in a single
driving period are separated by a time that is smaller than the coherence
time. The transitions induced by different driving cycles are separated by
more than the coherence time, and they therefore act as independent kicks.
As in the case of strong decoherence, a sufficiently large amplitude is
required to drive transitions between the two quantum states. However, if
the phase accumulated between the two interfering transitions corresponds to
destructive interference, the net effect of a full driving period will be no
mixing at all. As a result, we now find lines in the $A$--$\varepsilon _{0}$
plane where the TLS is driven back and forth across the avoided crossing but
the system remains in the lower level (in the fast-passage limit, the TLS
remains in the diabatic state that corresponds to the lower level for the
given value of $\varepsilon _{0}$). This situation is depicted in Figs.~\ref%
{Fig:low_freq_with_decoherence}(b) and \ref{Fig:high_freq_with_decoherence}%
(b).

We now increase the coherence time further, such that interference between
transitions from a few successive driving cycles occurs. [Related
discussions of the intracycle and intercycle interference were given in \cite%
{Ashhab:2007} and \cite{Arbo:2010}]. We now start to see the resonance lines
forming, or rather the mixing between the two states being suppressed
whenever the resonance condition is not satisfied (Figs.~\ref%
{Fig:low_freq_with_decoherence}(c) and \ref{Fig:high_freq_with_decoherence}%
(c)). Finally, in Figs.~\ref{Fig:low_freq_with_decoherence}(d) and \ref%
{Fig:high_freq_with_decoherence}(d) we take the case of negligible
decoherence, and we find that the resonance lines are now well defined.
Deviation from the resonance condition leads to the absence of any
substantial mixing between the two states.

%\begin{figure}[h]
%\includegraphics[width=8cm]{FiguresFromRobert/figure3_G1_0.007_G2_0.7500_w_7.500_d_0.025_reduced.eps}
%\includegraphics[width=8cm]{FiguresFromRobert/figure3_G1_0.007_G2_7.500_w_7.500_d_0.025_reduced.eps}
%\includegraphics[width=8cm]{FiguresFromRobert/figure3_G1_0.007_G2_15.000_w_7.500_d_0.025_reduced.eps}
%\includegraphics[width=8cm]{FiguresFromRobert/figure3_G1_0.007_G2_22.5000_w_7.500_d_0.250_reduced.eps}
%\caption{(Color online) Same as in ...}
%\end{figure}

\subsection{\textquotedblleft Breakdown\textquotedblright\ of the adiabatic
theorem}

The adiabatic theorem is used in various subfields of physics. The basic
idea of the theorem is the following \cite{Landau:1977}: if a physical
system starts in one of the eigenstates of the system's Hamiltonian, the
Hamiltonian is varied slowly enough and there are no degeneracies between
the eigenstates during the variation of the Hamiltonian, then the system
will end up in an eigenstate of the Hamiltonian whose order (starting from
the ground state counting upwards) is the same as that of the initial state.

As the term \textquotedblleft theorem\textquotedblright\ suggests, the above
statement is a rigorously provable statement. The tricky point is in
deciding how slow the Hamiltonian variation must be in order to ensure
adiabaticity. The most common form of the adiabatic condition is obtained by
inspecting the LZ formula (Eq.~(\ref{P=P_LZ})): $\delta \gg 1$, and slightly
generalized forms of this condition. If the condition $\delta \gg 1$ is
satisfied in the LZ problem, the system will almost certainly stay in the
ground state, assuming it started in the ground state. The problem is that
the above condition was derived for a specific problem, namely the simple LZ
problem, and not from general considerations that apply to any quantum
system with any form of Hamiltonian variation. This issue has been the
subject of a number of recent papers [\cite{Marzlin:2004}, \cite%
{Sarandy:2004}, \cite{Tong:2005, Tong:2007}, \cite{Amin:2009}]. For example,
if a many-level system starts in an eigenstate of the Hamiltonian, and then
during the variation of the Hamiltonian an enormously large number of
avoided level crossings are encountered (in what can colloquially be called
a \textquotedblleft spaghetti\textquotedblright\ energy-level structure), as
e.g. in [\cite{Pokrovsky:2002, Pokrovsky:2004}, \cite{Shytov:2004}, \cite%
{Garanin:2008}], knowing that each crossing depletes the population of the
initial eigenstate by a small amount does not guarantee that the total
depletion of the state from all the crossings will also be small.

Another example where the above simple condition does not guarantee
adiabaticity over the full time evolution occurs in the simple LZS
interferometry setup considered in this review. If the system is repeatedly
driven back and forth across the avoided-crossing point and the phase
accumulated between successive crossings is designed to produce constructive
interference, the small transition amplitudes of the different crossings
will add up to produce a large transition probability. For example, if we
take $\delta \gg 1$, $\varepsilon _{0}=0$ and $A/\omega =\pi /2+k\pi $
(i.e., we choose the parameters to satisfy the resonance condition, Eq.~(\ref%
{construct})), we know that it is only a matter of time until the
ground-state population is transferred completely to the excited state and
vice versa. Finally, it should be emphasized that one does not need to
satisfy any resonance condition of the kind given in Eq.~(\ref{res_cond}).
An arbitrary-looking driving signal that produces suitable phase
accumulation between successive crossings will also eventually result in
full population transfer between the ground and excited states. Even \textit{%
random} phases will result in diffusion-like dynamics that ultimately result
in large mixing between eigenstates.

When discussing the adiabatic theorem, it is worth digressing for a moment
to mention an idea that has received considerable attention recently, namely
adiabatic quantum computing (AQC) \cite{Farhi:2001}. The basic idea of AQC
is as follows: one subjects a physical system to a simple initial
Hamiltonian, such that it is guaranteed that the system will relax to its
ground state. One then adiabatically changes the Hamiltonian to reach a
final Hamiltonian whose ground state encodes the answer a given problem (the
specifics of the problem are encoded in the parameters of the final
Hamiltonian). Measuring the final state then provides the answer to the
problem. During the evolution from the initial to the final Hamiltonian, one
can expect that avoided crossing structures will be encountered, and the
size and number of these avoided crossings will determine the requirements
on the sweep rate. The smaller the energy gaps and the larger the number of
crossings, the more slowly one has to sweep the parameters of the
Hamiltonian. The traversal of the avoided crossings is governed by LZ
physics, and non-adiabatic LZ transitions play a destructive role and should
be avoided; this was studied for a superconducting qubit e.g. in \cite%
{Grajcar:2005}. It is currently unknown how the running time scales with
system size for different problems. The answer to this question will
determine whether AQC provides a computational advantage over classical
computers or not. In a study that comes a step closer to the subject of the
present article, it was argued in \cite{Ashhab:2006} that the interaction
between an AQC system and its surrounding environment could lead to a
situation where the environment effectively drives the system back and forth
across the avoided crossings, thus reducing the probability of remaining in
the ground state.

\subsection{Reversal of stimulated emission}

It is well known that when an atom is placed in a resonant cavity and the
atom is initialized in its excited state, it emits a photon into the cavity
with a rate that is proportional to the number of photons in the cavity
(more accurately, the emission rate is proportional to $n+1$, where $n$ is
the number of photons in the cavity). Recently, this situation was realized
in an experiment that uses a superconducting charge qubit whose energy-level
structure is essentially the one that we study here \cite{Astafiev:2007}.
The qubit is biased such that it experiences an effective inverted
relaxation taking it from the ground to the excited state. The qubit
therefore constantly emits photons into the resonant cavity, and a lasing
state is realized.

As discussed in \cite{Ashhab:2009}, for small photon numbers the emission
rate is proportional to the number of photons in the cavity, which can be
seen as being a result of the fact that the Rabi frequency is proportional
to the square-root of the photon number. It is interesting to consider what
happens if the photon number is so large that the cavity field effectively
drives the qubit back and forth across the quasi-crossing region. In this
case the Rabi frequency is no longer proportional to the driving amplitude
(which is proportional to the square-root of the photon number), but it
follows the Bessel-function dependence that is characteristic of LZS
interferometry (see Appendix C). One therefore has the situation where the
emission rate is no longer proportional to the photon number in the cavity $%
n $, but rather it is proportional to $J_{1}^{2}(\alpha \sqrt{n})$, where $%
\alpha $ is a conversion coefficient. Stimulated emission therefore exhibits
oscillatory behavior, where the emission rate has regions where it increases
with increasing photon number and regions where it decreases with increasing
photon number.

One consequence of the quasi-periodicity in the Bessel function is that,
with a suitable choice of parameters, there can exist several stable
steady-state photon numbers in the cavity. These solutions can be obtained
by comparing the emission rate into the cavity (which follows the
Bessel-function dependence) and the loss rate out of the cavity (which is
proportional to the photon number). This situation concerning the existence
of multiple steady-state solutions in this system was studied in \cite%
{Rodrigues:2007a, Rodrigues:2007b}.

\section{Recent experiments with strongly driven superconducting quantum
circuits}

\subsection{Multiphoton transitions in superconducting qubits}

If a quantum system is strongly driven, it can be excited by a multiphoton
process. One way to observe the multiphoton resonances is to consider the
frequency dependence, which is convenient for atoms exposed to a periodic
electromagnetic field [\cite{Gallagher:1994}, \cite{Fregenal:2004}, \cite%
{Forre:2004}, \cite{Maeda:2006}]. Another way is to tune the energy
difference at fixed driving frequency. In superconducting qubits the energy
difference is tuned by an external source. The observation of the resonances
allows for multiphoton spectroscopy. Furthermore, the strong driving
provides a tool to control the energy level population, with the driving
frequency being a fraction of the energy level difference. This is important
for operating superconducting qubits, where the energy difference is of the
order of $0.1$ to $10$ GHz. The multiphoton resonances have Lorentzian
shapes \cite{Tornes:2008}, as described by Eqs.~(\ref{P}) or (\ref%
{P_stat_@T=0}), which provide a way to estimate the relaxation rates in the
system.

Multiphoton resonances were observed in several superconducting qubits with
a small number of photons. They were observed in charge qubits \cite%
{Nakamura:2001}, where the energy-level difference is controlled by a gate
voltage, as well as in the charge-phase qubits, where the energy difference
is controlled by both a gate voltage and an external magnetic flux \cite%
{Shnyrkov:2006, Shnyrkov:2009}. Up to five-photon transitions were reported
for the phase qubit \cite{Wallraff:2003}, where the energy difference is
tuned by the bias current. Multiphoton transitions were also observed in a
flux qubit \cite{Saito:2004} controlled by an external magnetic flux.
Further improvement of the coherence characteristics of the superconducting
qubits allowed to observe resonant excitations with more photons, at higher
driving amplitudes. These results are presented below.

\subsection{LZS interferometry in superconducting qubits}

\begin{figure}[h]
\includegraphics[width=15.5cm]{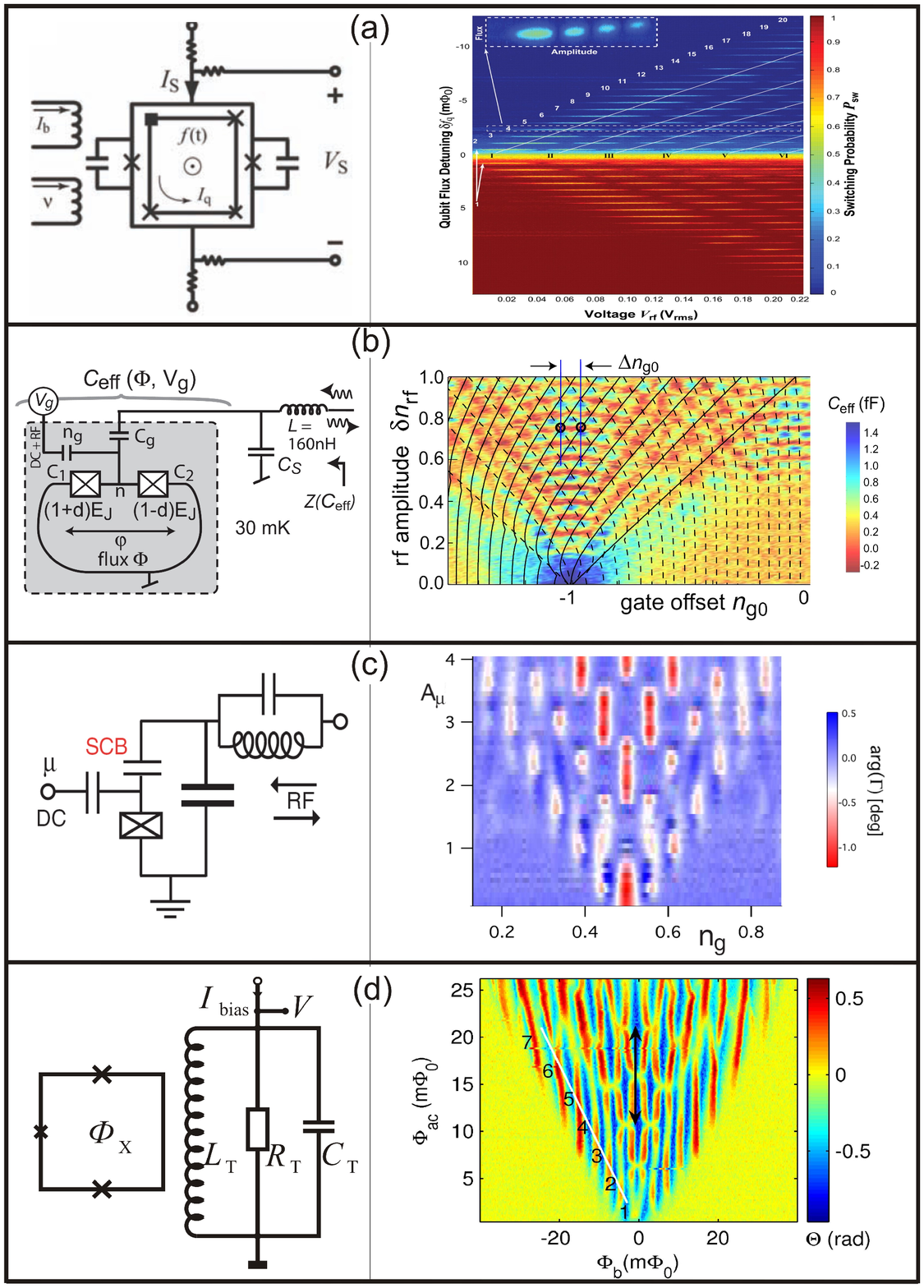}
\caption{(Color online) Experimentally realized Landau-Zener-St\"{u}ckelberg
(LZS) interferometry. The panels from top to bottom present the results of
the following articles: (a) \protect\cite{Oliver:2005}, (b) \protect\cite%
{Sillanpaa:2006}, (c) \protect\cite{Wilson:2007}, (d) \protect\cite%
{Izmalkov:2008}. Schematic diagrams of the circuits used are shown to the
left, while results for the LZS interferometry are presented to the right. A
more detailed description of the experiments can be found in the main text
and, of course, in the respective original articles. Figure (a) is reprinted
from \protect\cite{Oliver:2005} with permission from AAAS. Figure (b) is
reprinted from \protect\cite{Sillanpaa:2006} with permission; copyright
(2006) by APS. Figure (c) is reprinted from \protect\cite{Wilson:2007} with
permission; copyright (2007) by APS. Figure (d) is reprinted from
\protect\cite{Izmalkov:2008} with permission; copyright (2008) by APS.}
\label{experimt_LZI}
\end{figure}

The observation of LZ tunnelling in superconducting qubits was first
reported by \cite{Izmalkov:2004} and then by \cite{Ithier:2005,
Johansson:2009} (see also \cite{Ankerhold:2003} for the theoretical
background). Recently LZS interferometry was demonstrated by several groups
on different superconducting qubits with complementary measurement
techniques [\cite{Oliver:2005}, \cite{Sillanpaa:2006}, \cite{Wilson:2007},
\cite{Izmalkov:2008}, \cite{LaHaye:2009}]. Below we briefly describe these
works and present the major results in Fig.~\ref{experimt_LZI}. There, the
left column shows the circuits used in the experimental setups and the right
columns shows their corresponding experimental LZS interferometry. The
latter shows a measurable quantity versus driving power (amplitude) and
energy bias (which defines the energy level distance). Also recently the LZS
interferometry was demonstrated with the rf SQUID qubit in \cite{Sun:2009}
[see also \cite{Sun:2006} for experimental details]. One important
difference from the previous works is that the system is essentially a
multi-level system with several different avoided crossings involved (see
also next subsection). This allows for additional interesting and useful
phenomena such as controllable population inversion, as convincingly
demonstrated both experimentally and theoretically in \cite{Sun:2009}.

The parameters used in different experiments studying LZS interferometry
with driven superconducting qubits are given in Table \ref{table1}.
Remarkable is the fact that in all of these experiments the relation of the
characteristic time of the LZ transition (Eq.~(\ref{tLZ})) estimated with
maximal driving amplitude $A^{\max }$, $t_{\mathrm{LZ}}\sim 1/\sqrt{A^{\max
}\omega }$, to the driving period $T$ was almost identical: $t_{\mathrm{LZ}%
}/T\sim 0.05$. Note that the exact value of $t_{\mathrm{LZ}}/T$ does not
affect the analysis or results, provided that this ratio is much smaller
than one (see Eq.~(\ref{condition4LZSI})).

\begin{table}[tbp]
\caption{Parameters used in different experiments studying LZS
interferometry: tunnelling amplitude $\Delta $, maximal driving amplitude $%
A^{\max }$, and driving frequency $\protect\omega $ in the units GHz$\times 2%
\protect\pi $, minimal adiabaticity parameter $\protect\delta^{\min }=\Delta
^{2}/(4\protect\omega A^{\max })$, and maximal LZ probability $P_{\mathrm{LZ}%
}^{\max }=\exp (-2\protect\pi \protect\delta ^{\min })$.}
\label{table1}%
\begin{ruledtabular}
\begin{tabular}{cccccc}
    $ $&$\Delta$&$A^{\max}$&$\omega$&$\delta^{\min}$&$P_{\mathrm{LZ}}^{\max}$ \\
    \hline \cite{Oliver:2005} & 0.004 & 24 & 1.2 & $10^{-7}$ & 1 \\
    \hline \cite{Sillanpaa:2006} & 12.5 & 95 & 4 & 0.1 & 0.5 \\
    \hline \cite{Wilson:2007} & 2.6 & 62 & 7 & 0.004 & 0.98 \\
    \hline \cite{Izmalkov:2008} & 3.5 & 40 & 4 & 0.02 & 0.9 \\
\end{tabular}
\end{ruledtabular}
\end{table}

(1) In Ref. \cite{Oliver:2005} a superconducting flux qubit was subjected to
a strong microwave drive and LZS multiple interference fringes were
observed. The flux qubit was a niobium superconducting ring with three
Josephson junctions. The circuit and results are shown in Fig.~\ref%
{experimt_LZI}(a) to the left and to the right respectively. The qubit was
biased with both dc and microwave magnetic fluxes. The level population was
probed with a dc SQUID. The switching probability of the SQUID to the
resistive state measures the state of the qubit in the diabatic basis $%
\{\varphi _{\uparrow },\varphi _{\downarrow }\}$. These measurements showed
resonant peaks and dips when the energy of $k$ photons matched the qubit's
energy-level difference. The latter was controlled with the dc flux (flux
detuning) and multiphoton resonances with $k$ up to $20$ were observed.
Later the observation of multiphoton resonances of order up to $45$ was
reported \cite{Berns:2006}. The interference fringes shown in Fig.~\ref%
{experimt_LZI} are described by the theory presented in Sec. II.D.2; cf.
also Fig.~\ref{Fig:LZSI_for_MIT_and_IPHT}(a). The fringes exhibit the
Bessel-function dependence, according to Eqs.~(\ref{Delta_k}, \ref{P}), and
the steplike pattern in Fig.~\ref{experimt_LZI}(a) was called a
\textquotedblleft Bessel staircase\textquotedblright\ \cite{Oliver:2005};
the inclined white lines in Fig.~\ref{experimt_LZI}(a) show the maxima for $%
J_{k}^{2}\left( A/\omega \right) $ which were numbered with the Roman
numbers $\mathrm{I}$ to $\mathrm{VI}$. For more details see \cite%
{Rudner:2008}, \cite{Bylander:2009} and \cite{Oliver:2009}.

(2) In Ref. \cite{Sillanpaa:2006} an interferometer-type charge qubit was
studied. The qubit consisted of two nearby Josephson junctions embedded in a
loop (Fig.~\ref{experimt_LZI}(b)). The charge on the island between the
junctions (the so-called Cooper-pair box) was controlled via the applied
gate voltage $V_{g}=en_{g}/C_{g}$, $n_{g}(t)=n_{g0}+\delta n_{\mathrm{rf}%
}\sin (\omega _{\mathrm{rf}}t)$. The effective Josephson energy of the two
junctions was tuned using the applied magnetic flux $\Phi $ through the
loop. The energy levels of the qubit were controlled by the flux $\Phi $ and
by the dc component ($n_{g0}$) of the voltage, while the energy-level
populations were controlled by the ac component of the voltage. The qubit
was coupled in parallel to an $LC$-circuit. The effective capacitance $C_{%
\mathrm{eff}}$ of the qubit modifies the resonant frequency of the $LC$%
-circuit \cite{Sillanpaa:2005}. This provides a method for the
continuous-time monitoring of the qubit's effective capacitance, which is
related to the occupations of the qubit's energy eigenstates. The resulting
LZS interference pattern is shown to the right in Fig.~\ref{experimt_LZI}%
(b); this is described by the theory presented in Sec. II.D.1; cf. also Fig.~%
\ref{Fig:LZSI_for_HUT}. For more details see also \cite{Sillanpaa:2007,
Paila:2009}. Also very recently LZS interferometry with an
interferometer-type charge qubit was demonstrated in \cite{LaHaye:2009}. The
parameters of the qubit were close to the ones in Ref. \cite{Sillanpaa:2006}%
; consequently, LZS interference fringes were similar to the ones shown in
Fig.~\ref{experimt_LZI}(b). The important difference is in the measuring
device: the qubit was coupled to a nanoelectromechanical resonator, as
studied theoretically in \cite{Irish:2003}, and the nanoresonator frequency
shift was measured.

(3) References \cite{Wilson:2007, Wilson:2010} studied a single Cooper-pair
box (SCB), which was composed of an aluminium island connected to a
reservoir via a Josephson junction (Fig.~\ref{experimt_LZI}(c)). The state
of the qubit was controlled by the gate voltage $V_{g}$ ($%
n_{g}=C_{g}V_{g}/2e $) and by the magnetic field in the dc SQUID\ loop
connected to the island, which tunes the Josephson energy $E_{J}$. The qubit
was driven through the gate with the microwave amplitude $A_{\mu }$. The
qubit state was probed with an rf oscillator \cite{Duty:2005, Persson:2010}.
The color image in Fig.~\ref{experimt_LZI}(c) presents the phase of the rf
reflection coefficient $\Gamma $. The ground and excited states contributed
phase shifts with opposite signs. The results were interpreted in terms of
the dressed states of the qubit coupled to the driving microwave field \cite%
{Liu:2006}, \cite{Greenberg:2007}.

(4) In Ref. \cite{Izmalkov:2008} a superconducting flux qubit was driven by
an external magnetic flux, analogously to Ref. \cite{Oliver:2005}. The qubit
was weakly coupled to a classical resonant tank circuit, Fig.~\ref%
{experimt_LZI}(d). The tank circuit was biased with an rf current $I_{%
\mathrm{bias}}$ and the measurable quantity was the phase shift $\Theta $
between the voltage $V$ and the bias current in the tank circuit. The
inductive coupling of the qubit to the tank circuit resulted in a phase
shift determined by the effective inductance of the qubit \cite%
{Greenberg:2002, Ilichev:2009}. The resonant excitation of the flux qubit
changes the direction of the current in the qubit's loop, which resonantly
changes the effective inductance of the qubit. The respective resonances
were observed in the phase shift dependence as a sequence of the ridges and
troughs \cite{Shevchenko:2008}. The LZS interference pattern is shown in
Fig.~\ref{experimt_LZI}(d) in the dependence of the tank phase shift $\Theta
$ on the dc ($\Phi _{\mathrm{b}}$) and ac ($\Phi _{\mathrm{ac}}$) components
of the magnetic flux; the interference fringes are described by the theory
presented in Sec. II: see Fig.~\ref{Fig:LZSI_for_MIT_and_IPHT}(b).

%\begin{widetext}

%\end{widetext}

%\begin{widetext}

%\begin{figure}[h]
%\includegraphics[width=14.5cm]{experimt_LZI_2.eps}
%\caption{(Color online) Experimentally realized
%Landau-Zener-St\"{u}ckelberg (LZS) interferometry. The panels
%from top to bottom %
%present the results of the following articles: (a) \cite%
%{Oliver:2005}, (b) \cite{Sillanpaa:2006}, (c) \cite{Wilson:2007}, (d) \cite%
%{Izmalkov:2008}, (e) \cite%
%{Sun:2009}. Schematic diagrams of the circuits used are shown to
%the left, while results for the LZS interferometry
% are presented to the right. A more detailed description of
%the experiments can be found in the main text and, of course, in
%the respective original articles. \textit{}}
%\label{experimt_LZI_2}
%\end{figure}

%\end{widetext}

Thus, strongly driven superconducting qubits were studied by different,
complementary methods; their resonant characteristics were shown to be
periodic in the bias $\varepsilon_0$ and driving amplitude $A$ due to St\"{u}%
ckelberg oscillations. The different experimental measurements allowed to
obtain the following parameters: (i) those which characterize the qubit and
enter in the pseudo-spin Hamiltonian, (ii) the driving microwave amplitude
as felt by the qubit (which is the calibration of a generator signal), (iii)
the relaxation and dephasing rates, which describe the impact of the
environment on the qubit dynamics (via the Bloch equations).

%{Several issues are going to be described here: analogy to Mach-Zehnder interferometer <cite>Oliver:2005</cite>, difference from driving atomic systems <cite>Wilson:2007</cite>, etc.}

\subsection{Multiphoton resonances in multi-level systems}

Most qubits are formed by the lowest two levels in a multi-level structure.
Moreover, a qubit can be coupled to another quantum system, in particular,
to another qubit, so that the overall structure is also multi-level.
High-amplitude driving allows not only to reach the upper level of a TLS
(qubit) with the absorption of several photons, but also to involve more of
the \textit{qudit}'s energy levels in the process.

Multiphoton resonant excitations were studied in superconducting multi-level
systems. One-, two-, and three-photon transitions were driven in the flux
qubit not only between the two lowest levels (which form the qubit), but
rather between five levels in \cite{Yu:2005}. In this case multiphoton
spectroscopy allowed to obtain the system parameters and to visualize the
energy-level diagram. The low-frequency and extremely strong-driving regime
was studied in \cite{Berns:2008} [see also \cite{Wilhelm:2008} and \cite%
{Wen:2009}]. Berns et al. demonstrated that their method, which they called
amplitude spectroscopy, allows probing higher levels that are three orders
of magnitude higher in energy than the driving frequency. Recently
interference fringes associated with LZ transitions at two nearby avoided
crossings were also observed in a multi-level structure of the rf-SQUID
qubit driven by large-amplitude and high-frequency microwave field \cite%
{Wang:2010} [see also \cite{Sun:2009} for more details]. Also, multiphoton
multi-level Rabi oscillations were studied in a phase qubit \cite{Dutta:2008}%
. Multiphoton transitions in the multilevel system formed by the dressed
states of a qubit coupled to a quantum resonator were demonstrated in \cite%
{Fink:2009, Bishop:2009}. Another situation where a multi-level structure
appears, is a system of coupled qubits. Direct and ladder-type multiphoton
excitations were recently demonstrated in a two-flux-qubit system in \cite%
{Ilichev:2010}.

\section{Conclusions}

Strong driving of a TLS results in a periodic dependence of its response on
the bias, driving frequency and amplitude. This dependence was recently
observed in superconducting qubit systems by several groups using
complementary measurement techniques, both in flux qubits \cite{Oliver:2005}%
, \cite{Izmalkov:2008} and in charge qubits \cite{Sillanpaa:2006}, \cite%
{Wilson:2007}. Very recently LZS interferometry was also realized in a rf
SQUID qubit \cite{Sun:2009}. No related effect has been reported in
experiments using phase qubits. This is because it is difficult to produce
avoided energy-level structures in phase qubits without involving higher
levels.

LZS interferometry allows, first, to control the qubit state, and, second,
to obtain the system parameters. In particular, from the positions of the
multiphoton resonances, the qubit parameters can be obtained (spectroscopy);
the St\"{u}ckelberg oscillations can be used for the definition of the
driving amplitude (calibration of power); and the width and shape of the
interference fringes can be used to estimate the relaxation rates. These
properties make LZS interferometry a very powerful tool for studying TLSs
(see also \cite{LaHaye:2009}, where LZS interferometry was realized recently
for a superconducting qubit coupled to a nanoelectromechanical resonator).

We have reviewed recent experimental results on LZS interferometry, and we
have presented the theoretical description of this problem. In particular,
we have demonstrated the applicability of the adiabatic-impulse model for
both slow and fast driving regimes. This model is a powerful tool in the
study of non-adiabatic transitions at avoided crossings and the interference
between these transitions.

\section*{Acknowledgments}

We would like to thank A.N. Omelyanchouk, A.M. Zagoskin, A.A. Soroka and
D.A. Rodrigues for useful discussions. We would also like to thank J.R.
Johansson for useful discussions and for providing the numerical
calculations used in Figs.~\ref{Fig:From_slow_to_fast_passage}, \ref%
{Fig:low_freq_with_decoherence} and \ref{Fig:high_freq_with_decoherence}.
This work was supported in part by the National Security Agency (NSA),
Laboratory Physical Science (LPS), Army Research Office (ARO), National
Science Foundation (NSF) grant number 0726909, JSPS-RFBR grant 09-02-92114,
Fundamental Researches State Fund (grant number F28.2/019), and NAS of
Ukraine (project number 02/09-N).

%-----------------------------------------------------------
\appendix
%-----------------------------------------------------------

\section{Solution of the Landau-Zener problem}

\subsection{Adiabatic wave function}

In this Appendix we consider the driven two-level system in the adiabatic
representation, assuming the evolution to be adiabatic far from the avoided
level crossings, with non-adiabatic transitions occurring around these
points.

The wave function in the adiabatic approximation takes the form%
\begin{equation}
\psi =\left(
\begin{array}{c}
a_{1} \\
a_{2}%
\end{array}%
\right) ,\text{ }a_{1,2}=A_{1,2}\exp \left( -i\int E(t)dt\right) .
\end{equation}%
Then the Shr\"{o}dinger equation $H\psi =E\psi $ gives the instantaneous
eigenvalues, Eq.~(\ref{Energy(t)}), and eigenvectors:
\begin{subequations}
\label{psi}
\begin{eqnarray}
\psi _{\pm } &=&\varphi _{\pm }e^{\mp i(\zeta +\frac{\pi }{4})},\text{ }%
\zeta =\frac{1}{2}\int \Omega (t)dt, \\
\varphi _{\pm } &=&\left(
\begin{array}{c}
\beta _{\mp } \\
\mp \beta _{\pm }%
\end{array}%
\right) \text{, \ }\beta _{\pm }=\sqrt{\frac{\Omega \pm \varepsilon }{%
2\Omega }}.
\end{eqnarray}%
Here the term $\pi /4$ arises when the next term in the quasiclassical
(adiabatic) approximation is taken into account \cite{Delone:1985}.

Let us consider the original LZ problem described by the Hamiltonian (\ref%
{Hamiltonian}) with $\varepsilon =vt$\ \{see also Eq.~(\ref{H_linearzd}); we
consider $t$ instead of $t^{\prime }$\ in this Appendix\}:
\end{subequations}
\begin{equation}
H(t)=-\frac{\Delta }{2}\sigma _{x}-\frac{vt}{2}\sigma _{z}.
\end{equation}

Consider now the adiabatic wave function in the region far from the avoided
level crossing, where $\left\vert \varepsilon (t)\right\vert =v\left\vert
t\right\vert \gg \Delta $. Note that this latter condition is consistent
with the condition $\omega \left\vert t\right\vert \ll 1$ used for the
linearization of the harmonic-driving problem, Eq.~(\ref{e_linrzd}), at
large enough driving amplitudes. Then we have:
\begin{subequations}
\begin{eqnarray}
\varphi _{+} &\approx &\!\!\left(
\begin{array}{c}
1 \\
0%
\end{array}%
\right) \!\text{, }\ \varphi _{-}\approx \!\!\left(
\begin{array}{c}
0 \\
1%
\end{array}%
\right) \text{ if }t_{\text{$\mathrm{a}$}}<0, \\
\varphi _{+} &\approx &\!\!\left(
\begin{array}{c}
0 \\
-1%
\end{array}%
\right) \!\text{, }\ \varphi _{-}\approx \!\!\left(
\begin{array}{c}
1 \\
0%
\end{array}%
\right) \text{ if }t_{\text{$\mathrm{a}$}}>0,
\end{eqnarray}%
\end{subequations}
\begin{equation}
\zeta (\pm t_{\text{$\mathrm{a}$}})=\!\frac{1}{2}\int\limits_{0}^{\pm t_{%
\text{$\mathrm{a}$}}}\!\!\Omega (t)\,dt\approx \pm \left[ \Phi \left( \sqrt{%
\frac{v}{2}}t_{\text{$\mathrm{a}$}}\right) -\Phi _{\delta }\right] .
\label{zeta_prime}
\end{equation}%
The subscript \textquotedblleft $\mathrm{a}$\textquotedblright\ indicates
that we now consider the time within the asymptotic region, where $%
\left\vert \varepsilon (t)\right\vert \gg \Delta $, and the acquired phase
was split into time-dependent and independent parts:%
\begin{eqnarray}
\Phi (z_{\text{$\mathrm{a}$}}) &=&\frac{z_{\text{$\mathrm{a}$}}^{2}}{2}%
+\delta \ln (\sqrt{2}z_{\text{$\mathrm{a}$}}), \\
\Phi _{\delta } &=&\frac{1}{2}\delta (\ln \delta -1).
\end{eqnarray}

\subsection{Non-adiabatic transition (evolution matrix for the Landau-Zener
transition)}

Let us now consider the non-adiabatic transition between the energy levels
in the vicinity of the point of avoided level crossings, following \cite%
{Zener:1932}. Then the Schr\"{o}dinger equation $i\dot{\psi}=H\psi $ takes
the form:%
\begin{equation}
\left\{
\begin{array}{c}
i\dot{a}_{1}=-\frac{v}{2}ta_{1}-\frac{\Delta }{2}a_{2}, \\
i\dot{a}_{2}=-\frac{\Delta }{2}a_{1}+\frac{v}{2}ta_{2}.%
\end{array}%
\right.  \label{system}
\end{equation}%
This can be written in the form of the second-order Weber equations:%
\begin{equation}
\frac{d^{2}a_{1,2}}{dz^{2}}+(2\delta \mp i+z^{2})a_{1,2}=0,  \label{Weber}
\end{equation}%
\begin{equation}
z=\sqrt{\frac{v}{2}}t.
\end{equation}%
The solutions are combinations of parabolic cylinder functions \cite%
{Gradshteyn:1994}:
\begin{subequations}
\begin{eqnarray}
a_{1} &=&\sum_{\pm }A_{\pm }D_{-1-i\delta }(\pm \sqrt{2}e^{i\pi /4}z), \\
a_{2} &=&\sum_{\pm }B_{\pm }D_{-i\delta }(\pm \sqrt{2}e^{i\pi /4}z).
\end{eqnarray}%
We find the relation between the coefficients $A_{\pm }$ and $B_{\pm }$ from
the first-order equation (by inserting the solutions for $a_{1,2}$ in the
first equation of the system (\ref{system})) making use of the recurrence
formula:
\end{subequations}
\begin{equation}
B_{\pm }=\mp \frac{e^{-i\pi /4}}{\sqrt{\delta }}A_{\pm }.
\end{equation}

Now with the asymptotes of the parabolic cylinder functions \cite%
{Gradshteyn:1994}, we find $a_{1,2}$ far from the transition point (where $%
z=0$), at $z=\pm z_{\text{\textrm{a}}}$ with $z_{\text{\textrm{a}}}\gg 1$:
\begin{subequations}
\label{NA}
\begin{eqnarray}
\!\!a_{1}(-z_{\text{$\mathrm{a}$}})\!\!\! &\approx &\!\!\!A_{+}\Xi
_{1}e^{i\Phi (z_{\mathrm{a}})}\!, \\
\!\!a_{2}(-z_{\text{$\mathrm{a}$}})\!\!\! &\approx &\!\!\!(-e^{-\frac{\pi }{2%
}\delta }A_{+}+e^{\frac{\pi }{2}\delta }A_{-})\Xi _{2}e^{-i\Phi (z_{\text{$%
\mathrm{a}$}})}\!, \\
\!\!a_{1}(z_{\text{$\mathrm{a}$}})\!\!\! &\approx &\!\!\!A_{-}\Xi
_{1}e^{i\Phi (z_{\text{$\mathrm{a}$}})}\!, \\
\!\!a_{2}(z_{\text{$\mathrm{a}$}})\!\!\! &\approx &\!\!\!(-e^{\frac{\pi }{2}%
\delta }A_{+}+e^{-\frac{\pi }{2}\delta }A_{-})\Xi _{2}e^{-i\Phi (z_{\text{$%
\mathrm{a}$}})}\!,
\end{eqnarray}%
where
\end{subequations}
\begin{subequations}
\begin{eqnarray}
\Xi _{1} &\equiv &\frac{\sqrt{2\pi }}{\Gamma (1+i\delta )}\exp \left( -\frac{%
\pi }{4}\delta \right) , \\
\Xi _{2} &\equiv &\frac{1}{\sqrt{\delta }}\exp \left( -i\frac{\pi }{4}-\frac{%
\pi }{4}\delta \right) .
\end{eqnarray}%
Let us match this asymptotic solution with the adiabatic one, $\psi (t)=\sum
b_{\pm }\varphi _{\pm }$, from the previous subsection:
\end{subequations}
\begin{subequations}
\label{joining}
\begin{eqnarray}
\left(
\begin{array}{c}
b_{+} \\
b_{-}%
\end{array}%
\right) &=&\left(
\begin{array}{c}
a_{1} \\
a_{2}%
\end{array}%
\right) \text{ at }t_{\text{$\mathrm{a}$}}<0, \\
\left(
\begin{array}{c}
b_{-} \\
-b_{+}%
\end{array}%
\right) &=&\left(
\begin{array}{c}
a_{1} \\
a_{2}%
\end{array}%
\right) \text{ at }t_{\text{$\mathrm{a}$}}>0.
\end{eqnarray}%
Here the time dependence in the l.h.s. and in the r.h.s. is the same, see
respectively Eqs.~(\ref{zeta_prime}) and (\ref{NA}). This allows to describe
the evolution from $t=-t_{\text{\textrm{a}}}$ to $t_{\text{\textrm{a}}}$ as
a sequence of adiabatic evolutions from $t=-t_{\text{\textrm{a}}}$ to $t=-0$
and from $t=+0$ to $t=t_{\text{\textrm{a}}}$, with the nonadiabatic
transition in between (described by the time-independent matrix $N$),
namely:
\end{subequations}
\begin{equation}
\mathbf{b}(t_{\text{$\mathrm{a}$}})=U(t_{\text{$\mathrm{a}$}},+0)NU(-0,-t_{%
\text{$\mathrm{a}$}})\text{ }\mathbf{b}(-t_{\text{$\mathrm{a}$}%
})=e^{-i\sigma _{z}\zeta (t_{\text{$\mathrm{a}$}})}Ne^{-i\sigma _{z}\zeta
(t_{\text{$\mathrm{a}$}})}\mathbf{b}(-t_{\text{$\mathrm{a}$}}).
\end{equation}%
Then collecting this with Eqs.~(\ref{joining}) and (\ref{NA}) and equating
multipliers with $A_{+}$ and $A_{-}$, we obtain the expression (\ref{N}) for
the matrix $N$. This matrix describes the transition between the two
adiabatic states with the probability\ given by Eq.~(\ref{P=P_LZ}) and also
includes the phase jump $\widetilde{\varphi }_{\mathrm{S}}$. We note that
the matrix $N$ at $t=t_{2}$ is the same as at $t=t_{1}$ for the nonadiabatic
transitions.

The argument of the asymptotes of the gamma function $\Gamma $ are \cite%
{Gradshteyn:1994}:
\begin{equation}
\text{arg}\Gamma (1-i\delta )\approx \left\{
\begin{array}{c}
C\delta \text{, \ \ \ \ \ \ \ \ \ \ \ \ \ \ }\delta \ll 1, \\
\mathbf{-}\frac{\pi }{4}-\delta (\ln \delta -1)\text{, }\delta \gg 1,%
\end{array}%
\right.
\end{equation}%
where $C\approx 0.58$ is the Euler constant. Thus, the phase jump $\varphi _{%
\mathrm{S}}(\delta )$ is a monotonous function, which changes from $0$ in
the adiabatic limit ($\delta \gg 1$) to $\pi /4$ in the diabatic (fast
driving) limit ($\delta \ll 1$).

It is worth making a comment on bases here. In this review we have used the
energy eigenbasis for describing the dynamics of the TLS. Furthermore, we
have followed the reasonable convention where the ground state far to the
right of the avoided crossing region coincides with the excited state far to
the left of the avoided crossing region, and vice versa. In principle, we
could have defined these pairs of states such that they differ by some phase
factors. If we follow such a convention, there would be some asymmetry
between the different matrices (denoted by $N$) that describe the different
LZ transitions. However, since such a convention seems rather unnatural, we
do not comment on it further. A different convention from ours that appears
more frequently in the literature is the use of the diabatic states for
describing the dynamics. In this case the matrices that describe the
evolution from time $-t_{\text{\textrm{a}}}$ to time $t_{\text{\textrm{a}}}$
for a single passage depend on the signs that appear in the Hamiltonian,
Eq.~(\ref{H_linearzd}). The part describing the dynamical phase accumulated
during the adiabatic parts of the evolution is straightforward. The LZ
transition matrices are less obvious, and we give the rules for how to
determine them explicitly here. For the Hamiltonian in Eq.~(\ref{H_linearzd}%
) with negative signs in front of both terms in the Hamiltonian, the LZ
transition matrix takes the form:
\begin{equation}
\left(
\begin{array}{cc}
\sqrt{P_{\mathrm{LZ}}} & \sqrt{1-P_{\mathrm{LZ}}}e^{i\varphi _{\mathrm{S}}}
\\
-\sqrt{1-P_{\mathrm{LZ}}}e^{-i\varphi _{\mathrm{S}}} & \sqrt{P_{\mathrm{LZ}}}%
\end{array}%
\right) .  \label{Nnew}
\end{equation}%
If the sign in front of the second term in the Hamiltonian is reversed, the
off-diagonal matrix elements in Eq.~(\ref{Nnew}) are reversed. If the sign
in front of the first term in the Hamiltonian is reversed, only the minus
sign at the beginning of the bottom-left matrix element is moved to the
top-right matrix element. Combining the two rules above, if both signs in
the Hamiltonian are reversed, only the minus sign in front of the Stokes
phase is moved from the bottom-left matrix element to the top-right matrix
element. Note that in the diabatic basis there will always be two relevant
signs in front of the second term in the Hamiltonian, depending on whether
the avoided crossing is traversed from left to right or from right to left.
As a result, there will be two different matrices describing the different
crossings.

\section{Evolution of a periodically driven two-level system}

The time evolution of a periodically driven TLS is described by Eqs.~(\ref%
{(I)}) and (\ref{(II)}). Consider first the matrix for a one-period
evolution:
\begin{subequations}
\begin{equation}
NU_{2}NU_{1}=\left(
\begin{array}{cc}
\alpha & -\gamma ^{\ast } \\
\gamma & \alpha ^{\ast }%
\end{array}%
\right) ,
\end{equation}%
\end{subequations}
\begin{subequations}
\begin{eqnarray}
\alpha &=&(1-P_{\mathrm{LZ}})e^{-i\zeta _{+}}-P_{\mathrm{LZ}}e^{-i\zeta
_{-}}, \\
\gamma &=&\sqrt{P_{\mathrm{LZ}}(1-P_{\mathrm{LZ}})}e^{i\widetilde{\varphi }_{%
\mathrm{S}}}(e^{-i\zeta _{+}}+e^{-i\zeta _{-}}),
\end{eqnarray}%
\end{subequations}
\begin{subequations}
\label{zeta_pm}
\begin{eqnarray}
\zeta _{+} &=&\zeta _{1}+\zeta _{2}+2\widetilde{\varphi }_{\mathrm{S}}, \\
\zeta _{-} &=&\zeta _{1}-\zeta _{2}.
\end{eqnarray}

After diagonalizing the matrix, it is straightforward to write down its $n$%
-th power, e.g. \cite{Bychkov:1970}. This unitary evolution matrix $%
(NU_{2}NU_{1})^{n}$ has the form
\begin{equation}
(NU_{2}NU_{1})^{n}=\left(
\begin{array}{cc}
u_{11} & -u_{21}^{\ast } \\
u_{21} & u_{11}^{\ast }%
\end{array}%
\right) ,
\end{equation}%
with elements:
\end{subequations}
\begin{subequations}
\begin{eqnarray}
u_{11} &=&\cos n\phi +i(\text{Im}\alpha )\frac{\sin n\phi }{\sin \phi }, \\
u_{21} &=&\gamma \frac{\sin n\phi }{\sin \phi }, \\
\cos \phi &=&\text{Re}\alpha .
\end{eqnarray}%
Then with Eq.~(\ref{(I)}) for $\left( t-2\pi n/\omega \right) \in
(t_{1},t_{2})$, we obtain the upper-level occupation probability, $%
P_{+}=\left\vert b_{+}\right\vert ^{2}$,
\end{subequations}
\begin{equation}
P_{+}^{(\mathrm{I})}(n)=\left\vert \gamma \right\vert ^{2}\frac{\sin
^{2}n\phi }{\sin ^{2}\phi }.  \label{P_I(m)}
\end{equation}%
And for $\left( t-2\pi n/\omega \right) \in (t_{2},t_{1}+2\pi /\omega )$
with Eq.~(\ref{(II)}) we obtain:%
\begin{equation}
P_{+}^{(\mathrm{II})}(n)=2Q_{1}\frac{\sin ^{2}n\phi }{\sin ^{2}\phi }+Q_{2}%
\frac{\sin 2n\phi }{\sin \phi }+P_{\mathrm{LZ}}\cos 2n\phi ,  \label{P_II(m)}
\end{equation}%
\begin{subequations}
\begin{equation}
Q_{1}=P_{\mathrm{LZ}}\left[ P_{\mathrm{LZ}}\sin ^{2}\zeta _{-}+(1-P_{\mathrm{%
LZ}})(1+\cos \zeta _{+}\cos \zeta _{-})\right] ,
\end{equation}%
\end{subequations}
\begin{equation}
Q_{2}=2P_{\mathrm{LZ}}(1-P_{\mathrm{LZ}})\cos (\zeta _{1}+\widetilde{\varphi
}_{\mathrm{S}})\cos (\zeta _{2}+\widetilde{\varphi }_{\mathrm{S}}).
\end{equation}%
Now the time-averaging corresponds to averaging over $n\gg 1$ periods and,
from Eqs.~(\ref{P_I(m)}) and (\ref{P_II(m)}), we obtain:
\begin{equation}
\overline{P_{+}^{(\mathrm{I})}}=\frac{\left\vert \gamma \right\vert ^{2}}{%
2\sin ^{2}\phi }=\frac{1}{2}\frac{\left\vert \gamma \right\vert ^{2}}{%
\left\vert \gamma \right\vert ^{2}+(\text{Im}\alpha )^{2}},
\label{P_I_avrgd}
\end{equation}%
\begin{equation}
\overline{P_{+}^{(\mathrm{II})}}=\frac{Q_{1}}{\sin ^{2}\phi }.
\label{P_II_avrgd}
\end{equation}

From Eq.~(\ref{P_I_avrgd}) it follows that the upper-level occupation
probability is maximal at Im$\alpha =0$. This results in the general
resonance condition:%
\begin{equation}
(1-P_{\mathrm{LZ}})\sin \zeta _{+}-P_{\mathrm{LZ}}\sin \zeta _{-}=0.
\label{res_cond_gen}
\end{equation}%
In particular, in the slow and fast limits the resonance condition takes the
form:
\begin{subequations}
\begin{eqnarray}
\zeta _{1}+\zeta _{2} &=&k\pi \text{ for }2\pi \delta \gg 1,  \label{B12a} \\
\zeta _{1}-\zeta _{2} &=&k\pi \text{ for }2\pi \delta \ll 1.
\end{eqnarray}

The resonance condition (\ref{res_cond_gen}) was derived in \cite%
{Ashhab:2007} using an alternative, geometric visualization of the dynamics.
There, the evolution matrix for one full driving period, i.e.~$%
(NU_{2}NU_{1}) $, is decomposed into two rotations, a rotation around an
axis in the $xy$ plane followed by a rotation around the $z$-axis. The
resonance condition is then identified as the requirement that the $z$-axis
rotation does not affect the dynamics, i.e.~the $z$-axis rotation is a
rotation by a multiple of $2\pi $. Straightforward algebra then gives Eq.~(%
\ref{res_cond_gen}). A related alternative derivation of the resonance
condition relies on visualizing the evolution matrix for one full driving
period as a single rotation: on resonance the rotation axis lies in the $xy$%
-plane

First, let us consider the case of the \textit{slow-passage limit}, where $%
\delta\gg 1$ and $P_{\mathrm{LZ}}\ll 1$. We now neglect the difference
between $\overline{P_{+}^{(\mathrm{I})}}$ and $\overline{P_{+}^{(\mathrm{II}%
)}}$ and, from Eq.~(\ref{P_II_avrgd}) to the lowest order approximation in $%
P_{\mathrm{LZ}}$, we obtain:

\end{subequations}
\begin{equation}
\overline{P_{+}}=\frac{P_{\mathrm{LZ}}(1+\cos \zeta _{+}\cos \zeta _{-})}{%
\sin ^{2}\zeta _{+}+2P_{\mathrm{LZ}}(1+\cos \zeta _{+}\cos \zeta _{-})}.
\label{ShIF}
\end{equation}

For small offset $\varepsilon _{0}/A\ll 1$ the expressions for $\zeta _{\pm
} $ can be simplified: $\zeta _{1}+\zeta _{2}\approx 2A/\omega $, $\zeta
_{1}-\zeta _{2}\approx \pi \varepsilon _{0}/\omega $. This makes Eq.~(\ref%
{ShIF}) analogous to the result obtained in \cite{Shytov:2003} (their
Eq.~(20) at $u=w\rightarrow 0$). But for arbitrary (not small) offset $%
\varepsilon _{0}$ one has to make use of Eq.~(\ref{ShIF}) with $\zeta _{\pm
} $ given by Eqs.~(\ref{zeta_pm}) and (\ref{zeta_1_2}); the result, such as
Fig.~\ref{Fig:LZSI_for_HUT} is significantly dependent on this correction.

One can also obtain an idea about the form of the resonance condition in the
zeroth in $\Delta/ \varepsilon_0$ approximation:
\begin{subequations}
\begin{eqnarray}
\zeta_1 & \approx & \frac{1}{2} \int_{t_1}^{t_2} \varepsilon(t) dt  \notag \\
& = & \frac{\pi \varepsilon_0}{\omega} - \frac{\varepsilon_0}{\omega}
\cos^{-1}\frac{\varepsilon_0}{A} + \frac{\sqrt{A^2-\varepsilon_0^2}}{\omega}
,   \label{B14a} \\
\zeta_2 & \approx & - \frac{\varepsilon_0}{\omega} \cos^{-1}\frac{%
\varepsilon_0}{A} + \frac{\sqrt{A^2-\varepsilon_0^2}}{\omega}.  \label{B14b}
\end{eqnarray}
Equation (\ref{B12a}) then gives
\end{subequations}
\begin{equation}
\frac{\varepsilon_0}{\omega} + 2 \frac{\sqrt{A^2-\varepsilon_0^2}}{\pi \omega%
} - 2 \frac{\varepsilon_0}{\pi \omega} \cos^{-1} \frac{\varepsilon_0}{A} = k.
\end{equation}
We emphasize, however, that the above equation should be seen as a limiting
case for small $\Delta$ that demonstrates the complicated form of the
resonance condition.

Now consider the case of \textit{fast passage}, where $\left( 1-P_{\mathrm{LZ%
}}\right) \approx 2\pi \delta \ll 1$. The fast limit ($\delta \ll 1$)
assumes a small value of $\Delta $. In this limit there is a large
probability ($P_{\mathrm{LZ}}\sim 1$) for transitions between the adiabatic
states in a single-passage; while the transition probability between
diabatic states is small, $\left( 1-P_{\mathrm{LZ}}\right) \ll 1$. Hence, we
will consider the time-averaged probability of the upper diabatic state $P_{%
\mathrm{up}}$. Note that upper diabatic state is $\varphi _{\uparrow }$ for $%
\varepsilon _{0}<0$ and $\varphi _{\downarrow }$ for $\varepsilon _{0}>0$.
Let us consider the latter case, as shown in Fig.~\ref{E(e)}. Then we can
relate $P_{\mathrm{up}}$ to $P_{+}^{(I)}$ and $P_{+}^{(II)}$ as follows:
\begin{equation}
P_{\mathrm{up}}\!\approx \!P_{+}^{(I)}\text{ \ for }\left( t-2\pi n/\omega
\right) \in (t_{1},t_{2}),
\end{equation}%
\begin{equation}
P_{\mathrm{up}}\!\approx \!1-P_{+}^{(II)}\text{\ for }\left( t-2\pi n/\omega
\right) \in (t_{2},t_{1}+2\pi /\omega ).
\end{equation}%
Then we note that $1-P_{+}^{(II)}\approx P_{+}^{(I)}$ and obtain:
\begin{equation}
\overline{P}_{\mathrm{up}}=\frac{1}{2}\frac{4\cos ^{2}(\zeta _{2}-\pi /4)}{%
\sin ^{2}\zeta _{-}+4\cos ^{2}(\zeta _{2}-\pi /4)},
\end{equation}%
where $\zeta_{2}$ is given by Eq.~(\ref{B14b}). On resonance $\zeta
_{-}=k\pi $ and $\overline{P}_{\mathrm{up}}=1/2$, then with $\zeta
_{-}\approx \pi \varepsilon _{0}/\omega $ for small offset ($\varepsilon
_{0}/A\ll 1$) we obtain the resonance frequency $\omega ^{(k)}=\varepsilon
_{0}/k$. Consider now the vicinity of the $k$-th resonance, introducing a
small frequency detuning as follows
\begin{equation}
\delta \omega =k\omega -\varepsilon _{0}\text{.}
\end{equation}%
Then to first approximation in $\varepsilon _{0}/A$, we obtain:
\begin{subequations}
\label{Pup_in_AI_model}
\begin{eqnarray}
\overline{P}_{\mathrm{up}}^{(k)} &=&\frac{1}{2}\frac{\Delta _{k}^{2}}{\Delta
_{k}^{2}+\delta \omega ^{2}}, \\
\Delta _{k} &=&\Delta \sqrt{\frac{2\omega }{\pi A}}\cos \left( \frac{A}{%
\omega }-\frac{\pi }{4}(2k+1)\right) .
\end{eqnarray}
\end{subequations}

The total probability $\overline{P}_{\mathrm{up}}$ is obtained as
the sum of the partial contributions
$\overline{P}_{\mathrm{up}}^{(k)}$. This result, obtained in the
adiabatic-impulse model, coincides with the one obtained within
the RWA, Eqs.~(\ref{P}-\ref{bessel}).

\section{Rotating-wave approximation}

\subsection{Hamiltonian in the RWA-form}

Following \cite{Pegg:1973}, \cite{Lopez-Castillo:1992}, \cite{Oliver:2005}
and \cite{Ashhab:2007} we consider the rotating-wave approximation (RWA).

First, we split the Hamiltonian into time-independent and time-dependent
parts:

%\end{subequations}

\begin{equation}
H=-\frac{\Delta }{2}\sigma _{x}-\frac{\varepsilon _{0}+A\cos \omega t}{2}%
\sigma _{z}=H_{0}+V(t),  \label{Ham_with_cos}
\end{equation}%
\begin{subequations}
\begin{eqnarray}
H_{0} &=&-\frac{\Delta }{2}\sigma _{x}-\frac{\varepsilon _{0}}{2}\sigma _{z},
\\
V(t) &=&-\frac{A\cos \omega t}{2}\sigma _{z}.
\end{eqnarray}%
Here we consider the time-dependent term with a cosine function (in contrast
to the sine function used in Eq.~(\ref{eps(t)})) because this choice agrees
with the standard convention in the literature. The difference between the
two choices is simply a shift of $\omega t$ by $\pi /2$, and it does not
change the time-averaged probabilities which are of most interest to us.

Then we make a transformation to a rotating frame with the operator $W$:
\end{subequations}
\begin{equation}
W(t)=\exp \left( -i\int V(t)dt\right) =\exp \left( i\frac{\eta (t)}{2}\sigma
_{z}\right) ,  \label{U(t)}
\end{equation}%
\begin{equation}
\eta (t)=\frac{A}{\omega }\sin \omega t.
\end{equation}%
This operator matches the wave function in the reference frame $\psi $ with
the wave function in the rotating frame $\psi ^{\prime }$: $\psi =W(t)\psi
^{\prime }$. Then the Shr\"{o}dinger equation in the rotating frame reads: $i%
\dot{\psi}^{\prime }=H^{\prime }\psi ^{\prime }$ with
\begin{equation}
H^{\prime }=W^{+}H_{0}W=-\frac{\Delta }{2}(e^{-i\eta (t)}\sigma _{+}+h.c.)-%
\frac{\varepsilon _{0}}{2}\sigma _{z},
\end{equation}%
$\sigma _{+}=\frac{1}{2}\left( \sigma _{x}+i\sigma _{y}\right) $. Making use
of the formula
\begin{equation}
e^{ix\sin \tau }=\sum\limits_{n=-\infty }^{\infty }J_{n}(x)\;e^{in\tau },
\end{equation}%
the new Hamiltonian takes the form%
\begin{equation}
H^{\prime }=-\sum\limits_{n=-\infty }^{\infty }\frac{\Delta _{n}}{2}\left(
e^{-in\omega t}\sigma _{+}+h.c.\right) -\frac{\varepsilon _{0}}{2}\sigma
_{z},  \label{H_prime}
\end{equation}%
\begin{equation}
\Delta _{n}=\Delta J_{n}(A/\omega ).
\end{equation}%
Note that the unitary transformation $W$ does not change the level
occupation probabilities (the absolute values of the spinor components):%
\begin{equation}
\psi =\binom{\psi _{1}}{\psi _{2}}=W\psi ^{\prime }=e^{i\frac{\eta }{2}%
\sigma _{z}}\binom{\psi _{1}^{\prime }}{\psi _{2}^{\prime }}=\binom{e^{i%
\frac{\eta }{2}}\psi _{1}^{\prime }}{e^{-i\frac{\eta }{2}}\psi _{2}^{\prime }%
}.
\end{equation}

\subsection{Solving the Schr\"{o}dinger equation in the absence of relaxation%
}

Let us now study first the system without relaxation. We now look for a
solution of the Schr\"{o}dinger equation $i\dot{\psi}^{\prime }=H^{\prime
}\psi ^{\prime }$ in the form:
\begin{equation}
\psi ^{\prime }=\binom{\psi _{1}^{\prime \prime }\exp (-i\frac{k\omega t}{2})%
}{\psi _{2}^{\prime \prime }\exp (i\frac{k\omega t}{2})}.
\end{equation}%
Actually, this corresponds to the transformation: $\psi ^{\prime \prime }=%
\widetilde{W}(t)\psi ^{\prime }$ with $\widetilde{W}(t)=\exp (i\frac{k\omega
t}{2}\sigma _{z})$.

Consider now parameters close to the $k$-photon resonance, where $k\omega
\approx \Delta E=\left\vert \varepsilon _{0}\right\vert $. In other words,
we now consider the frequency detuning $\delta \omega =k\omega -\varepsilon
_{0}$ to be small. The RWA consists in the assumption that, in the vicinity
of the $k$-th resonance, fast oscillating terms with $n\neq k$ can be
neglected. Then the Schr\"{o}dinger equation can be readily solved. The
probability of the upper diabatic state (if the system was initially in the
lower diabatic state) is given by:
\begin{equation}
\overline{P}_{\mathrm{up}}^{(k)}(t)\!\!=\!\!\left\vert \psi
_{2}(t)\right\vert ^{2}\!\!=\!\!\frac{1}{2}\frac{\Delta _{k}^{2}}{\delta
\omega ^{2}+\Delta _{k}^{2}}\left( 1-\cos \sqrt{\delta \omega ^{2}+\Delta
_{k}^{2}}t\right) .
\end{equation}

\subsection{Solving the Bloch equations with relaxation}

Now let us take into account relaxation by using the Bloch equations, which
include two phenomenological relaxation times, $T_{1}$ and $T_{2}$ \cite%
{Abragam:1961}. A convenient parametrization for the density matrix is the
following:
\begin{equation}
\rho =\frac{1}{2}\left(
\begin{array}{cc}
1+Z & X-iY \\
X+iY & 1-Z%
\end{array}%
\right) .
\end{equation}%
Here we note that the Bloch equations have to be written in the adiabatic
(energy) representation, so that the constants $T_{1}$ and $T_{2}$ describe
relaxation to the equilibrium values: $Z\rightarrow Z_{0}$ and $\rho
_{21}=X+iY\rightarrow 0$. This representation differs from the diabatic
representation\ (with the states $\{\varphi _{\uparrow },\varphi
_{\downarrow }\}$) by a small rotation with the angle $\Delta /\varepsilon
_{0}\ll 1$. Here we ignore this difference in the bases and consider the
diabatic representation. Then for $Z$ and $\rho _{21}=X+iY$, the Bloch
equations with the Hamiltonian $H^{\prime }$ read:
\begin{subequations}
\label{Bloch}
\begin{eqnarray}
\dot{Z} &=&-\Delta _{k}\text{Im}\left( e^{-ik\omega t}\rho _{21}\right) -%
\frac{Z-Z_{0}}{T_{1}}, \\
\dot{\rho}_{21} &=&i\Delta _{k}e^{ik\omega t}Z-i\varepsilon _{0}\rho _{21}-%
\frac{\rho _{21}}{T_{2}}.
\end{eqnarray}%
(Again, we have neglected fast oscillating terms with $n\neq k$.) Here $%
Z_{0}=\tanh \frac{\Delta E}{2T}$ describes the equilibrium energy level
distribution at temperature $T$. After the substitution $\widetilde{\rho }%
_{21}=\rho _{21}\exp (-ik\omega t)=\widetilde{X}+i\widetilde{Y}$, we obtain
the system of equations:
\end{subequations}
\begin{subequations}
\begin{eqnarray}
\overset{.}{\widetilde{X}} &=&\delta \omega \widetilde{Y}-\frac{\widetilde{X}%
}{T_{2}}, \\
\overset{.}{\widetilde{Y}} &=&-\delta \omega \widetilde{X}+\Delta _{k}Z-%
\frac{\widetilde{Y}}{T_{2}}, \\
\dot{Z} &=&-\Delta _{k}\widetilde{Y}-\frac{Z-Z_{0}}{T_{1}}.
\end{eqnarray}%
The stationary solution of these equations is obtained after substituting $%
\overset{.}{\widetilde{X}}=\overset{.}{\widetilde{Y}}=\dot{Z}=0$. Then the
stationary value for the probability of the upper diabatic state is given by
\end{subequations}
\begin{equation}
\overline{P}_{\mathrm{up}}^{(k)}=\overline{\rho }_{22}=\frac{1}{2}(1-%
\overline{Z}),  \label{P_stat}
\end{equation}%
\begin{equation}
\overline{Z}=\frac{1+T_{2}^{2}\delta \omega ^{2}}{1+T_{2}^{2}\delta \omega
^{2}+T_{1}T_{2}\Delta _{k}^{2}}Z_{0}.  \label{Z_stat}
\end{equation}%
At low temperatures we can substitute $Z_{0}\approx 1$ and obtain from Eqs.~(%
\ref{P_stat}-\ref{Z_stat}) the following formula%
\begin{equation}
\overline{P}_{\mathrm{up}}=\sum_{k}\overline{P}_{\mathrm{up}}^{(k)}=\frac{1}{%
2}\sum_{k}\frac{\Delta _{k}^{2}}{\frac{1}{T_{1}T_{2}}+\frac{T_{2}}{T_{1}}%
\delta \omega ^{2}+\Delta _{k}^{2}},  \label{P_with_T1_2}
\end{equation}%
which is useful for the description of multiphoton resonances at $\Delta
/\varepsilon _{0}\ll 1$.

\section{Floquet theory}

Floquet theory can be applied to any quantum system governed by a
time-periodic Hamiltonian. The main idea of applying this theoretical
approach is as follows: the time-dependent problem is turned into a
time-independent one at the expense of increasing the effective number of
states in the Hilbert space to an infinite number of states. Approximations
are then made to reduce the problem into a finite-dimensional one, and
analytic or numerical results are obtained. One advantage of Floquet theory
is that it allows for the inclusion of dissipative processes relatively
easily into the analysis. In this Appendix we briefly outline the
application of Floquet theory to the problem of LZS interferometry, closely
following \cite{Son:2009}. More detailed discussions can be found in \cite%
{Grifoni:1998}, \cite{Chu:2004}, and \cite{Son:2009}.

From Floquet's theorem we know that, since the Hamiltonian given by Eqs.~(%
\ref{Hamiltonian}) and (\ref{eps(t)}) is periodic, there exist two solutions
to the Schr\"{o}dinger equation with the property that
\begin{equation}
\psi _{j}(t)=e^{-i\widetilde{\epsilon }_{j}t}\widetilde{\psi }_{j}(t)
\end{equation}%
and $\widetilde{\psi }_{j}(t)$ is a periodic function of $t$ with the same
period as the Hamiltonian, i.e.,~$2\pi /\omega $. Furthermore, these two
quasi-periodic solutions are orthogonal and form a complete basis. As a
result, knowledge of these states and of the quasi-energies $\widetilde{%
\epsilon }$ allows one to describe the dynamics for any given initial state.

Finding the Floquet states and quasi-energies is typically achieved using
Fourier analysis. Writing
\begin{equation}
\widetilde{\psi }_{j}(t)=\sum_{n=-\infty }^{+\infty }\xi _{j,n}e^{in\omega
t},
\end{equation}%
where each $\xi _{j,n}=(\xi _{j,n,+},\xi _{j,n,-})^{T}$ is a two-component
vector (with components $\xi _{j,n,+}$ and $\xi _{j,n,-}$), and writing the
Hamiltonian (\ref{Ham_with_cos}) as

\begin{equation}
H(t)=-\frac{\Delta }{2}\sigma _{x}-\frac{\varepsilon _{0}}{2}\sigma _{z}-%
\frac{A}{4}\sigma _{z}\left( e^{i\omega t}+e^{-i\omega t}\right) ,
\end{equation}%
one obtains the following set of equations governing $\xi _{jn}$:
\begin{equation}
\widetilde{\varepsilon }\xi _{j,n}=\left( -\frac{\Delta }{2}\sigma _{x}-%
\frac{\varepsilon _{0}}{2}\sigma _{z}+n\omega \right) \xi _{jn}-\frac{A}{4}%
\sigma _{z}\xi _{j,n-1}-\frac{A}{2}\sigma _{z}\xi _{j,n+1}.
\end{equation}%
The task is now to find $\xi _{j,n}$ for all values of $n$ (noting that $\xi
_{j,n}$ is composed of the components $\xi _{j,n,-}$ and $\xi _{j,n,+}$) and
along with them the quasi-energy $\widetilde{\varepsilon }_{j}$ (note that
the above equations apply to both values of $j$). If one constructs the
infinite-dimensional vector $\widetilde{\xi }_{j}=\{...,\xi _{j,n-1,-},\xi
_{j,n-1,+},\xi _{j,n,-},\xi _{j,n,+},\xi _{j,n+1,-},\xi _{j,n+1,+},...\}$,
then the above set of equations governing $\xi _{j,n}$ turns into a single
eigenvalue problem
\begin{equation}
\widetilde{\varepsilon }_{j}\widetilde{\xi }_{j}=H_{F}\widetilde{\xi }_{j}
\end{equation}%
where %\begin{widetext}
\begin{equation}
H_{F}=\left(
\begin{array}{cccccccc}
\ddots &  &  &  &  &  &  &  \\
& -\frac{\varepsilon _{0}}{2}+(n-1)\omega & -\frac{\Delta }{2} & -\frac{A}{4}
& 0 & 0 & 0 &  \\
& -\frac{\Delta }{2} & \frac{\varepsilon _{0}}{2}+(n-1)\omega & 0 & \frac{A}{%
4} & 0 & 0 &  \\
& -\frac{A}{4} & 0 & -\frac{\varepsilon _{0}}{2}+n\omega & -\frac{\Delta }{2}
& -\frac{A}{4} & 0 &  \\
& 0 & \frac{A}{4} & -\frac{\Delta }{2} & \frac{\varepsilon _{0}}{2}+n\omega
& 0 & \frac{A}{4} &  \\
& 0 & 0 & -\frac{A}{4} & 0 & -\frac{\varepsilon _{0}}{2}+(n+1)\omega & -%
\frac{\Delta }{2} &  \\
& 0 & 0 & 0 & \frac{A}{4} & -\frac{\Delta }{2} & \frac{\varepsilon _{0}}{2}%
+(n+1)\omega &  \\
&  &  &  &  &  &  & \ddots \\
&  &  &  &  &  &  &
\end{array}%
\right) .  \label{H_Floquet}
\end{equation}%
%
%
%
%
%
%
%
%
%
%
%
%
%
%
%
%
%
%
%
%
%
%
%
%
%
%
%\end{widetext}
This eigenvalue problem has an infinite number of solutions, even though we
started by looking for only two solutions. This issue is resolved by noting
that this infinite number of solutions can be divided into two groups, each
of which contains an infinite number of equivalent solutions; clearly, if
there is a solution to the original Floquet problem with quasi-energy $%
\widetilde{\varepsilon }$, then essentially the same solution can be
re-written such that the quasi-energy is given by $\widetilde{\varepsilon }%
+n\omega $, with any integer $n$.

The eigenvalue problem derived above cannot be solved in analytic closed
form, and approximations have to be made. It is common to treat $\Delta $ as
a small parameter in a perturbation-theory-like calculation. Taking $\Delta
=0$, one finds the (unperturbed) eigenvectors \cite{Son:2009}
\begin{equation}
\widetilde{\xi }_{k,-}^{(0)}:\left\{
\begin{array}{lcl}
& \vdots &  \\
\xi _{k,-,n+k-1,-} & = & J_{1}\left( \frac{A}{2\omega }\right) \\
\xi _{k,-,n+k-1,+} & = & 0 \\
\xi _{k,-,n+k,-} & = & J_{0}\left( \frac{A}{2\omega }\right) \\
\xi _{k,-,n+k,+} & = & 0 \\
\xi _{k,-,n+k+1,-} & = & J_{1}\left( \frac{A}{2\omega }\right) \\
\xi _{k,-,n+k+1,+} & = & 0 \\
& \vdots &
\end{array}%
\right. \text{ \ \ \ \ \ \ \ \ \ \ \ \ \ \ \ \ \ }\widetilde{\xi }%
_{k,+}^{(0)}:\left\{
\begin{array}{lcl}
& \vdots &  \\
\xi _{k,+,n+k-1,-} & = & 0 \\
\xi _{k,+,n+k-1,+} & = & J_{1}\left( -\frac{A}{2\omega }\right) \\
\xi _{k,+,n+k,-} & = & 0 \\
\xi _{k,+,n+k,+} & = & J_{0}\left( -\frac{A}{2\omega }\right) \\
\xi _{k,+,n+k+1,-} & = & 0 \\
\xi _{k,+,n+k+1,+} & = & J_{1}\left( -\frac{A}{2\omega }\right) \\
& \vdots &
\end{array}%
\right. ,  \notag
\end{equation}%
with quasienergies $\widetilde{\varepsilon }_{k,\pm }=\pm \varepsilon
_{0}/2-k\omega $ (note that the above eigenvectors are normalized to unity).
When two quasienergies are almost equal (specifically $\widetilde{%
\varepsilon }_{0,-}$ and $\widetilde{\varepsilon }_{k,+}$), and assuming
that $\Delta /\omega $ is small, all others are far off resonance from these
two, one can make a rotating-wave approximation and construct an effective
two-level problem that involves the two respective Floquet states. Using the
relation
\begin{equation}
\sum_{n=-\infty }^{\infty }J_{n}(x)J_{k-n}(x)=J_{k}(2x),
\end{equation}%
one can calculate the effective coupling strength between the two relevant
states. The resulting $2\times 2$ effective Hamiltonian is given by
\begin{equation}
H_{F,RWA}=\frac{1}{2}\left(
\begin{array}{cc}
-\varepsilon _{0} & -\Delta J_{-k}\left( A/\omega \right) \\
-\Delta J_{-k}\left( A/\omega \right) & \varepsilon _{0}-2k\omega%
\end{array}%
\right) .
\end{equation}%
Using this effective Hamiltonian, it is now straightforward to find the
Floquet states and quasi-energies. The separation between the quasienergies
on resonance (i.e.,~when $\varepsilon _{0}=k\omega $) is given by $|\Delta
J_{k}(A/\omega )|$ and determines the oscillation frequency between the two
states of the TLS, in clear analogy to the RWA calculation of Appendix C.
One can proceed with the calculation in order to obtain more accurate
approximations to the Floquet states and quasienergies \cite{Son:2009}, but
we shall not do this here.

In this appendix we have re-derived results that we had already derived
using the adiabatic-impulse model and the RWA. The use of Floquet theory in
this context can therefore be seen as just an alternative approach that can
be used to obtain results that can be obtained using other methods. One
advantage that the Floquet approach has over the adiabatic-impulse model and
the RWA is that decoherence can be introduced into the problem in the
Floquet approach following a standard formalism \cite{Grifoni:1998}, whereas
the Bloch equations that we have used in Sec.~II can be seen as a
phenomenological approach to including decoherence in the theoretical
analysis. The question of the effects of decoherence on LZS interferometry
in the Floquet approach is currently under study \cite{Hausinger:2010}.

\section{Dressed-state picture: quantized driving field}

In all our discussion above, we have treated the driving field as an
externally applied, classical driving signal. An alternative theoretical
approach that can be used to study the problem of a driven quantum system
treats the driving field quantum mechanically. The quantum system under
study is then enlarged from simply a two-level system to a two-level system
and a harmonic oscillator that can have any number of excitations (i.e.,
photons). As in the Floquet-theory approach, a major advantage that
justifies the expansion of the Hilbert space is the simplification
associated with turning the time-dependent Hamiltonian into a
time-independent one. In fact, the algebra encountered in the Floquet-theory
calculation is identical to the one encountered in the dressed-state
analysis. The difference between these two approaches is therefore simply
the physical interpretation of the mechanisms at play during the system
dynamics.

The equivalence between the Floquet approach and the dressed-state approach
can be seen by considering the Hamiltonian of the dressed-state picture:
\begin{equation}
H=-\frac{\Delta }{2}\sigma _{x}-\frac{\varepsilon _{0}}{2}\sigma _{z}+\omega
a^{\dagger }a-\frac{A}{2}\sigma _{z}\left( a+a^{\dagger }\right) ,
\end{equation}%
where $a$ and $a^{\dagger }$ are the annihilation and creation operators for
photons in the electromagnetic field. When written in matrix form, the above
Hamiltonian is given by Eq.~(\ref{H_Floquet}) derived in the Floquet
approach. More detailed studies of the dressed-state picture can be found in
[\cite{Liu:2006}, \cite{Greenberg:2007}, \cite{Wilson:2007, Wilson:2010}].

%\bibliography{nc,cds,sw,connes}

%\bibliography{rmp-sample}

\begin{thebibliography}{42}
\expandafter\ifx\csname
natexlab\endcsname\relax\def\natexlab#1{#1}\fi
\expandafter\ifx\csname bibnamefont\endcsname\relax
  \def\bibnamefont#1{#1}\fi
\expandafter\ifx\csname bibfnamefont\endcsname\relax
  \def\bibfnamefont#1{#1}\fi
\expandafter\ifx\csname citenamefont\endcsname\relax
  \def\citenamefont#1{#1}\fi
\expandafter\ifx\csname url\endcsname\relax
  \def\url#1{\texttt{#1}}\fi
\expandafter\ifx\csname
urlprefix\endcsname\relax\def\urlprefix{URL }\fi
\providecommand{\bibinfo}[2]{#2}
\providecommand{\eprint}[2][]{\url{#2}}

\bibitem[Abragam(1961)Abragam]{Abragam:1961} \bibinfo{author}{%
\bibnamefont{Abragam}, \bibfnamefont{A.}}, \bibinfo{year}{1961}, \emph{%
\bibinfo{title}{The Principles of Nuclear Magnetism}} (%
\bibinfo{publisher}{Clarendon, Oxford}).

\bibitem[Amin(2009)Amin]{Amin:2009}
\bibinfo{author}{\bibnamefont{Amin},
\bibfnamefont{M. H. S.}}, \bibinfo{year}{2009},
\bibinfo{journal}{Phys. Rev.
Lett.} \textbf{\bibinfo{volume}{102}}, \bibinfo{pages}{122001}.

\bibitem[Ankerhold and Grabert(2003)Ankerhold and Grabert]{Ankerhold:2003} %
\bibinfo{author}{\bibnamefont{Ankerhold}, \bibfnamefont{J.}} and %
\bibinfo{author}{\bibfnamefont{H.}~\bibnamefont{Grabert}}, %
\bibinfo{year}{2003}, \bibinfo{journal}{Phys. Rev. Lett.} \textbf{%
\bibinfo{volume}{91}}, \bibinfo{pages}{016803}.

\bibitem[Ao and Rammer(1991)Ao and Rammer]{Ao:1991} \bibinfo{author}{%
\bibnamefont{Ao}, \bibfnamefont{P.}} and \bibinfo{author}{\bibfnamefont{J.}~%
\bibnamefont{Rammer}}, \bibinfo{year}{1991}, \bibinfo{journal}{Phys. Rev. B}
\textbf{\bibinfo{volume}{43}}, \bibinfo{pages}{5397}.

\bibitem[Aravind and Hirschfelder(1984)Aravind and Hirschfelder]%
{Aravind:1984} \bibinfo{author}{\bibnamefont{Aravind},
\bibfnamefont{P. K.}} and
\bibinfo{author}{\bibfnamefont{J.
O.}~\bibnamefont{Hirschfelder}}, \bibinfo{year}{1984},
\bibinfo{journal}{J.
Phys. Chem.} \textbf{\bibinfo{volume}{88}}, \bibinfo{pages}{4788}.

\bibitem[Arbo \emph{et~al.}(2010)Arbo, Ishikawa, Schiessl, Persson, and
Bugdorfer]{Arbo:2010}
\bibinfo{author}{\bibnamefont{Arbo},
\bibfnamefont{D. G.}}, \bibinfo{author}{\bibfnamefont{K.
L.}~\bibnamefont{Ishikawa}}, \bibinfo{author}{\bibfnamefont{K.}~%
\bibnamefont{Schiessl}}, \bibinfo{author}{\bibfnamefont{E.}~%
\bibnamefont{Persson}}, and \bibinfo{author}{\bibfnamefont{J.}~%
\bibnamefont{Bugdorfer}}, \bibinfo{year}{2010}, \bibinfo{journal}{Phys. Rev.
A} \textbf{\bibinfo{volume}{81}}, \bibinfo{pages}{021403}.

\bibitem[Ashhab \emph{et~al.}(2006)Ashhab]{Ashhab:2006} \bibinfo{author}{%
\bibnamefont{Ashhab}, \bibfnamefont{S.}},
\bibinfo{author}{\bibfnamefont{J.
R.}~\bibnamefont{Johansson}}, and \bibinfo{author}{\bibfnamefont{F.}~%
\bibnamefont{Nori}}, \bibinfo{year}{2006}, \bibinfo{journal}{Phys. Rev. A}
\textbf{\bibinfo{volume}{74}}, \bibinfo{pages}{052330}.

\bibitem[Ashhab \emph{et~al.}(2007)Ashhab]{Ashhab:2007} \bibinfo{author}{%
\bibnamefont{Ashhab}, \bibfnamefont{S.}},
\bibinfo{author}{\bibfnamefont{J.
R.}~\bibnamefont{Johansson}},
\bibinfo{author}{\bibfnamefont{A.
M.}~\bibnamefont{Zagoskin}}, and \bibinfo{author}{\bibfnamefont{F.}~%
\bibnamefont{Nori}}, \bibinfo{year}{2007}, \bibinfo{journal}{Phys. Rev. A}
\textbf{\bibinfo{volume}{75}}, \bibinfo{pages}{063414}.

\bibitem[Ashhab \emph{et~al.}(2009)Ashhab]{Ashhab:2009} \bibinfo{author}{%
\bibnamefont{Ashhab}, \bibfnamefont{S.}},
\bibinfo{author}{\bibfnamefont{J.
R.}~\bibnamefont{Johansson}},
\bibinfo{author}{\bibfnamefont{A.
M.}~\bibnamefont{Zagoskin}}, and \bibinfo{author}{\bibfnamefont{F.}~%
\bibnamefont{Nori}}, \bibinfo{year}{2009}, \bibinfo{journal}{New. J. Phys.}
\textbf{\bibinfo{volume}{11}}, \bibinfo{pages}{023030}.

\bibitem[Astafiev \emph{et~al.}(2007)Astafiev]{Astafiev:2007} %
\bibinfo{author}{\bibnamefont{Astafiev}, \bibfnamefont{O.}}, %
\bibinfo{author}{\bibfnamefont{K.}~\bibnamefont{Inomata}}, %
\bibinfo{author}{\bibfnamefont{A. O.}~\bibnamefont{Niskanen}}, %
\bibinfo{author}{\bibfnamefont{T.}~\bibnamefont{Yamamoto}}, %
\bibinfo{author}{\bibfnamefont{Yu.}~\bibnamefont{Pashkin}}, %
\bibinfo{author}{\bibfnamefont{Y.}~\bibnamefont{Nakamura}}, %
\bibinfo{author}{\bibfnamefont{J. S.}~\bibnamefont{Tsai}}, %
\bibinfo{year}{2007}, \bibinfo{journal}{Nature} \textbf{\bibinfo{volume}{449}%
}, \bibinfo{pages}{588}.

\bibitem[Autler and Townes(1955)Autler and Townes]{Autler:1955} %
\bibinfo{author}{\bibnamefont{Autler}, \bibfnamefont{S. H.}} and %
\bibinfo{author}{\bibfnamefont{C. F.}~\bibnamefont{Townes}}, %
\bibinfo{year}{1955}, \bibinfo{journal}{Phys. Rev.} \textbf{%
\bibinfo{volume}{100}}, \bibinfo{pages}{703}.

\bibitem[Averbukh and Perel'man(1985)Averbukh and Perel'man]{Averbukh:1985} %
\bibinfo{author}{\bibnamefont{Averbukh}, \bibfnamefont{I. Sh.}} and %
\bibinfo{author}{\bibfnamefont{I. F.}~\bibnamefont{Perel'man}}, %
\bibinfo{year}{1985}, \bibinfo{journal}{Sov. Phys. JETP} \textbf{%
\bibinfo{volume}{61}}, \bibinfo{pages}{665}.

\bibitem[Banerjee and Yakovenko(2008)Banerjee and Yakovenko]{Banerjee:2008} %
\bibinfo{author}{\bibnamefont{Banerjee}, \bibfnamefont{A.}} and %
\bibinfo{author}{\bibfnamefont{V. M.}~\bibnamefont{Yakovenko}}, %
\bibinfo{year}{2008}, \bibinfo{journal}{Phys. Rev. B} \textbf{%
\bibinfo{volume}{78}}, \bibinfo{pages}{125404}.

\bibitem[Barone \emph{et~al.}(1977)Barone, Narcowich, and Narcowich]%
{Barone:1977} \bibinfo{author}{\bibnamefont{Barone}, \bibfnamefont{S. R.}}, %
\bibinfo{author}{\bibfnamefont{M. A.}~\bibnamefont{Narcowich}}, and %
\bibinfo{author}{\bibfnamefont{F. J.}~\bibnamefont{Narcowich}}, %
\bibinfo{year}{1977}, \bibinfo{journal}{Phys. Rev. A} \textbf{%
\bibinfo{volume}{15}}, \bibinfo{pages}{1109}.

\bibitem[Baruch and Gallagher(1992)Baruch and Gallagher]{Baruch:1992} %
\bibinfo{author}{\bibnamefont{Baruch}, \bibfnamefont{M. C.}} and %
\bibinfo{author}{\bibfnamefont{T. F.}~\bibnamefont{Gallagher}}, %
\bibinfo{year}{1992}, \bibinfo{journal}{Phys. Rev. Lett.} \textbf{%
\bibinfo{volume}{68}}, \bibinfo{pages}{3515}.

\bibitem[Benderskii \emph{et~al.}(2003)Benderskii, Vetoshkin, and Kats]%
{Benderskii:2003}
\bibinfo{author}{\bibnamefont{Benderskii},
\bibfnamefont{V. A.}},
\bibinfo{author}{\bibfnamefont{E.
V.}~\bibnamefont{Vetoshkin}}, and
\bibinfo{author}{\bibfnamefont{E.
I.}~\bibnamefont{Kats}}, \bibinfo{year}{2003}, \bibinfo{journal}{JETP}
\textbf{\bibinfo{volume}{97}}, \bibinfo{pages}{232}.

\bibitem[Berns \emph{et~al.}(2006)Berns, Oliver, Valenzuela, Shytov,
Berggren, Levitov, and Orlando]{Berns:2006} \bibinfo{author}{%
\bibnamefont{Berns}, \bibfnamefont{D. M.}}, \bibinfo{author}{%
\bibfnamefont{W. D.}~\bibnamefont{Oliver}}, \bibinfo{author}{%
\bibfnamefont{S. O.}~\bibnamefont{Valenzuela}}, \bibinfo{author}{%
\bibfnamefont{A. V.}~\bibnamefont{Shytov}}, \bibinfo{author}{%
\bibfnamefont{K. K.}~\bibnamefont{Berggren}}, \bibinfo{author}{%
\bibfnamefont{L. S.}~\bibnamefont{Levitov}}, and \bibinfo{author}{%
\bibfnamefont{T. P.}~\bibnamefont{Orlando}}, \bibinfo{year}{2006}, %
\bibinfo{journal}{Phys. Rev. Lett.} \textbf{\bibinfo{volume}{97}}, %
\bibinfo{pages}{150502}.

\bibitem[Berns \emph{et~al.}(2008)Berns, Rudner, Valenzuela, Berggren,
Oliver, Levitov, and Orlando]{Berns:2008} \bibinfo{author}{%
\bibnamefont{Berns}, \bibfnamefont{D. M.}}, \bibinfo{author}{%
\bibfnamefont{M. S.}~\bibnamefont{Rudner}}, \bibinfo{author}{%
\bibfnamefont{S. O.}~\bibnamefont{Valenzuela}}, \bibinfo{author}{%
\bibfnamefont{K. K.}~\bibnamefont{Berggren}}, \bibinfo{author}{%
\bibfnamefont{W. D.}~\bibnamefont{Oliver}}, \bibinfo{author}{%
\bibfnamefont{L. S.}~\bibnamefont{Levitov}}, and \bibinfo{author}{%
\bibfnamefont{T. P.}~\bibnamefont{Orlando}}, \bibinfo{year}{2008}, %
\bibinfo{journal}{Nature} \textbf{\bibinfo{volume}{455}}, \bibinfo{pages}{51}%
.

\bibitem[Bishop \emph{et~al.}(2009)Bishop]{Bishop:2009} \bibinfo{author}{%
\bibnamefont{Bishop}, \bibfnamefont{L. S.}}, \bibinfo{author}{%
\bibfnamefont{J. M.}~\bibnamefont{Chow}}, \bibinfo{author}{%
\bibfnamefont{J.}~\bibnamefont{Koch}}, \bibinfo{author}{\bibfnamefont{A.
A.}~\bibnamefont{Houck}}, \bibinfo{author}{\bibfnamefont{M.
H.}~\bibnamefont{Devoret}}, \bibinfo{author}{\bibfnamefont{E.}~%
\bibnamefont{Thuneberg}}, \bibinfo{author}{\bibfnamefont{S.
M.}~\bibnamefont{Girvin}}, \bibinfo{author}{\bibfnamefont{R.
J.}~\bibnamefont{Schoelkopf}}, \bibinfo{year}{2009}, %
\bibinfo{journal}{Nature Phys.} \textbf{\bibinfo{volume}{5}}, %
\bibinfo{pages}{105}.

\bibitem[Burkard(2010)Burkard]{Burkard:2010}
\bibinfo{author}{\bibnamefont{Burkard},
\bibfnamefont{G.}}, \bibinfo{year}{2010}, \bibinfo{journal}{Science} \textbf{%
\bibinfo{volume}{327}}, \bibinfo{pages}{651}.

\bibitem[Bychkov and Dykhne(1970)Bychkov and Dykhne]{Bychkov:1970} %
\bibinfo{author}{\bibnamefont{Bychkov}, \bibfnamefont{Yu. A.}} and %
\bibinfo{author}{\bibfnamefont{A. M.}~\bibnamefont{Dykhne}}, %
\bibinfo{year}{1970}, \bibinfo{journal}{Sov. Phys. JETP} \textbf{%
\bibinfo{volume}{31}}, \bibinfo{pages}{928}.

\bibitem[Bylander \emph{et~al.}(2009)Bylander, Rudner, Shytov, Valenzuela,
Berns, Berggren, Levitov, and Oliver]{Bylander:2009} \bibinfo{author}{%
\bibnamefont{Bylander}, \bibfnamefont{J.}}, \bibinfo{author}{%
\bibfnamefont{M. S.}~\bibnamefont{Rudner}}, \bibinfo{author}{%
\bibfnamefont{A. V.}~\bibnamefont{Shytov}}, \bibinfo{author}{%
\bibfnamefont{S. O.}~\bibnamefont{Valenzuela}}, \bibinfo{author}{%
\bibfnamefont{D. M.}~\bibnamefont{Berns}},
\bibinfo{author}{\bibfnamefont{K.
K.}~\bibnamefont{Berggren}},
\bibinfo{author}{\bibfnamefont{L.
S.}~\bibnamefont{Levitov}}, and
\bibinfo{author}{\bibfnamefont{W.
D.}~\bibnamefont{Oliver}}, \bibinfo{year}{2009},
\bibinfo{journal}{Phys.
Rev. B} \textbf{\bibinfo{volume}{80}}, \bibinfo{pages}{220506}.

\bibitem[Calero \emph{et~al.}(2005)Calero, Chudnovsky, and Garanin]%
{Calero:2005} \bibinfo{author}{\bibnamefont{Calero}, \bibfnamefont{C.}}, %
\bibinfo{author}{\bibfnamefont{E. M.}~\bibnamefont{Chudnovsky}}, and %
\bibinfo{author}{\bibfnamefont{D. A.}~\bibnamefont{Garanin}}, %
\bibinfo{year}{2005}, \bibinfo{journal}{Phys. Rev. B} \textbf{%
\bibinfo{volume}{72}}, \bibinfo{pages}{024409}.

\bibitem[Child(1974)Child]{Child:1974}
\bibinfo{author}{\bibnamefont{Child},
\bibfnamefont{M. S.}}, \bibinfo{year}{1974}, \emph{%
\bibinfo{title}{Molecular
Collision Theory}} (\bibinfo{publisher}{Academic Press,
London}).

\bibitem[Chu and Tel'nov(2004)Chu and Tel'nov]{Chu:2004} \bibinfo{author}{%
\bibnamefont{Chu}, \bibfnamefont{S.-I.}} and \bibinfo{author}{%
\bibfnamefont{D. A.}~\bibnamefont{Tel'nov}}, \bibinfo{year}{2004}, %
\bibinfo{journal}{Phys. Rep.} \textbf{\bibinfo{volume}{390}}, %
\bibinfo{pages}{1}.

\bibitem[Clarke and Wilhelm(2008)Clarke and Wilhelm]{Clarke:2008} %
\bibinfo{author}{\bibnamefont{Clarke}, \bibfnamefont{J.}} and %
\bibinfo{author}{\bibfnamefont{F. K.}~\bibnamefont{Wilhelm}}, %
\bibinfo{year}{2008}, \bibinfo{journal}{Nature} \textbf{\bibinfo{volume}{453}%
}, \bibinfo{pages}{1031}.

\bibitem[Coffey \emph{et~al.}(1969)Coffey, Lorents, and Smith]{Coffey:1969} %
\bibinfo{author}{\bibnamefont{Coffey}, \bibfnamefont{D., Jr.}}, %
\bibinfo{author}{\bibfnamefont{D. C.}~\bibnamefont{Lorents}}, and %
\bibinfo{author}{\bibfnamefont{F. T.}~\bibnamefont{Smith}}, %
\bibinfo{year}{1969}, \bibinfo{journal}{Phys. Rev.} \textbf{%
\bibinfo{volume}{187}}, \bibinfo{pages}{201}.

\bibitem[Cooper and Yakovenko(2006)Cooper and Yakovenko]{Cooper:2006} %
\bibinfo{author}{\bibnamefont{Cooper}, \bibfnamefont{B. K.}} and %
\bibinfo{author}{\bibfnamefont{V. M.}~\bibnamefont{Yakovenko}}, %
\bibinfo{year}{2006}, \bibinfo{journal}{Phys. Rev. Lett.} \textbf{%
\bibinfo{volume}{96}}, \bibinfo{pages}{037001}.

\bibitem[Damski(2005)Damski]{Damski:2005}
\bibinfo{author}{\bibnamefont{Damski},
\bibfnamefont{B.}}, \bibinfo{year}{2005},
\bibinfo{journal}{Phys. Rev.
Lett.} \textbf{\bibinfo{volume}{95}}, \bibinfo{pages}{035701}.

\bibitem[Damski and Zurek(2006)Damski and Zurek]{Damski:2006} %
\bibinfo{author}{\bibnamefont{Damski}, \bibfnamefont{B.}} and %
\bibinfo{author}{\bibfnamefont{W. H.}~\bibnamefont{Zurek}}, %
\bibinfo{year}{2006}, \bibinfo{journal}{Phys. Rev. A} \textbf{%
\bibinfo{volume}{73}}, \bibinfo{pages}{063405}.

\bibitem[Delone and Krainov(1985)Delone and Krainov]{Delone:1985} %
\bibinfo{author}{\bibnamefont{Delone}, \bibfnamefont{N. B.}} and %
\bibinfo{author}{\bibfnamefont{V. P.}~\bibnamefont{Krainov}}, %
\bibinfo{year}{1985}, \emph{\bibinfo{title}{Atoms in Strong Light Fields}} (%
\bibinfo{publisher}{Springer-Verlag, Berlin}),
\bibinfo{journal}{Springer
Series in Chemical Physics} \textbf{\bibinfo{volume}{28}}.

\bibitem[Delos and Thorson(1972)Delos and Thorson]{Delos:1972} %
\bibinfo{author}{\bibnamefont{Delos}, \bibfnamefont{J. B.}} and %
\bibinfo{author}{\bibfnamefont{W. R.}~\bibnamefont{Thorson}}, %
\bibinfo{year}{1972}, \bibinfo{journal}{Phys. Rev. A} \textbf{%
\bibinfo{volume}{6}}, \bibinfo{pages}{728}.

\bibitem[Devoret and Martinis(2004)Devoret and Martinis]{Devoret:2004} %
\bibinfo{author}{\bibnamefont{Devoret}, \bibfnamefont{M. H.}} and %
\bibinfo{author}{\bibfnamefont{J. M.}~\bibnamefont{Martinis}}, %
\bibinfo{year}{2004}, \bibinfo{journal}{Quant. Inf. Proc.} \textbf{%
\bibinfo{volume}{3}}, \bibinfo{pages}{163}.

\bibitem[Di Giacomo and Nikitin(2005)Di Giacomo and Nikitin]{Giacomo:2005} %
\bibinfo{author}{\bibnamefont{Di Giacomo}, \bibfnamefont{F.}} and %
\bibinfo{author}{\bibfnamefont{E.}~\bibnamefont{Nikitin}}, %
\bibinfo{year}{2005}, \bibinfo{journal}{Sov. Phys. Uspekhi} \textbf{%
\bibinfo{volume}{48}}, \bibinfo{pages}{515}.

\bibitem[Dutta \emph{et~al.}(2008)Dutta, Strauch, Lewis, Mitra, Paik,
Palomaki, Tiesinga, Anderson, Dragt, Lobb, Wellstood]{Dutta:2008} %
\bibinfo{author}{\bibnamefont{Dutta}, \bibfnamefont{S. K.}}, %
\bibinfo{author}{\bibfnamefont{F. W.}~\bibnamefont{Strauch}}, %
\bibinfo{author}{\bibfnamefont{R. M.}~\bibnamefont{Lewis}}, %
\bibinfo{author}{\bibfnamefont{K.}~\bibnamefont{Mitra}}, \bibinfo{author}{%
\bibfnamefont{H.}~\bibnamefont{Paik}},
\bibinfo{author}{\bibfnamefont{T.
A.}~\bibnamefont{Palomaki}}, \bibinfo{author}{\bibfnamefont{E.}~%
\bibnamefont{Tiesinga}},
\bibinfo{author}{\bibfnamefont{J.
R.}~\bibnamefont{Anderson}},
\bibinfo{author}{\bibfnamefont{A.
J.}~\bibnamefont{Dragt}},
\bibinfo{author}{\bibfnamefont{C.
J.}~\bibnamefont{Lobb}}, and
\bibinfo{author}{\bibfnamefont{F.
C.}~\bibnamefont{Wellstood}}, \bibinfo{year}{2008},
\bibinfo{journal}{Phys.
Rev. B} \textbf{\bibinfo{volume}{78}}, \bibinfo{pages}{104510}.

\bibitem[Duty \emph{et~al.}(2005)Duty, Johansson, Bladh, Gunnarsson, Wilson,
and Delsing]{Duty:2005}
\bibinfo{author}{\bibnamefont{Duty},
\bibfnamefont{T.}}, \bibinfo{author}{\bibfnamefont{G.}~%
\bibnamefont{Johansson}}, \bibinfo{author}{\bibfnamefont{K.}~%
\bibnamefont{Bladh}}, \bibinfo{author}{\bibfnamefont{D.}~%
\bibnamefont{Gunnarsson}},
\bibinfo{author}{\bibfnamefont{C.
M.}~\bibnamefont{Wilson}}, and \bibinfo{author}{\bibfnamefont{P.}~%
\bibnamefont{Delsing}}, \bibinfo{year}{2005},
\bibinfo{journal}{Phys. Rev.
Lett.} \textbf{\bibinfo{volume}{95}}, \bibinfo{pages}{206807}.

\bibitem[Dziarmaga(2009)Dziarmaga]{Dziarmaga:2009} \bibinfo{author}{%
\bibnamefont{Dziarmaga}, \bibfnamefont{J.}}, \bibinfo{year}{2009}, %
\bibinfo{journal}{arXiv:0912.4034}
%\textbf{\bibinfo{volume}{ }}, \bibinfo{pages}{ }.

\bibitem[Eckel \emph{et~al.}(2009)Eckel, Reina, and Thorwart]{Eckel:2009} %
\bibinfo{author}{\bibnamefont{Eckel}, \bibfnamefont{J.}}, %
\bibinfo{author}{\bibfnamefont{J. H.}~\bibnamefont{Reina}}, and %
\bibinfo{author}{\bibfnamefont{M.}~\bibnamefont{Thorwart}}, %
\bibinfo{year}{2009}, \bibinfo{journal}{New J. Phys.} \textbf{%
\bibinfo{volume}{11}}, \bibinfo{pages}{085001}.

\bibitem[Farhi \emph{et~al.}(2001)Farhi, Goldstone, Gutmann, Lapan,
Lundgren, and Preda]{Farhi:2001}
\bibinfo{author}{\bibnamefont{Farhi},
\bibfnamefont{E.}}, \bibinfo{author}{\bibfnamefont{J.}~%
\bibnamefont{Goldstone}}, \bibinfo{author}{\bibfnamefont{S.}~%
\bibnamefont{Gutmann}}, \bibinfo{author}{\bibfnamefont{J.}~%
\bibnamefont{Lapan}}, \bibinfo{author}{\bibfnamefont{A.}~%
\bibnamefont{Lundgren}}, and \bibinfo{author}{\bibfnamefont{D.}~%
\bibnamefont{Preda}}, \bibinfo{year}{2001}, \bibinfo{journal}{Science}
\textbf{\bibinfo{volume}{292}}, \bibinfo{pages}{472}.

\bibitem[Fink \emph{et~al.}(2009)Fink, Baur, Bianchetti, Filipp, Goppl,
Leek, Steffen, Blais, and Wallraff]{Fink:2009}
\bibinfo{author}{\bibnamefont{Fink}, \bibfnamefont{J.
M.}}, \bibinfo{author}{\bibfnamefont{M.}~\bibnamefont{Baur}}, %
\bibinfo{author}{\bibfnamefont{R.}~\bibnamefont{Bianchetti}}, %
\bibinfo{author}{\bibfnamefont{S.}~\bibnamefont{Filipp}}, %
\bibinfo{author}{\bibfnamefont{M.}~\bibnamefont{G\"{o}ppl}}, %
\bibinfo{author}{\bibfnamefont{R. J.}~\bibnamefont{Leek}}, %
\bibinfo{author}{\bibfnamefont{L.}~\bibnamefont{Steffen}}, %
\bibinfo{author}{\bibfnamefont{A.}~\bibnamefont{Blais}}, and %
\bibinfo{author}{\bibfnamefont{A.}~\bibnamefont{Wallraff}}, %
\bibinfo{year}{2009}, \bibinfo{journal}{Phys. Scr. T} \textbf{%
\bibinfo{volume}{137}}, \bibinfo{pages}{014013}.

\bibitem[F\"{o}ldi \emph{et~al.}(2007)F\"{o}ldi, Benedict, Milton Pereira,
and Peeters]{Foldi:2007}
\bibinfo{author}{\bibnamefont{F\"{o}ldi},
\bibfnamefont{P.}},
\bibinfo{author}{\bibfnamefont{M.
G.}~\bibnamefont{Benedict}}, \bibinfo{author}{\bibfnamefont{J.}~%
\bibnamefont{Milton Pereira, Jr.}}, and
\bibinfo{author}{\bibfnamefont{F.
M.}~\bibnamefont{Peeters}}, \bibinfo{year}{2007},
\bibinfo{journal}{Phys.
Rev. B} \textbf{\bibinfo{volume}{75}}, \bibinfo{pages}{104430}.

\bibitem[F\"{o}ldi \emph{et~al.}(2008)F\"{o}ldi, Benedict, and Peeters]%
{Foldi:2008} \bibinfo{author}{\bibnamefont{F\"{o}ldi}, \bibfnamefont{P.}}, %
\bibinfo{author}{\bibfnamefont{M. G.}~\bibnamefont{Benedict}}, and %
\bibinfo{author}{\bibfnamefont{F. M.}~\bibnamefont{Peeters}}, %
\bibinfo{year}{2008}, \bibinfo{journal}{Phys. Rev. B} \textbf{%
\bibinfo{volume}{77}}, \bibinfo{pages}{013406}.

\bibitem[F\o rre(2004)F\o rre]{Forre:2004} \bibinfo{author}{F\o rre, M.}, %
\bibinfo{year}{2004}, \bibinfo{journal}{Phys.
Rev. A} \textbf{\bibinfo{volume}{70}}, \bibinfo{pages}{013406}.

\bibitem[Frasca(2003)Frasca]{Frasca:2003} \bibinfo{author}{%
\bibnamefont{Frasca}, \bibfnamefont{M.}}, \bibinfo{year}{2003}, %
\bibinfo{journal}{Ann. Phys.} \textbf{\bibinfo{volume}{306}}, %
\bibinfo{pages}{193}.

\bibitem[Fregenal \emph{et~al.}(2004)Fregenal, Horsdal-Pedersen, Madsen, F
\o rre, Hansen, Ostrovsky]{Fregenal:2004} \bibinfo{author}{%
\bibnamefont{Fregenal}, \bibfnamefont{D.}}, \bibinfo{author}{%
\bibfnamefont{E.}~\bibnamefont{Horsdal-Pedersen}}, \bibinfo{author}{%
\bibfnamefont{L. B.}~\bibnamefont{Madsen}}, \bibinfo{author}{%
\bibfnamefont{M.}~\bibnamefont{F\o rre}},
\bibinfo{author}{\bibfnamefont{J.
P.}~\bibnamefont{Hansen}}, and
\bibinfo{author}{\bibfnamefont{V.
N.}~\bibnamefont{Ostrovsky}}, \bibinfo{year}{2004},
\bibinfo{journal}{Phys.
Rev. A} \textbf{\bibinfo{volume}{69}}, \bibinfo{pages}{031401}.

\bibitem[Fuchs \emph{et~al.}(2009)Fuchs, Dobrovitski, Toyli, Heremans, and
Awschalom]{Fuchs:2009}
\bibinfo{author}{\bibnamefont{Fuchs},
\bibfnamefont{G. D.}},
\bibinfo{author}{\bibfnamefont{V.
V.}~\bibnamefont{Dobrovitski}},
\bibinfo{author}{\bibfnamefont{D.
M.}~\bibnamefont{Toyli}},
\bibinfo{author}{\bibfnamefont{F.
J.}~\bibnamefont{Heremans}}, and
\bibinfo{author}{\bibfnamefont{D.
D.}~\bibnamefont{Awschalom}}, \bibinfo{year}{2009}, %
\bibinfo{journal}{Science} \textbf{\bibinfo{volume}{326}}, %
\bibinfo{pages}{1520}.

\bibitem[Gaitan(2003)Gaitan]{Gaitan:2003} \bibinfo{author}{%
\bibnamefont{Gaitan}, \bibfnamefont{F.}}, \bibinfo{year}{2003}, %
\bibinfo{journal}{Phys. Rev. A} \textbf{\bibinfo{volume}{68}}, %
\bibinfo{pages}{052314}.

\bibitem[Gallagher(1994)Gallagher]{Gallagher:1994} \bibinfo{author}{%
\bibnamefont{Gallagher}, \bibfnamefont{T.}}, \bibinfo{year}{1994}, \emph{%
\bibinfo{title}{Rydberg Atoms}} (%
\bibinfo{publisher}{Cambridge University
Press, Cambridge 1994}).

\bibitem[Garanin(2004)Garanin]{Garanin:2004} \bibinfo{author}{%
\bibnamefont{Garanin}, \bibfnamefont{D. A.}}, \bibinfo{year}{2004}, %
\bibinfo{journal}{Phys. Rev. B} \textbf{\bibinfo{volume}{70}}, %
\bibinfo{pages}{212403}.

\bibitem[Garanin \emph{et~al.}(2008)Garanin, Neb, and Schilling]%
{Garanin:2008} \bibinfo{author}{\bibnamefont{Garanin}, \bibfnamefont{D. A.}}%
, \bibinfo{author}{\bibfnamefont{R.}~\bibnamefont{Neb}}, and %
\bibinfo{author}{\bibfnamefont{R.}~\bibnamefont{Schilling}}, %
\bibinfo{year}{2008}, \bibinfo{journal}{Phys. Rev. B} \textbf{%
\bibinfo{volume}{78}}, \bibinfo{pages}{094405}.

\bibitem[Garraway and Vitanov(1997)Garraway and Vitanov]{Garraway:1997} %
\bibinfo{author}{\bibnamefont{Garraway}, \bibfnamefont{B. M.}} and %
\bibinfo{author}{\bibfnamefont{N. V.}~\bibnamefont{Vitanov}}, %
\bibinfo{year}{1997}, \bibinfo{journal}{Phys. Rev. A} \textbf{%
\bibinfo{volume}{55}}, \bibinfo{pages}{4418}.

\bibitem[Gorelik \emph{et~al.}(1998)Gorelik, Lundin, Shumeiko, Shekhter, and
Jonson]{Gorelik:1998}
\bibinfo{author}{\bibnamefont{Gorelik},
\bibfnamefont{L. Y.}},
\bibinfo{author}{\bibfnamefont{N.
I.}~\bibnamefont{Lundin}},
\bibinfo{author}{\bibfnamefont{V.
S.}~\bibnamefont{Shumeiko}},
\bibinfo{author}{\bibfnamefont{R.
I.}~\bibnamefont{Shekhter}}, and \bibinfo{author}{\bibfnamefont{M.}~%
\bibnamefont{Jonson}}, \bibinfo{year}{1998},
\bibinfo{journal}{Phys. Rev.
Lett.} \textbf{\bibinfo{volume}{81}}, \bibinfo{pages}{2538}.

\bibitem[Goswami(2003)Goswami]{Goswami:2003} \bibinfo{author}{%
\bibnamefont{Goswami}, \bibfnamefont{D.}}, \bibinfo{year}{2003}, %
\bibinfo{journal}{Phys. Rep.} \textbf{\bibinfo{volume}{374}}, %
\bibinfo{pages}{385}.

\bibitem[Gradshteyn and Ryzhik(1994)Gradshteyn and Ryzhik]{Gradshteyn:1994} %
\bibinfo{author}{\bibnamefont{Gradshteyn}, \bibfnamefont{I. S.}} and %
\bibinfo{author}{\bibfnamefont{I. M.}~\bibnamefont{Ryzhik}}, %
\bibinfo{year}{1994}, \emph{%
\bibinfo{title}{Table of Integrals, Series, and
Products}} (\bibinfo{publisher}{Academic Press, New York}).

\bibitem[Grajcar \emph{et~al.}(2005)Grajcar, Izmalkov, and Il'ichev]%
{Grajcar:2005} \bibinfo{author}{\bibnamefont{Grajcar}, \bibfnamefont{M.}}, %
\bibinfo{author}{\bibfnamefont{A.}~\bibnamefont{Izmalkov}}, and %
\bibinfo{author}{\bibfnamefont{E.}~\bibnamefont{Il'ichev}}, %
\bibinfo{year}{2005}, \bibinfo{journal}{Phys. Rev. B} \textbf{%
\bibinfo{volume}{71}}, \bibinfo{pages}{144501}.

\bibitem[Greenberg \emph{et~al.}(2002)Greenberg, Izmalkov, Grajcar,
Il'ichev, Krech, Meyer, Amin, and Maassen van den Brink]{Greenberg:2002}
\bibinfo{author}{\bibnamefont{Greenberg},
\bibfnamefont{Ya. S.}}, \bibinfo{author}{\bibfnamefont{A.}~%
\bibnamefont{Izmalkov}}, \bibinfo{author}{\bibfnamefont{M.}~%
\bibnamefont{Grajcar}}, \bibinfo{author}{\bibfnamefont{E.}~%
\bibnamefont{Il'ichev}}, \bibinfo{author}{\bibfnamefont{W.}~%
\bibnamefont{Krech}}, \bibinfo{author}{\bibfnamefont{H.-G.}~%
\bibnamefont{Meyer}},
\bibinfo{author}{\bibfnamefont{M. H.
S.}~\bibnamefont{Amin}}, and \bibinfo{author}{\bibfnamefont{A.}~%
\bibnamefont{Maassen van den Brink}}, \bibinfo{year}{2002}, %
\bibinfo{journal}{Phys. Rev. B} \textbf{\bibinfo{volume}{66}}, %
\bibinfo{pages}{214525}.

\bibitem[Greenberg(2007)Greenberg]{Greenberg:2007}
\bibinfo{author}{\bibnamefont{Greenberg},
\bibfnamefont{Ya. S.}}, \bibinfo{year}{2007}, \bibinfo{journal}{Phys. Rev. B}
\textbf{\bibinfo{volume}{76}}, \bibinfo{pages}{104520}.

\bibitem[Grifoni and H\"{a}nggi(1998)Grifoni and H\"{a}nggi]{Grifoni:1998} %
\bibinfo{author}{\bibnamefont{Grifoni}, \bibfnamefont{M.}} and %
\bibinfo{author}{\bibfnamefont{P.}~\bibnamefont{H\"{a}nggi}}, %
\bibinfo{year}{1998}, \bibinfo{journal}{Phys. Rep.} \textbf{%
\bibinfo{volume}{304}}, \bibinfo{pages}{229}.

\bibitem[Grossmann \emph{et~al.}(1991a)Grossmann, Dittrich, Jung, and H\"{a}%
nggi]{Grossmann:1991a}
\bibinfo{author}{\bibnamefont{Grossmann},
\bibfnamefont{F.}}, \bibinfo{author}{\bibfnamefont{T.}~%
\bibnamefont{Dittrich}}, \bibinfo{author}{\bibfnamefont{P.}~%
\bibnamefont{Jung}}, and \bibinfo{author}{\bibfnamefont{P.}~\bibnamefont{H%
\"{a}nggi}}, \bibinfo{year}{1991}, \bibinfo{journal}{Phys. Rev. Lett.}
\textbf{\bibinfo{volume}{67}}, \bibinfo{pages}{516}.

\bibitem[Grossmann \emph{et~al.}(1991b)Grossmann, Jung, Dittrich, and H\"{a}%
nggi]{Grossmann:1991b}
\bibinfo{author}{\bibnamefont{Grossmann},
\bibfnamefont{F.}}, \bibinfo{author}{\bibfnamefont{P.}~\bibnamefont{Jung}}, %
\bibinfo{author}{\bibfnamefont{T.}~\bibnamefont{Dittrich}}, and %
\bibinfo{author}{\bibfnamefont{P.}~\bibnamefont{H\"{a}nggi}}, %
\bibinfo{year}{1991}, \bibinfo{journal}{Z. Physik B} \textbf{%
\bibinfo{volume}{84}}, \bibinfo{pages}{315}.

\bibitem[Grossmann and H\"{a}nggi(1992)Grossmann and H\"{a}nggi]%
{Grossmann:1992} \bibinfo{author}{\bibnamefont{Grossmann}, \bibfnamefont{F.}}
and \bibinfo{author}{\bibfnamefont{P.}~\bibnamefont{H\"{a}nggi}}, %
\bibinfo{year}{1992}, \bibinfo{journal}{Europhys. Lett.} \textbf{%
\bibinfo{volume}{18}}, \bibinfo{pages}{571}.

\bibitem[Hausinger and Grifoni(2010)Hausinger and Grifoni]{Hausinger:2010} %
\bibinfo{author}{\bibnamefont{Hausinger}, \bibfnamefont{J.}} and %
\bibinfo{author}{\bibfnamefont{M.}~\bibnamefont{Grifoni}}, %
\bibinfo{year}{2010}, \bibinfo{journal}{Phys. Rev. A} \textbf{%
\bibinfo{volume}{81}}, \bibinfo{pages}{022117}.

\bibitem[Heinrich \emph{et~al.}(2010)Heinrich, Harris, and Marquardt]%
{Heinrich:2010} \bibinfo{author}{\bibnamefont{Heinrich},
\bibfnamefont{G.}},
\bibinfo{author}{\bibfnamefont{J. G.
E.}~\bibnamefont{Harris}}, and \bibinfo{author}{\bibfnamefont{F.}~%
\bibnamefont{Marquardt}}, \bibinfo{year}{2010},
\bibinfo{journal}{Phys. Rev.
A} \textbf{\bibinfo{volume}{81}}, \bibinfo{pages}{011801}.

\bibitem[Henry and Lang(1977)Henry and Lang]{Henry:1977} \bibinfo{author}{%
\bibnamefont{Henry}, \bibfnamefont{C. H.}} and \bibinfo{author}{%
\bibfnamefont{D. V.}~\bibnamefont{Lang}}, \bibinfo{year}{1977}, %
\bibinfo{journal}{Phys. Rev. B} \textbf{\bibinfo{volume}{15}}, %
\bibinfo{pages}{989}.

\bibitem[Ho \emph{et~al.}(2009)Ho, Hung, Chen, and Chu]{Ho:2009} %
\bibinfo{author}{\bibnamefont{Ho}, \bibfnamefont{T.-S.}}, %
\bibinfo{author}{\bibfnamefont{S.-H.}~\bibnamefont{Hung}}, %
\bibinfo{author}{\bibfnamefont{H.-T.}~\bibnamefont{Chen}}, and %
\bibinfo{author}{\bibfnamefont{S.-I.}~\bibnamefont{Chu}}, %
\bibinfo{year}{2009}, \bibinfo{journal}{Phys. Rev. B} \textbf{%
\bibinfo{volume}{79}}, \bibinfo{pages}{235323}.

\bibitem[Il'ichev \emph{et~al.}(2009)Il'ichev, van der Ploeg, Grajcar, and
Meyer]{Ilichev:2009}
\bibinfo{author}{\bibnamefont{Il'ichev},
\bibfnamefont{E.}},
\bibinfo{author}{\bibfnamefont{S. H.
W.}~\bibnamefont{van der Ploeg}}, \bibinfo{author}{\bibfnamefont{M.}~%
\bibnamefont{Grajcar}}, and \bibinfo{author}{\bibfnamefont{H.-G.}~%
\bibnamefont{Meyer}}, \bibinfo{year}{2009},
\bibinfo{journal}{Quantum Inf.
Process.} \textbf{\bibinfo{volume}{8}}, \bibinfo{pages}{133}.

\bibitem[Il'ichev \emph{et~al.}(2010)Il'ichev, Shevchenko, van der Ploeg,
Grajcar, Temchenko, Omelyanchouk and Meyer]{Ilichev:2010}
\bibinfo{author}{\bibnamefont{Il'ichev},
\bibfnamefont{E.}},
\bibinfo{author}{\bibfnamefont{S.
N.}~\bibnamefont{Shevchenko}},
\bibinfo{author}{\bibfnamefont{S. H.
W.}~\bibnamefont{van der Ploeg}}, \bibinfo{author}{\bibfnamefont{M.}~%
\bibnamefont{Grajcar}},
\bibinfo{author}{\bibfnamefont{E.
A.}~\bibnamefont{Temchenko}},
\bibinfo{author}{\bibfnamefont{A.
N.}~\bibnamefont{Omelyanchouk}}, and \bibinfo{author}{\bibfnamefont{H.-G.}~%
\bibnamefont{Meyer}}, \bibinfo{year}{2010}, \bibinfo{journal}{Phys. Rev. B}
\textbf{\bibinfo{volume}{81}}, \bibinfo{pages}{012506}.

\bibitem[Irish and Schwab(2003)Irish and Schwab]{Irish:2003} %
\bibinfo{author}{\bibnamefont{Irish}, \bibfnamefont{E. K.}} and %
\bibinfo{author}{\bibfnamefont{K. C.}~\bibnamefont{Schwab}}, %
\bibinfo{year}{2003}, \bibinfo{journal}{Phys. Rev. B} \textbf{%
\bibinfo{volume}{68}}, \bibinfo{pages}{155311}.

\bibitem[Ithier \emph{et~al.}(2005)Ithier, Collin, Joyez, Vion, Esteve,
Ankerhold, and Grabert]{Ithier:2005}
\bibinfo{author}{\bibnamefont{Ithier},
\bibfnamefont{G.}}, \bibinfo{author}{\bibfnamefont{E.}~\bibnamefont{Collin}}%
, \bibinfo{author}{\bibfnamefont{P.}~\bibnamefont{Joyez}}, %
\bibinfo{author}{\bibfnamefont{D.}~\bibnamefont{Vion}}, \bibinfo{author}{%
\bibfnamefont{D.}~\bibnamefont{Esteve}}, \bibinfo{author}{\bibfnamefont{J.}~%
\bibnamefont{Ankerhold}}, and \bibinfo{author}{\bibfnamefont{H.}~%
\bibnamefont{Grabert}}, \bibinfo{year}{2005},
\bibinfo{journal}{Phys. Rev.
Lett.} \textbf{\bibinfo{volume}{94}}, \bibinfo{pages}{057004}.

\bibitem[Izmalkov \emph{et~al.}(2004)Izmalkov, Grajcar, Il'ichev, Oukhanski,
Wagner, Meyer, Krech, Amin, Maassen van den Brink, and Zagoskin]%
{Izmalkov:2004} \bibinfo{author}{\bibnamefont{Izmalkov}, \bibfnamefont{A.}}, %
\bibinfo{author}{\bibfnamefont{M.}~\bibnamefont{Grajcar}}, %
\bibinfo{author}{\bibfnamefont{E.}~\bibnamefont{Il'ichev}}, %
\bibinfo{author}{\bibfnamefont{N.}~\bibnamefont{Oukhanski}}, %
\bibinfo{author}{\bibfnamefont{Th.}~\bibnamefont{Wagner}}, %
\bibinfo{author}{\bibfnamefont{H.-G.}~\bibnamefont{Meyer}}, %
\bibinfo{author}{\bibfnamefont{W.}~\bibnamefont{Krech}}, \bibinfo{author}{%
\bibfnamefont{M. H. S.}~\bibnamefont{Amin}}, \bibinfo{author}{%
\bibfnamefont{A.}~\bibnamefont{Maassen van den Brink}}, and %
\bibinfo{author}{\bibfnamefont{A. M.}~\bibnamefont{Zagoskin}}, %
\bibinfo{year}{2004}, \bibinfo{journal}{Europhys. Lett.} \textbf{%
\bibinfo{volume}{65}}, \bibinfo{pages}{844}.

\bibitem[Izmalkov \emph{et~al.}(2008)Izmalkov, van der Ploeg, Shevchenko,
Grajcar, Il'ichev, H\"{u}bner, Omelyanchouk, and Meyer]{Izmalkov:2008} %
\bibinfo{author}{\bibnamefont{Izmalkov}, \bibfnamefont{A.}}, %
\bibinfo{author}{\bibfnamefont{S. H. W.}~\bibnamefont{van der Ploeg}}, %
\bibinfo{author}{\bibfnamefont{S. N.}~\bibnamefont{Shevchenko}}, %
\bibinfo{author}{\bibfnamefont{M.}~\bibnamefont{Grajcar}}, %
\bibinfo{author}{\bibfnamefont{E.}~\bibnamefont{Il'ichev}}, %
\bibinfo{author}{\bibfnamefont{U.}~\bibnamefont{H\"{u}bner}}, %
\bibinfo{author}{\bibfnamefont{A. N.}~\bibnamefont{Omelyanchouk}}, and %
\bibinfo{author}{\bibfnamefont{H.-G.}~\bibnamefont{Meyer}}, %
\bibinfo{year}{2008}, \bibinfo{journal}{Phys. Rev. Lett.} \textbf{%
\bibinfo{volume}{101}}, \bibinfo{pages}{017003}.

\bibitem[Ji \emph{et~al.}(2003)Ji, Chung, Sprinzak, Heiblum, Mahalu, and
Shtrikman]{Ji:2003} \bibinfo{author}{\bibnamefont{Ji}, \bibfnamefont{Y.}}, %
\bibinfo{author}{\bibfnamefont{Y.}~\bibnamefont{Chung}}, \bibinfo{author}{%
\bibfnamefont{D.}~\bibnamefont{Sprinzak}}, \bibinfo{author}{%
\bibfnamefont{M.}~\bibnamefont{Heiblum}}, \bibinfo{author}{%
\bibfnamefont{D.}~\bibnamefont{Mahalu}}, and \bibinfo{author}{%
\bibfnamefont{H.}~\bibnamefont{Shtrikman}}, \bibinfo{year}{2003}, %
\bibinfo{journal}{Nature} \textbf{\bibinfo{volume}{422}}, %
\bibinfo{pages}{415}.

\bibitem[Johansson \emph{et~al.}(2009)Johansson, Amin, Berkley, Bunyuk,
Choi, Harris, Johnson, Lantig, Lloyd, and Rose]{Johansson:2009} %
\bibinfo{author}{\bibnamefont{Johansson}, \bibfnamefont{J.}}, %
\bibinfo{author}{\bibfnamefont{M. H. S.}~\bibnamefont{Amin}}, %
\bibinfo{author}{\bibfnamefont{A. J.}~\bibnamefont{Berkley}}, %
\bibinfo{author}{\bibfnamefont{P.}~\bibnamefont{Bunyk}}, \bibinfo{author}{%
\bibfnamefont{V.}~\bibnamefont{Choi}}, \bibinfo{author}{\bibfnamefont{R.}~%
\bibnamefont{Harris}},
\bibinfo{author}{\bibfnamefont{M.
W.}~\bibnamefont{Johnson}},
\bibinfo{author}{\bibfnamefont{T.
M.}~\bibnamefont{Lanting}}, \bibinfo{author}{\bibfnamefont{S.}~%
\bibnamefont{Lloyd}}, and \bibinfo{author}{\bibfnamefont{G.}~%
\bibnamefont{Rose}}, \bibinfo{year}{2009}, \bibinfo{journal}{Phys. Rev. B}
\textbf{\bibinfo{volume}{80}}, \bibinfo{pages}{012507}.

\bibitem[Kayanuma(1994)Kayanuma]{Kayanuma:1994} \bibinfo{author}{%
\bibnamefont{Kayanuma}, \bibfnamefont{Y.}}, \bibinfo{year}{1994}, %
\bibinfo{journal}{Phys. Rev. A} \textbf{\bibinfo{volume}{50}}, %
\bibinfo{pages}{843}.

\bibitem[Kayanuma(1997)Kayanuma]{Kayanuma:1997} \bibinfo{author}{%
\bibnamefont{Kayanuma}, \bibfnamefont{Y.}}, \bibinfo{year}{1997}, %
\bibinfo{journal}{Phys. Rev. A} \textbf{\bibinfo{volume}{55}}, %
\bibinfo{pages}{R2495}.

\bibitem[Kayanuma and Saito(2008)Kayanuma and Saito]{Kayanuma:2008}
\bibinfo{author}{\bibnamefont{Kayanuma},
\bibfnamefont{Y.}}, and \bibinfo{author}{\bibfnamefont{K.}~%
\bibnamefont{Saito}}, \bibinfo{year}{2008}, \bibinfo{journal}{Phys. Rev. A}
\textbf{\bibinfo{volume}{77}}, \bibinfo{pages}{010101(R)}.

\bibitem[Krainov and Yakovlev(1980)Krainov and Yakovlev]{Krainov:1980} %
\bibinfo{author}{\bibnamefont{Krainov}, \bibfnamefont{V. P.}} and %
\bibinfo{author}{\bibfnamefont{V. P.}~\bibnamefont{Yakovlev}}, %
\bibinfo{year}{1980}, \bibinfo{journal}{Sov. Phys. JETP} \textbf{%
\bibinfo{volume}{51}}, \bibinfo{pages}{1104}.

\bibitem[Kral \emph{et~al.}(2007)Kral, Thanopulos, and Shapiro]{Kral:2007} %
\bibinfo{author}{\bibnamefont{Kral}, \bibfnamefont{P.}}, \bibinfo{author}{%
\bibfnamefont{I.}~\bibnamefont{Thanopulos}}, and \bibinfo{author}{%
\bibfnamefont{M.}~\bibnamefont{Shapiro}}, \bibinfo{year}{2007}, %
\bibinfo{journal}{Rev. Mod. Phys.} \textbf{\bibinfo{volume}{79}}, %
\bibinfo{pages}{53}.

\bibitem[LaHaye \emph{et~al.}(2009)LaHaye, Suh, Echternach, Schwab, and
Roukes]{LaHaye:2009}
\bibinfo{author}{\bibnamefont{LaHaye},
\bibfnamefont{M. D.}}, \bibinfo{author}{\bibfnamefont{J.}~\bibnamefont{Suh}}%
, \bibinfo{author}{\bibfnamefont{P. M.}~\bibnamefont{Echternach}}, %
\bibinfo{author}{\bibfnamefont{K. C.}~\bibnamefont{Schwab}}, and %
\bibinfo{author}{\bibfnamefont{M. L.}~\bibnamefont{Roukes}}, %
\bibinfo{year}{2009}, \bibinfo{journal}{Nature} \textbf{\bibinfo{volume}{459}%
}, \bibinfo{pages}{960}.

\bibitem[Landau(1932a)Landau]{Landau:1932a} \bibinfo{author}{%
\bibnamefont{Landau}, \bibfnamefont{L.}}, \bibinfo{year}{1932}, %
\bibinfo{journal}{Phys. Z. Sowjetunion} \textbf{\bibinfo{volume}{1}}, %
\bibinfo{pages}{88}.

\bibitem[Landau(1932b)Landau]{Landau:1932b} \bibinfo{author}{%
\bibnamefont{Landau}, \bibfnamefont{L.}}, \bibinfo{year}{1932}, %
\bibinfo{journal}{Phys. Z. Sowjetunion} \textbf{\bibinfo{volume}{2}}, %
\bibinfo{pages}{46}.

\bibitem[Landau and Lifshitz(1977)Landau and Lifshitz]{Landau:1977} %
\bibinfo{author}{\bibnamefont{Landau}, \bibfnamefont{L. D.}} and %
\bibinfo{author}{\bibfnamefont{E. M.}~\bibnamefont{Lifshitz}}, %
\bibinfo{year}{1977}, \emph{%
\bibinfo{title}{Quantum Mechanics:
Non-Relativistic Theory}} (\bibinfo{publisher}{Pergamon Press,
Oxford}).

\bibitem[Leggett \emph{et~al.}(1987)Leggett, Chakravarty, Dorsey, Fisher,
Garg, and Zwerger]{Leggett:1987}
\bibinfo{author}{\bibnamefont{Leggett},
\bibfnamefont{A. J.}}, \bibinfo{author}{\bibfnamefont{S.}~%
\bibnamefont{Chakravarty}},
\bibinfo{author}{\bibfnamefont{A.
T.}~\bibnamefont{Dorsey}},
\bibinfo{author}{\bibfnamefont{M. P.
A.}~\bibnamefont{Fisher}}, \bibinfo{author}{\bibfnamefont{A.}~%
\bibnamefont{Garg}}, and \bibinfo{author}{\bibfnamefont{W.}~%
\bibnamefont{Zwerger}}, \bibinfo{year}{1987},
\bibinfo{journal}{Rev. Mod.
Phys.} \textbf{\bibinfo{volume}{59}}, \bibinfo{pages}{1}.

\bibitem[Li \emph{et~al.}(2009)Li, Hoover, and Gaitan]{Li:2009} %
\bibinfo{author}{\bibnamefont{Li}, \bibfnamefont{R.}}, \bibinfo{author}{%
\bibfnamefont{M.}~\bibnamefont{Hoover}}, and \bibinfo{author}{%
\bibfnamefont{F.}~\bibnamefont{Gaitan}}, \bibinfo{year}{2009}, %
\bibinfo{journal}{Quantum Info. and Computation} \textbf{\bibinfo{volume}{9}}%
, \bibinfo{pages}{290}.

\bibitem[Liao \emph{et~al.}(2010)Liao, Gong, Zhou, Liu, Sun, and Nori]%
{Liao:2010} \bibinfo{author}{\bibnamefont{Liao},
\bibfnamefont{J.-Q.}}, \bibinfo{author}{\bibfnamefont{Z.
R.}~\bibnamefont{Gong}}, \bibinfo{author}{\bibfnamefont{L.}~%
\bibnamefont{Zhou}}, \bibinfo{author}{\bibfnamefont{Y. X.}~\bibnamefont{Liu}}%
, \bibinfo{author}{\bibfnamefont{C. P.}~\bibnamefont{Sun}}, and %
\bibinfo{author}{\bibfnamefont{F.}~\bibnamefont{Nori}}, \bibinfo{year}{2010}%
, \bibinfo{journal}{Phys. Rev. A} \textbf{\bibinfo{volume}{81}}, %
\bibinfo{pages}{042304}.

\bibitem[Liu \emph{et~al.}(2006)Liu, Sun, and Nori]{Liu:2006} %
\bibinfo{author}{\bibnamefont{Liu}, \bibfnamefont{Y. X.}}, %
\bibinfo{author}{\bibfnamefont{C. P.}~\bibnamefont{Sun}}, and %
\bibinfo{author}{\bibfnamefont{F.}~\bibnamefont{Nori}}, \bibinfo{year}{2006}%
, \bibinfo{journal}{Phys. Rev. A} \textbf{\bibinfo{volume}{74}}, %
\bibinfo{pages}{052321}.

\bibitem[Lopez-Castillo \emph{et~al.}(1992)Lopez-Castillo, Filali-Mouhim,
and Jay-Gerin]{Lopez-Castillo:1992} \bibinfo{author}{%
\bibnamefont{Lopez-Castillo}, \bibfnamefont{J.-M.}}, \bibinfo{author}{%
\bibfnamefont{A.}~\bibnamefont{Filali-Mouhim}}, and \bibinfo{author}{%
\bibfnamefont{J.-P.}~\bibnamefont{Jay-Gerin}}, \bibinfo{year}{1992}, %
\bibinfo{journal}{J. Chem. Phys.} \textbf{\bibinfo{volume}{97}}, %
\bibinfo{pages}{1905}.

\bibitem[Maeda \emph{et~al.}(2006)Maeda, Gurian, Norum, and Gallagher]%
{Maeda:2006} \bibinfo{author}{\bibnamefont{Maeda}, \bibfnamefont{H.}}, %
\bibinfo{author}{\bibfnamefont{J. H.}~\bibnamefont{Gurian}}, %
\bibinfo{author}{\bibfnamefont{D. V. L.}~\bibnamefont{Norum}}, and %
\bibinfo{author}{\bibfnamefont{T. F.}~\bibnamefont{Gallagher}}, %
\bibinfo{year}{2006}, \bibinfo{journal}{Phys. Rev. Lett.} \textbf{%
\bibinfo{volume}{96}}, \bibinfo{pages}{073002}.

\bibitem[Majorana(1932)Majorana]{Majorana:1932} \bibinfo{author}{%
\bibnamefont{Majorana}, \bibfnamefont{E.}}, \bibinfo{year}{1932}, %
\bibinfo{journal}{Nuovo Cimento} \textbf{\bibinfo{volume}{9}}, %
\bibinfo{pages}{43}.

\bibitem[Makhlin \emph{et~al.}(2001)Makhlin, Schon, and Shnirman]%
{Makhlin:2001} \bibinfo{author}{\bibnamefont{Makhlin}, \bibfnamefont{Yu.}}, %
\bibinfo{author}{\bibfnamefont{G.}~\bibnamefont{Sch\"{o}n}}, and %
\bibinfo{author}{\bibfnamefont{A.}~\bibnamefont{Shnirman}}, %
\bibinfo{year}{2001}, \bibinfo{journal}{Rev. Mod. Phys.} \textbf{%
\bibinfo{volume}{73}}, \bibinfo{pages}{357}.

\bibitem[Mark \emph{et~al.}(2007a)Mark, Kraemer, Waldburger, Herbig, Chin, N%
\"{a}gerl, and Grimm]{Mark:2007a}
\bibinfo{author}{\bibnamefont{Mark},
\bibfnamefont{M.}}, \bibinfo{author}{\bibfnamefont{T.}~\bibnamefont{Kraemer}}%
, \bibinfo{author}{\bibfnamefont{P.}~\bibnamefont{Waldburger}}, %
\bibinfo{author}{\bibfnamefont{J.}~\bibnamefont{Herbig}}, %
\bibinfo{author}{\bibfnamefont{C.}~\bibnamefont{Chin}}, \bibinfo{author}{%
\bibfnamefont{H.-C.}~\bibnamefont{N\"{a}gerl}}, and \bibinfo{author}{%
\bibfnamefont{R.}~\bibnamefont{Grimm}}, \bibinfo{year}{2007}, %
\bibinfo{journal}{Phys. Rev. Lett.} \textbf{\bibinfo{volume}{99}}, %
\bibinfo{pages}{113201}.

\bibitem[Mark \emph{et~al.}(2007b)Mark, Ferlaino, Knoop, Danzl, Kraemer,
Chin, N\"{a}gerl, and Grimm]{Mark:2007b}
\bibinfo{author}{\bibnamefont{Mark},
\bibfnamefont{M.}}, \bibinfo{author}{\bibfnamefont{F.}~%
\bibnamefont{Ferlaino}}, \bibinfo{author}{\bibfnamefont{S.}~%
\bibnamefont{Knoop}},
\bibinfo{author}{\bibfnamefont{J.
G.}~\bibnamefont{Danzl}}, \bibinfo{author}{\bibfnamefont{T.}~%
\bibnamefont{Kraemer}}, \bibinfo{author}{\bibfnamefont{C.}~%
\bibnamefont{Chin}}, \bibinfo{author}{\bibfnamefont{H.-C.}~\bibnamefont{N%
\"{a}gerl}}, and \bibinfo{author}{\bibfnamefont{R.}~\bibnamefont{Grimm}}, %
\bibinfo{year}{2007}, \bibinfo{journal}{Phys. Rev. A} \textbf{%
\bibinfo{volume}{76}}, \bibinfo{pages}{042514}.

\bibitem[Marzlin and Sanders(2004)Marzlin and Sanders]{Marzlin:2004} %
\bibinfo{author}{\bibnamefont{Marzlin}, \bibfnamefont{K.-P.}} and %
\bibinfo{author}{\bibfnamefont{B. C.}~\bibnamefont{Sanders}}, %
\bibinfo{year}{2004}, \bibinfo{journal}{Phys. Rev. Lett.} \textbf{%
\bibinfo{volume}{93}}, \bibinfo{pages}{160408}.

\bibitem[McDermott(2009)McDermott]{McDermott:2009} \bibinfo{author}{%
\bibnamefont{McDermott}, \bibfnamefont{R.}}, \bibinfo{year}{2009}, %
\bibinfo{journal}{IEEE Trans. Appl. Supercond} \textbf{\bibinfo{volume}{19}}%
, \bibinfo{pages}{2}.

\bibitem[Mukherjee and Dutta(2009)Mukherjee and Dutta]{Mukherjee:2009} %
\bibinfo{author}{\bibnamefont{Mukherjee}, \bibfnamefont{V.}} and %
\bibinfo{author}{\bibfnamefont{A.}~\bibnamefont{Dutta}}, \bibinfo{year}{2009}%
, \bibinfo{journal}{J. Stat. Mech.} \bibinfo{pages}{P05005}.

\bibitem[Mullen \emph{et~al.}(1989)Mullen, Ben-Jacob, Gefen, and Schuss]%
{Mullen:1989} \bibinfo{author}{\bibnamefont{Mullen}, \bibfnamefont{K.}}, %
\bibinfo{author}{\bibfnamefont{E.}~\bibnamefont{Ben-Jacob}}, %
\bibinfo{author}{\bibfnamefont{Yu.}~\bibnamefont{Gefen}}, and %
\bibinfo{author}{\bibfnamefont{Z.}~\bibnamefont{Schuss}}, %
\bibinfo{year}{1989}, \bibinfo{journal}{Phys. Rev. Lett.} \textbf{%
\bibinfo{volume}{62}}, \bibinfo{pages}{2543}.

\bibitem[Nagaya \emph{et~al.}(2007)Nagaya, Zhu, and Lin]{Nagaya:2007} %
\bibinfo{author}{\bibnamefont{Nagaya}, \bibfnamefont{K.}}, %
\bibinfo{author}{\bibfnamefont{C.}~\bibnamefont{Zhu}}, and %
\bibinfo{author}{\bibfnamefont{S. H.}~\bibnamefont{Lin}}, %
\bibinfo{year}{2007}, \bibinfo{journal}{J. Chem. Phys.} \textbf{%
\bibinfo{volume}{127}}, \bibinfo{pages}{094304}.

\bibitem[Nakamura \emph{et~al.}(2001)Nakamura, Pashkin, and Tsai]%
{Nakamura:2001} \bibinfo{author}{\bibnamefont{Nakamura}, \bibfnamefont{Y.}}, %
\bibinfo{author}{\bibfnamefont{Yu. A.}~\bibnamefont{Pashkin}}, and %
\bibinfo{author}{\bibfnamefont{J. S.}~\bibnamefont{Tsai}}, %
\bibinfo{year}{2001}, \bibinfo{journal}{Phys. Rev. Lett.} \textbf{%
\bibinfo{volume}{87}}, \bibinfo{pages}{246601}.

\bibitem[Nalbach and Thorwart(2009)Nalbach and Thorwart]{Nalbach:2009}
\bibinfo{author}{\bibnamefont{Nalbach},
\bibfnamefont{P.}} and \bibinfo{author}{\bibfnamefont{M.}~%
\bibnamefont{Thorwart}}, \bibinfo{year}{2009},
\bibinfo{journal}{Phys. Rev.
Lett.} \textbf{\bibinfo{volume}{103}}, \bibinfo{pages}{220401}.

\bibitem[Nikitin(1996)Nikitin]{Nikitin:1996} \bibinfo{author}{%
\bibnamefont{Nikitin}, \bibfnamefont{E. E.}}, \bibinfo{year}{1996}, \emph{%
\bibinfo{title}{Atomic, Molecular and Optical Physics Handbook (Ed. G.W.F.
Drake)}} (\bibinfo{publisher}{Woodbury, NY: Americal Institute of Physics}), %
\bibinfo{pages}{561}.

\bibitem[Nikitin and Ovchinnikova(1972)Nikitin and Ovchinnikova]%
{Nikitin:1972} \bibinfo{author}{\bibnamefont{Nikitin},
\bibfnamefont{E. E.}} and
\bibinfo{author}{\bibfnamefont{M.
Ya.}~\bibnamefont{Ovchinnikova}}, \bibinfo{year}{1972}, %
\bibinfo{journal}{Sov. Phys. Uspekhi} \textbf{\bibinfo{volume}{14}}, %
\bibinfo{pages}{394}.

\bibitem[Nikitin and Umanskii(1984)Nikitin and Umanskii]{Nikitin:1984} %
\bibinfo{author}{\bibnamefont{Nikitin}, \bibfnamefont{E. E.}} and %
\bibinfo{author}{\bibfnamefont{S. Ya.}~\bibnamefont{Umanskii}}, %
\bibinfo{year}{1984}, \emph{\bibinfo{title}{Theory of Slow Atomic Collisions}%
} (\bibinfo{publisher}{Springer,
Heidelberg}).

\bibitem[Oh \emph{et~al.}(2008)Oh, Huang, Peskin, and Kais]{Oh:2008}
\bibinfo{author}{\bibnamefont{Oh},
\bibfnamefont{S.}}, \bibinfo{author}{\bibfnamefont{Z.}~\bibnamefont{Huang}}, %
\bibinfo{author}{\bibfnamefont{U.}~\bibnamefont{Peskin}}, and %
\bibinfo{author}{\bibfnamefont{S.}~\bibnamefont{Kais}}, \bibinfo{year}{2008}%
, \bibinfo{journal}{Phys. Rev. A} \textbf{\bibinfo{volume}{78}}, %
\bibinfo{pages}{062106}.

\bibitem[Oliver and Valenzuela(2009)Oliver and Valenzuela]{Oliver:2009}
\bibinfo{author}{\bibnamefont{Oliver},
\bibfnamefont{W. D.}} and
\bibinfo{author}{\bibfnamefont{S.
O.}~\bibnamefont{Valenzuela}}, \bibinfo{year}{2009}, %
\bibinfo{journal}{Quantum Inf. Process.} \textbf{\bibinfo{volume}{8}}, %
\bibinfo{pages}{261}.

\bibitem[Oliver \emph{et~al.}(2005)Oliver, Yu, Lee, Berggren, Levitov, and
Orlando]{Oliver:2005}
\bibinfo{author}{\bibnamefont{Oliver},
\bibfnamefont{W. D.}}, \bibinfo{author}{\bibfnamefont{Ya.}~\bibnamefont{Yu}}%
, \bibinfo{author}{\bibfnamefont{J. C.}~\bibnamefont{Lee}}, %
\bibinfo{author}{\bibfnamefont{K. K.}~\bibnamefont{Berggren}}, %
\bibinfo{author}{\bibfnamefont{L. S.}~\bibnamefont{Levitov}}, and %
\bibinfo{author}{\bibfnamefont{T. P.}~\bibnamefont{Orlando}}, %
\bibinfo{year}{2005}, \bibinfo{journal}{Science} \textbf{%
\bibinfo{volume}{310}}, \bibinfo{pages}{1653}.

\bibitem[Paila \emph{et~al.}(2009)Paila, Tuorila, Sillanp\"{a}\"{a},
Gunnarsson, Sarkar, Makhlin, Thuneberg, and Hakonen]{Paila:2009}
\bibinfo{author}{\bibnamefont{Paila},
\bibfnamefont{A.}}, \bibinfo{author}{\bibfnamefont{J.}~\bibnamefont{Tuorila}}%
, \bibinfo{author}{\bibfnamefont{M.}~\bibnamefont{Sillanp\"{a}\"{a}}}, %
\bibinfo{author}{\bibfnamefont{D.}~\bibnamefont{Gunnarsson}}, %
\bibinfo{author}{\bibfnamefont{J.}~\bibnamefont{Sarkar}}, %
\bibinfo{author}{\bibfnamefont{Yu.}~\bibnamefont{Makhlin}}, %
\bibinfo{author}{\bibfnamefont{E.}~\bibnamefont{Thuneberg}}, and %
\bibinfo{author}{\bibfnamefont{P.}~\bibnamefont{Hakonen}}, %
\bibinfo{year}{2009}, \bibinfo{journal}{Quantum Inf. Process.} \textbf{%
\bibinfo{volume}{8}}, \bibinfo{pages}{245}.

\bibitem[Pegg and Series(1973)Pegg and Series]{Pegg:1973}
\bibinfo{author}{\bibnamefont{Pegg},
\bibfnamefont{D. T.}} and
\bibinfo{author}{\bibfnamefont{G.
W.}~\bibnamefont{Series}}, \bibinfo{year}{1973},
\bibinfo{journal}{Proc. R.
Soc. A} \textbf{\bibinfo{volume}{332}}, \bibinfo{pages}{281}.

\bibitem[Persson \emph{et~al.}(2010)Persson, Wilson, Sandberg, Johansson,
and Delsing]{Persson:2010}
\bibinfo{author}{\bibnamefont{Persson},
\bibfnamefont{F.}},
\bibinfo{author}{\bibfnamefont{C.
M.}~\bibnamefont{Wilson}}, \bibinfo{author}{\bibfnamefont{M.}~%
\bibnamefont{Sandberg}}, \bibinfo{author}{\bibfnamefont{G.}~%
\bibnamefont{Johansson}}, and \bibinfo{author}{\bibfnamefont{P.}~%
\bibnamefont{Delsing}}, \bibinfo{year}{2010}, \bibinfo{journal}{Nano Lett.}
\textbf{\bibinfo{volume}{10}}, \bibinfo{pages}{953}.

\bibitem[Petta \emph{et~al.}(2010)Petta, Lu, and Gossard]{Petta:2010} %
\bibinfo{author}{\bibnamefont{Petta}, \bibfnamefont{J. R.}}, %
\bibinfo{author}{\bibfnamefont{H.}~\bibnamefont{Lu}}, and %
\bibinfo{author}{\bibfnamefont{A. C.}~\bibnamefont{Gossard}}, %
\bibinfo{year}{2010}, \bibinfo{journal}{Science} \textbf{%
\bibinfo{volume}{327}}, \bibinfo{pages}{669}.

\bibitem[Pezze \emph{et~al.}(2007)Pezze, Smerzi, Khoury, Hodelin, and
Bouwmeester]{Pezze:2007}
\bibinfo{author}{\bibnamefont{Pezze},
\bibfnamefont{L.}}, \bibinfo{author}{\bibfnamefont{A.}~\bibnamefont{Smerzi}}%
, \bibinfo{author}{\bibfnamefont{G.}~\bibnamefont{Khoury}}, %
\bibinfo{author}{\bibfnamefont{J. F.}~\bibnamefont{Hodelin}}, and %
\bibinfo{author}{\bibfnamefont{D.}~\bibnamefont{Bouwmeester}}, %
\bibinfo{year}{2007}, \bibinfo{journal}{Phys. Rev. Lett.} \textbf{%
\bibinfo{volume}{99}}, \bibinfo{pages}{223602}.

\bibitem[Pokrovsky and Sinitsyn(2002)Pokrovsky and Sinitsyn]{Pokrovsky:2002}
\bibinfo{author}{\bibnamefont{Pokrovsky},
\bibfnamefont{V. L.}}, and
\bibinfo{author}{\bibfnamefont{N.
A.}~\bibnamefont{Sinitsyn}}, \bibinfo{year}{2002},
\bibinfo{journal}{Phys. Rev.
B} \textbf{\bibinfo{volume}{65}}, \bibinfo{pages}{153105}.

\bibitem[Pokrovsky and Sinitsyn(2004)Pokrovsky and Sinitsyn]{Pokrovsky:2004}
\bibinfo{author}{\bibnamefont{Pokrovsky},
\bibfnamefont{V. L.}}, and
\bibinfo{author}{\bibfnamefont{N.
A.}~\bibnamefont{Sinitsyn}}, \bibinfo{year}{2004},
\bibinfo{journal}{Phys. Rev.
B} \textbf{\bibinfo{volume}{69}}, \bibinfo{pages}{104414}.

\bibitem[Rabi(1937)Rabi]{Rabi:1937}
\bibinfo{author}{\bibnamefont{Rabi},
\bibfnamefont{I. I.}}, \bibinfo{year}{1937}, \bibinfo{journal}{Phys. Rev.}
\textbf{\bibinfo{volume}{51}}, \bibinfo{pages}{652}.

\bibitem[Ribeiro and Burkard(2009)Ribeiro and Burkard]{Ribeiro:2009}
\bibinfo{author}{\bibnamefont{Ribeiro},
\bibfnamefont{H.}}, and \bibinfo{author}{\bibfnamefont{G.}~%
\bibnamefont{Burkard}}, \bibinfo{year}{2009},
\bibinfo{journal}{Phys. Rev.
Lett.} \textbf{\bibinfo{volume}{102}}, \bibinfo{pages}{216802}.

\bibitem[Ribeiro \emph{et~al.}(2010)Ribeiro, Petta, and Burkard]%
{Ribeiro:2010} \bibinfo{author}{\bibnamefont{Ribeiro}, \bibfnamefont{H.}}, %
\bibinfo{author}{\bibfnamefont{J. R.}~\bibnamefont{Petta}}, and %
\bibinfo{author}{\bibfnamefont{G.}~\bibnamefont{Burkard}}, %
\bibinfo{year}{2010}, \bibinfo{journal}{arXiv:1002.4630}%
% \textbf{\bibinfo{volume}{187}}, \bibinfo{pages}{201}.

\bibitem[Ritus(1967)Ritus]{Ritus:1967}
\bibinfo{author}{\bibnamefont{Ritus},
\bibfnamefont{V. I.}}, \bibinfo{year}{1967},
\bibinfo{journal}{Sov. Phys.
JETP} \textbf{\bibinfo{volume}{24}}, \bibinfo{pages}{1041}.

\bibitem[Rodrigues \emph{et~al.}(2007a)Rodrigues, Imbers, and Armour]%
{Rodrigues:2007a}
\bibinfo{author}{\bibnamefont{Rodrigues}, \bibfnamefont{D.
A.}}, \bibinfo{author}{\bibfnamefont{J.}~\bibnamefont{Imbers}}, and %
\bibinfo{author}{\bibfnamefont{A. D.}~\bibnamefont{Armour}}, %
\bibinfo{year}{2007}, \bibinfo{journal}{Phys. Rev. Lett.} \textbf{%
\bibinfo{volume}{98}}, \bibinfo{pages}{067204}.

\bibitem[Rodrigues \emph{et~al.}(2007b)Rodrigues, Imbers, and Armour]%
{Rodrigues:2007b}
\bibinfo{author}{\bibnamefont{Rodrigues}, \bibfnamefont{D.
A.}}, \bibinfo{author}{\bibfnamefont{J.}~\bibnamefont{Imbers}}, %
\bibinfo{author}{\bibfnamefont{T. J.}~\bibnamefont{Harvey}}, and %
\bibinfo{author}{\bibfnamefont{A. D.}~\bibnamefont{Armour}}, %
\bibinfo{year}{2007}, \bibinfo{journal}{New J. Phys.} \textbf{%
\bibinfo{volume}{9}}, \bibinfo{pages}{84}.

\bibitem[Rotvig \emph{et~al.}(1995)Rotvig, Jauho, and Smith]{Rotvig:1995} %
\bibinfo{author}{\bibnamefont{Rotvig}, \bibfnamefont{J.}}, %
\bibinfo{author}{\bibfnamefont{A. P.}~\bibnamefont{Jauho}}, and %
\bibinfo{author}{\bibfnamefont{H.}~\bibnamefont{Smith}}, \bibinfo{year}{1995}%
, \bibinfo{journal}{Phys. Rev. Lett.} \textbf{\bibinfo{volume}{74}}, %
\bibinfo{pages}{1831}.

\bibitem[Rotvig \emph{et~al.}(1996)Rotvig, Jauho, and Smith]{Rotvig:1996} %
\bibinfo{author}{\bibnamefont{Rotvig}, \bibfnamefont{J.}}, %
\bibinfo{author}{\bibfnamefont{A. P.}~\bibnamefont{Jauho}}, and %
\bibinfo{author}{\bibfnamefont{H.}~\bibnamefont{Smith}}, \bibinfo{year}{1996}%
, \bibinfo{journal}{Phys. Rev. B} \textbf{\bibinfo{volume}{54}}, %
\bibinfo{pages}{17691}.

\bibitem[Rudner \emph{et~al.}(2008)Rudner, Shytov, Levitov, Berns, Oliver,
Valenzuela, and Orlando]{Rudner:2008}
\bibinfo{author}{\bibnamefont{Rudner},
\bibfnamefont{M. S.}},
\bibinfo{author}{\bibfnamefont{A.
V.}~\bibnamefont{Shytov}},
\bibinfo{author}{\bibfnamefont{L.
S.}~\bibnamefont{Levitov}},
\bibinfo{author}{\bibfnamefont{D.
M.}~\bibnamefont{Berns}},
\bibinfo{author}{\bibfnamefont{W.
D.}~\bibnamefont{Oliver}},
\bibinfo{author}{\bibfnamefont{S.
O.}~\bibnamefont{Valenzuela}}, and
\bibinfo{author}{\bibfnamefont{T.
P.}~\bibnamefont{Orlando}}, \bibinfo{year}{2008},
\bibinfo{journal}{Phys.
Rev. Lett.} \textbf{\bibinfo{volume}{101}}, \bibinfo{pages}{190502}.

\bibitem[Saito \emph{et~al.}(2004)Saito, Thorwart, Tanaka, Ueda, Nakano,
Semba, and Takayanagi]{Saito:2004}
\bibinfo{author}{\bibnamefont{Saito},
\bibfnamefont{S.}}, \bibinfo{author}{\bibfnamefont{M.}~%
\bibnamefont{Thorwart}}, \bibinfo{author}{\bibfnamefont{H.}~%
\bibnamefont{Tanaka}}, \bibinfo{author}{\bibfnamefont{M.}~\bibnamefont{Ueda}}%
, \bibinfo{author}{\bibfnamefont{H.}~\bibnamefont{Nakano}}, %
\bibinfo{author}{\bibfnamefont{K.}~\bibnamefont{Semba}}, and %
\bibinfo{author}{\bibfnamefont{H.}~\bibnamefont{Takayanagi}}, %
\bibinfo{year}{2004}, \bibinfo{journal}{Phys. Rev. Lett.} \textbf{%
\bibinfo{volume}{93}}, \bibinfo{pages}{037001}.

\bibitem[Saito and Kayanuma(2002)Saito and Kayanuma]{Saito:2002}
\bibinfo{author}{\bibnamefont{Saito},
\bibfnamefont{K.}}, and \bibinfo{author}{\bibfnamefont{Y.}~%
\bibnamefont{Kayanuma}}, \bibinfo{year}{2002},
\bibinfo{journal}{Phys. Rev.
A} \textbf{\bibinfo{volume}{65}}, \bibinfo{pages}{033407}.

\bibitem[Saito \emph{et~al.}(2007)Saito, Wubs, Kohler, Kayanuma, and Hanggi]%
{Saito:2007} \bibinfo{author}{\bibnamefont{Saito},
\bibfnamefont{K.}}, \bibinfo{author}{\bibfnamefont{M.}~\bibnamefont{Wubs}}, %
\bibinfo{author}{\bibfnamefont{S.}~\bibnamefont{Kohler}}, %
\bibinfo{author}{\bibfnamefont{Y.}~\bibnamefont{Kayanuma}}, and %
\bibinfo{author}{\bibfnamefont{P.}~\bibnamefont{H\"{a}nggi}}, %
\bibinfo{year}{2007}, \bibinfo{journal}{Phys. Rev. B} \textbf{%
\bibinfo{volume}{75}}, \bibinfo{pages}{214308}.

\bibitem[Sambe(1973)Sambe]{Sambe:1973}
\bibinfo{author}{\bibnamefont{Sambe},
\bibfnamefont{H.}}, \bibinfo{year}{1973}, \bibinfo{journal}{Phys. Rev. A}
\textbf{\bibinfo{volume}{7}}, \bibinfo{pages}{2203}.

\bibitem[Sarandy \emph{et~al.}(2004)Sarandy, Wu, and Lidar]{Sarandy:2004} %
\bibinfo{author}{\bibnamefont{Sarandy}, \bibfnamefont{M. S.}}, %
\bibinfo{author}{\bibfnamefont{L.-A.}~\bibnamefont{Wu}}, and %
\bibinfo{author}{\bibfnamefont{D. A.}~\bibnamefont{Lidar}}, %
\bibinfo{year}{2004}, \bibinfo{journal}{Quantum Inf. Process.} \textbf{%
\bibinfo{volume}{3}}, \bibinfo{pages}{331}.

\bibitem[Shevchenko \emph{et~al.}(2005)Shevchenko, Kiyko, Omelyanchouk, and
Krech]{Shevchenko:2005}
\bibinfo{author}{\bibnamefont{Shevchenko},
\bibfnamefont{S. N.}},
\bibinfo{author}{\bibfnamefont{A.
S.}~\bibnamefont{Kiyko}},
\bibinfo{author}{\bibfnamefont{A.
N.}~\bibnamefont{Omelyanchouk}}, and \bibinfo{author}{\bibfnamefont{W.}~%
\bibnamefont{Krech}}, \bibinfo{year}{2005},
\bibinfo{journal}{Low Temp.
Phys.} \textbf{\bibinfo{volume}{31}}, \bibinfo{pages}{569}.

\bibitem[Shevchenko and Omelyanchouk(2006)Shevchenko and Omelyanchouk]%
{Shevchenko:2006}
\bibinfo{author}{\bibnamefont{Shevchenko},
\bibfnamefont{S. N.}} and
\bibinfo{author}{\bibfnamefont{A.
N.}~\bibnamefont{Omelyanchouk}}, \bibinfo{year}{2006},
\bibinfo{journal}{Low
Temp. Phys.} \textbf{\bibinfo{volume}{32}}, \bibinfo{pages}{973}.

\bibitem[Shevchenko \emph{et~al.}(2008)Shevchenko, van der Ploeg, Grajcar,
Il'ichev, Omelyanchouk, and Meyer]{Shevchenko:2008}
\bibinfo{author}{\bibnamefont{Shevchenko},
\bibfnamefont{S. N.}},~%
\bibinfo{author}{\bibfnamefont{S. H.
W.}~\bibnamefont{van der Ploeg}},~\bibinfo{author}{\bibfnamefont{M.}~%
\bibnamefont{Grajcar}},~\bibinfo{author}{\bibfnamefont{E.}~%
\bibnamefont{Il'ichev}},~%
\bibinfo{author}{\bibfnamefont{A.
N.}~\bibnamefont{Omelyanchouk}},~\bibinfo{author}{\bibfnamefont{H.-G.}~%
\bibnamefont{Meyer}}, \bibinfo{year}{2008}, \bibinfo{journal}{Phys. Rev. B}
\textbf{\bibinfo{volume}{78}}, \bibinfo{pages}{174527}.

\bibitem[Shimshoni and Gefen(1991)Shimshoni and Gefen]{Shimshoni:1991} %
\bibinfo{author}{\bibnamefont{Shimshoni}, \bibfnamefont{E.}} and %
\bibinfo{author}{\bibfnamefont{Y.}~\bibnamefont{Gefen}}, \bibinfo{year}{1991}%
, \bibinfo{journal}{Ann. Phys.} \textbf{\bibinfo{volume}{210}}, %
\bibinfo{pages}{16}.

\bibitem[Shirley(1965)Shirley]{Shirley:1965} \bibinfo{author}{%
\bibnamefont{Shirley}, \bibfnamefont{J. H.}}, \bibinfo{year}{1965}, %
\bibinfo{journal}{Phys. Rev.} \textbf{\bibinfo{volume}{138}}, %
\bibinfo{pages}{B979}.

\bibitem[Shnyrkov \emph{et~al.}(2006)Shnyrkov, Wagner, Born, Shevchenko,
Krech, Omelyanchouk, Il'ichev, and Meyer]{Shnyrkov:2006} \bibinfo{author}{%
\bibnamefont{Shnyrkov}, \bibfnamefont{V. I.}}, \bibinfo{author}{%
\bibfnamefont{Th.}~\bibnamefont{Wagner}}, \bibinfo{author}{%
\bibfnamefont{D.}~\bibnamefont{Born}},
\bibinfo{author}{\bibfnamefont{S.
N.}~\bibnamefont{Shevchenko}}, \bibinfo{author}{\bibfnamefont{W.}~%
\bibnamefont{Krech}},
\bibinfo{author}{\bibfnamefont{A.
N.}~\bibnamefont{Omelyanchouk}}, \bibinfo{author}{\bibfnamefont{E.}~%
\bibnamefont{Il'ichev}}, and \bibinfo{author}{\bibfnamefont{H.-G.}~%
\bibnamefont{Meyer}}, \bibinfo{year}{2006}, \bibinfo{journal}{Phys. Rev. B}
\textbf{\bibinfo{volume}{73}}, \bibinfo{pages}{024506}.

\bibitem[Shnyrkov \emph{et~al.}(2009)Shnyrkov, Born, Soroka, and Krech]%
{Shnyrkov:2009}
\bibinfo{author}{\bibnamefont{Shnyrkov}, \bibfnamefont{V.
I.}}, \bibinfo{author}{\bibfnamefont{D.}~\bibnamefont{Born}}, %
\bibinfo{author}{\bibfnamefont{A. A.}~\bibnamefont{Soroka}}, and %
\bibinfo{author}{\bibfnamefont{W.}~\bibnamefont{Krech}}, \bibinfo{year}{2009}%
, \bibinfo{journal}{Phys. Rev. B} \textbf{\bibinfo{volume}{79}}, %
\bibinfo{pages}{184522}.

\bibitem[Shytov \emph{et~al.}(2003)Shytov, Ivanov, and Feigel'man]%
{Shytov:2003} \bibinfo{author}{\bibnamefont{Shytov}, \bibfnamefont{A. V.}}, %
\bibinfo{author}{\bibfnamefont{D. A.}~\bibnamefont{Ivanov}}, and %
\bibinfo{author}{\bibfnamefont{M. V.}~\bibnamefont{Feigel'man}}, %
\bibinfo{year}{2003}, \bibinfo{journal}{Eur. Phys. J. B} \textbf{%
\bibinfo{volume}{36}}, \bibinfo{pages}{263}.

\bibitem[Shytov(2004)Shytov]{Shytov:2004} \bibinfo{author}{%
\bibnamefont{Shytov}, \bibfnamefont{A. V.}}, \bibinfo{year}{2004}, %
\bibinfo{journal}{Phys. Rev. A} \textbf{\bibinfo{volume}{70}}, %
\bibinfo{pages}{052708}.

\bibitem[Sillanp\"{a}\"{a} \emph{et~al.}(2005)Sillanp\"{a}\"{a}, Lehtinen,
Paila, Makhlin, Roschier, and Hakonen]{Sillanpaa:2005} \bibinfo{author}{%
\bibnamefont{Sillanp\"{a}\"{a}}, \bibfnamefont{M.}}, \bibinfo{author}{%
\bibfnamefont{T.}~\bibnamefont{Lehtinen}}, \bibinfo{author}{%
\bibfnamefont{A.}~\bibnamefont{Paila}}, \bibinfo{author}{\bibfnamefont{Yu.}~%
\bibnamefont{Makhlin}}, \bibinfo{author}{\bibfnamefont{L.}~%
\bibnamefont{Roschier}}, and \bibinfo{author}{\bibfnamefont{P.}~%
\bibnamefont{Hakonen}}, \bibinfo{year}{2005},
\bibinfo{journal}{Phys. Rev.
Lett.} \textbf{\bibinfo{volume}{95}}, \bibinfo{pages}{206806}.

\bibitem[Sillanp\"{a}\"{a} \emph{et~al.}(2006)Sillanp\"{a}\"{a}, Lehtinen,
Paila, Makhlin, and Hakonen]{Sillanpaa:2006} \bibinfo{author}{%
\bibnamefont{Sillanp\"{a}\"{a}}, \bibfnamefont{M.}}, \bibinfo{author}{%
\bibfnamefont{T.}~\bibnamefont{Lehtinen}}, \bibinfo{author}{%
\bibfnamefont{A.}~\bibnamefont{Paila}}, \bibinfo{author}{\bibfnamefont{Yu.}~%
\bibnamefont{Makhlin}}, and \bibinfo{author}{\bibfnamefont{P.}~%
\bibnamefont{Hakonen}}, \bibinfo{year}{2006},
\bibinfo{journal}{Phys. Rev.
Lett.} \textbf{\bibinfo{volume}{96}}, \bibinfo{pages}{187002}.

\bibitem[Sillanp\"{a}\"{a} \emph{et~al.}(2007)Sillanp\"{a}\"{a}, Lehtinen,
Paila, Makhlin, and Hakonen]{Sillanpaa:2007} \bibinfo{author}{%
\bibnamefont{Sillanp\"{a}\"{a}}, \bibfnamefont{M.}}, \bibinfo{author}{%
\bibfnamefont{T.}~\bibnamefont{Lehtinen}}, \bibinfo{author}{%
\bibfnamefont{A.}~\bibnamefont{Paila}}, \bibinfo{author}{\bibfnamefont{Yu.}~%
\bibnamefont{Makhlin}}, and \bibinfo{author}{\bibfnamefont{P.}~%
\bibnamefont{Hakonen}}, \bibinfo{year}{2007},
\bibinfo{journal}{J. Low Temp.
Phys.} \textbf{\bibinfo{volume}{146}}, \bibinfo{pages}{253}.

\bibitem[Son \emph{et~al.}(2009)Son, Han, and Chu]{Son:2009} %
\bibinfo{author}{\bibnamefont{Son}, \bibfnamefont{S.-K.}}, %
\bibinfo{author}{\bibfnamefont{S.}~\bibnamefont{Han}}, and %
\bibinfo{author}{\bibfnamefont{S.-I.}~\bibnamefont{Chu}}, %
\bibinfo{year}{2009}, \bibinfo{journal}{Phys. Rev. A} \textbf{%
\bibinfo{volume}{79}}, \bibinfo{pages}{032301}.

\bibitem[St\"{u}ckelberg(1932)St\"{u}ckelberg]{Stueckelberg:1932} %
\bibinfo{author}{\bibnamefont{St\"{u}ckelberg}, \bibfnamefont{E. C. G.}}, %
\bibinfo{year}{1932}, \bibinfo{journal}{Helv. Phys. Acta} \textbf{%
\bibinfo{volume}{5}}, \bibinfo{pages}{369}.

\bibitem[Sun \emph{et~al.}(2006)Sun, Chen, Ji, Xu, Kang, Wu, Dong, Mao, Zhu,
and Xing]{Sun:2006} \bibinfo{author}{\bibnamefont{Sun}, \bibfnamefont{G.}}, %
\bibinfo{author}{\bibfnamefont{J.}~\bibnamefont{Chen}}, \bibinfo{author}{%
\bibfnamefont{Z.}~\bibnamefont{Ji}}, \bibinfo{author}{\bibfnamefont{W.}~%
\bibnamefont{Xu}}, \bibinfo{author}{\bibfnamefont{L.}~\bibnamefont{Kang}}, %
\bibinfo{author}{\bibfnamefont{P.}~\bibnamefont{Wu}}, \bibinfo{author}{%
\bibfnamefont{N.}~\bibnamefont{Dong}}, \bibinfo{author}{\bibfnamefont{G.}~%
\bibnamefont{Mao}}, \bibinfo{author}{\bibfnamefont{Y.}~\bibnamefont{Yu}},
and \bibinfo{author}{\bibfnamefont{D.}~\bibnamefont{Xing}}, %
\bibinfo{year}{2006}, \bibinfo{journal}{Appl. Phys.
Lett.} \textbf{\bibinfo{volume}{89}}, \bibinfo{pages}{082516}.

\bibitem[Sun \emph{et~al.}(2009)Sun, Wen, Wang, Cong, Chen, Kang, Xu, Yu,
Han, and Wu]{Sun:2009} \bibinfo{author}{\bibnamefont{Sun}, \bibfnamefont{G.}}%
, \bibinfo{author}{\bibfnamefont{X.}~\bibnamefont{Wen}}, \bibinfo{author}{%
\bibfnamefont{Y.}~\bibnamefont{Wang}}, \bibinfo{author}{\bibfnamefont{S.}~%
\bibnamefont{Cong}}, \bibinfo{author}{\bibfnamefont{J.}~\bibnamefont{Chen}}, %
\bibinfo{author}{\bibfnamefont{L.}~\bibnamefont{Kang}}, \bibinfo{author}{%
\bibfnamefont{W.}~\bibnamefont{Xu}}, \bibinfo{author}{\bibfnamefont{Y.}~%
\bibnamefont{Yu}}, \bibinfo{author}{\bibfnamefont{S.}~\bibnamefont{Han}},
and \bibinfo{author}{\bibfnamefont{P.}~\bibnamefont{Wu}}, %
\bibinfo{year}{2009}, \bibinfo{journal}{Appl. Phys. Lett.} \textbf{%
\bibinfo{volume}{94}}, \bibinfo{pages}{102502}.

\bibitem[Teranishi and Nakamura(1998)Teranishi and Nakamura]{Teranishi:1998}
\bibinfo{author}{\bibnamefont{Teranishi},
\bibfnamefont{Y.}} and \bibinfo{author}{\bibfnamefont{H.}~%
\bibnamefont{Nakamura}}, \bibinfo{year}{1998},
\bibinfo{journal}{Phys. Rev.
Lett.} \textbf{\bibinfo{volume}{81}}, \bibinfo{pages}{2032}.

\bibitem[Thorwart \emph{et~al.}(2000)Thorwart, Grifoni, and H\"{a}nggi]%
{Thorwart:2000} \bibinfo{author}{\bibnamefont{Thorwart}, \bibfnamefont{M.}}, %
\bibinfo{author}{\bibfnamefont{M.}~\bibnamefont{Grifoni}}, and %
\bibinfo{author}{\bibfnamefont{P.}~\bibnamefont{H\"{a}nggi}}, %
\bibinfo{year}{2000}, \bibinfo{journal}{Phys. Rev. Lett.} \textbf{%
\bibinfo{volume}{85}}, \bibinfo{pages}{860}.

\bibitem[Thorwart \emph{et~al.}(2001)Thorwart, Grifoni, and H\"{a}nggi]%
{Thorwart:2001} \bibinfo{author}{\bibnamefont{Thorwart}, \bibfnamefont{M.}}, %
\bibinfo{author}{\bibfnamefont{M.}~\bibnamefont{Grifoni}}, and %
\bibinfo{author}{\bibfnamefont{P.}~\bibnamefont{H\"{a}nggi}}, %
\bibinfo{year}{2001}, \bibinfo{journal}{Ann. Phys.} \textbf{%
\bibinfo{volume}{293}}, \bibinfo{pages}{15}.

\bibitem[Tong \emph{et~al.}(2005)Tong, Singh, Kwek, and Oh]{Tong:2005}
\bibinfo{author}{\bibnamefont{Tong},
\bibfnamefont{D. M.}}, \bibinfo{author}{\bibfnamefont{K.}~%
\bibnamefont{Singh}},
\bibinfo{author}{\bibfnamefont{L.
C.}~\bibnamefont{Kwek}} and
\bibinfo{author}{\bibfnamefont{C.
H.}~\bibnamefont{Oh}}, \bibinfo{year}{2005},
\bibinfo{journal}{Phys. Rev.
Lett.} \textbf{\bibinfo{volume}{95}}, \bibinfo{pages}{110407}.

\bibitem[Tong \emph{et~al.}(2007)Tong, Singh, Kwek, and Oh]{Tong:2007}
\bibinfo{author}{\bibnamefont{Tong},
\bibfnamefont{D. M.}}, \bibinfo{author}{\bibfnamefont{K.}~%
\bibnamefont{Singh}},
\bibinfo{author}{\bibfnamefont{L.
C.}~\bibnamefont{Kwek}} and
\bibinfo{author}{\bibfnamefont{C.
H.}~\bibnamefont{Oh}}, \bibinfo{year}{2007},
\bibinfo{journal}{Phys. Rev.
Lett.} \textbf{\bibinfo{volume}{98}}, \bibinfo{pages}{150402}.

\bibitem[Tornes and Stroud(2008)Tornes and Stroud]{Tornes:2008}
\bibinfo{author}{\bibnamefont{Tornes},
\bibfnamefont{I.}} and \bibinfo{author}{\bibfnamefont{D.}~%
\bibnamefont{Stroud}}, \bibinfo{year}{2008}, \bibinfo{journal}{Phys. Rev. B}
\textbf{\bibinfo{volume}{77}}, \bibinfo{pages}{224513}.

\bibitem[van Ditzhuijzen \emph{et~al.}(2009)van Ditzhuijzen, Tauschinsky,
and van Linden van den Heuvell]{van Ditzhuijzen:2009} \bibinfo{author}{%
\bibnamefont{van Ditzhuijzen}, \bibfnamefont{C. S. E.}}, \bibinfo{author}{%
\bibfnamefont{A.}~\bibnamefont{Tauschinsky}}, and \bibinfo{author}{%
\bibfnamefont{H. B.}~\bibnamefont{van Linden van den Heuvell}}, %
\bibinfo{year}{2009}, \bibinfo{journal}{Phys. Rev. A} \textbf{%
\bibinfo{volume}{80}}, \bibinfo{pages}{063407}.

\bibitem[Vitanov and Garraway(1996)Vitanov and Garraway]{Vitanov:1996} %
\bibinfo{author}{\bibnamefont{Vitanov}, \bibfnamefont{N. V.}} and %
\bibinfo{author}{\bibfnamefont{B. M.}~\bibnamefont{Garraway}}, %
\bibinfo{year}{1996}, \bibinfo{journal}{Phys. Rev. A} \textbf{%
\bibinfo{volume}{53}}, \bibinfo{pages}{4288}.

\bibitem[Vitanov(1999)Vitanov]{Vitanov:1999} \bibinfo{author}{%
\bibnamefont{Vitanov}, \bibfnamefont{N. V.}}, \bibinfo{year}{1999}, %
\bibinfo{journal}{Phys. Rev. A} \textbf{\bibinfo{volume}{59}}, %
\bibinfo{pages}{988}.

\bibitem[Vitanov \emph{et~al.}(2001)Vitanov, Halfmann, Shore, and Bergmann]%
{Vitanov:2001} \bibinfo{author}{\bibnamefont{Vitanov}, \bibfnamefont{N. V.}}%
, \bibinfo{author}{\bibfnamefont{Th.}~\bibnamefont{Halfmann}}, %
\bibinfo{author}{\bibfnamefont{B. W.}~\bibnamefont{Shore}}, and %
\bibinfo{author}{\bibfnamefont{K.}~\bibnamefont{Bergmann}}, %
\bibinfo{year}{2001}, \bibinfo{journal}{Annu. Rev. Phys. Chem.} \textbf{%
\bibinfo{volume}{52}}, \bibinfo{pages}{763}.

\bibitem[Vitanov \emph{et~al.}(2003)Vitanov, Yatsenko, and Bergmann]%
{Vitanov:2003} \bibinfo{author}{\bibnamefont{Vitanov}, \bibfnamefont{N. V.}}%
, \bibinfo{author}{\bibfnamefont{L. P.}~\bibnamefont{Yatsenko}}, and %
\bibinfo{author}{\bibfnamefont{K.}~\bibnamefont{Bergmann}}, %
\bibinfo{year}{2003}, \bibinfo{journal}{Phys. Rev. A} \textbf{%
\bibinfo{volume}{68}}, \bibinfo{pages}{043401}.

\bibitem[Wallraff \emph{et~al.}(2003)Wallraff, Duty, Lukashenko, and Ustinov]%
{Wallraff:2003} \bibinfo{author}{\bibnamefont{Wallraff}, \bibfnamefont{A.}}, %
\bibinfo{author}{\bibfnamefont{T.}~\bibnamefont{Duty}}, \bibinfo{author}{%
\bibfnamefont{A.}~\bibnamefont{Lukashenko}}, and \bibinfo{author}{%
\bibfnamefont{A. V.}~\bibnamefont{Ustinov}}, \bibinfo{year}{2003}, %
\bibinfo{journal}{Phys. Rev. Lett.} \textbf{\bibinfo{volume}{90}}, %
\bibinfo{pages}{037003}.

\bibitem[Wang \emph{et~al.}(2010)Wang, Cong, Wen, Pan, Sun, Chen, Kang, Xu,
Yu, and Wu]{Wang:2010}
\bibinfo{author}{\bibnamefont{Wang},
\bibfnamefont{Y.}}, \bibinfo{author}{\bibfnamefont{S.}~\bibnamefont{Cong}}, %
\bibinfo{author}{\bibfnamefont{X.}~\bibnamefont{Wen}}, \bibinfo{author}{%
\bibfnamefont{C.}~\bibnamefont{Pan}}, \bibinfo{author}{\bibfnamefont{G.}~%
\bibnamefont{Sun}}, \bibinfo{author}{\bibfnamefont{J.}~\bibnamefont{Chen}}, %
\bibinfo{author}{\bibfnamefont{L.}~\bibnamefont{Kang}}, \bibinfo{author}{%
\bibfnamefont{W.}~\bibnamefont{Xu}}, \bibinfo{author}{\bibfnamefont{Y.}~%
\bibnamefont{Yu}}, and \bibinfo{author}{\bibfnamefont{P.}~\bibnamefont{Wu}}, %
\bibinfo{year}{2010}, \bibinfo{journal}{Phys. Rev. B} \textbf{%
\bibinfo{volume}{81}}, \bibinfo{pages}{144505}.

\bibitem[Wei \emph{et~al.}(2008)Wei, Johansson, Cen, Ashhab, and Nori]%
{Wei:2008} \bibinfo{author}{\bibnamefont{Wei}, \bibfnamefont{L. F.}}, %
\bibinfo{author}{\bibfnamefont{J. R.}~\bibnamefont{Johansson}}, %
\bibinfo{author}{\bibfnamefont{L. X.}~\bibnamefont{Cen}}, %
\bibinfo{author}{\bibfnamefont{S.}~\bibnamefont{Ashhab}}, and %
\bibinfo{author}{\bibfnamefont{F.}~\bibnamefont{Nori}}, \bibinfo{year}{2008}%
, \bibinfo{journal}{Phys. Rev.
Lett.} \textbf{\bibinfo{volume}{100}}, \bibinfo{pages}{113601}.

\bibitem[Wen and Yu(2009)Wen and Yu]{Wen:2009} \bibinfo{author}{%
\bibnamefont{Wen}, \bibfnamefont{X.}} and \bibinfo{author}{%
\bibfnamefont{Y.}~\bibnamefont{Yu}}, \bibinfo{year}{2009}, %
\bibinfo{journal}{Phys. Rev. B} \textbf{\bibinfo{volume}{79}}, %
\bibinfo{pages}{094529}.

\bibitem[Wendin and Shumeiko(2007)Wendin and Shumeiko]{Wendin:2007} %
\bibinfo{author}{\bibnamefont{Wendin}, \bibfnamefont{G.}} and %
\bibinfo{author}{\bibfnamefont{V. S.}~\bibnamefont{Shumeiko}}, %
\bibinfo{year}{2007}, \bibinfo{journal}{Low Temp. Phys.} \textbf{%
\bibinfo{volume}{33}}, \bibinfo{pages}{724}.

\bibitem[Wernsdorfer \emph{et~al.}(2000)Wernsdorfer, Sessoli, Caneshi,
Gatteschi, and Cornia]{Wernsdorfer:2000}
\bibinfo{author}{\bibnamefont{Wernsdorfer},
\bibfnamefont{W.}}, \bibinfo{author}{\bibfnamefont{R.}~\bibnamefont{Sessoli}}%
, \bibinfo{author}{\bibfnamefont{A.}~\bibnamefont{Caneshi}}, %
\bibinfo{author}{\bibfnamefont{D.}~\bibnamefont{Gatteschi}}, and %
\bibinfo{author}{\bibfnamefont{A.}~\bibnamefont{Cornia}}, %
\bibinfo{year}{2000}, \bibinfo{journal}{Europhys. Lett.} \textbf{%
\bibinfo{volume}{50}}, \bibinfo{pages}{552}.

\bibitem[Wilhelm(2008)Wilhelm]{Wilhelm:2008}
\bibinfo{author}{\bibnamefont{Wilhelm},
\bibfnamefont{F. K.}}, \bibinfo{year}{2008}, \bibinfo{journal}{Nature}
\textbf{\bibinfo{volume}{455}}, \bibinfo{pages}{41}.

\bibitem[Wilson \emph{et~al.}(2007)Wilson, Duty, Persson, Sandberg,
Johansson, and Delsing]{Wilson:2007}
\bibinfo{author}{\bibnamefont{Wilson},
\bibfnamefont{C. M.}}, \bibinfo{author}{\bibfnamefont{T.}~\bibnamefont{Duty}}%
, \bibinfo{author}{\bibfnamefont{F.}~\bibnamefont{Persson}}, %
\bibinfo{author}{\bibfnamefont{M.}~\bibnamefont{Sandberg}}, %
\bibinfo{author}{\bibfnamefont{G.}~\bibnamefont{Johansson}}, and %
\bibinfo{author}{\bibfnamefont{P.}~\bibnamefont{Delsing}}, %
\bibinfo{year}{2007}, \bibinfo{journal}{Phys. Rev. Lett.} \textbf{%
\bibinfo{volume}{98}}, \bibinfo{pages}{257003}.

\bibitem[Wilson \emph{et~al.}(2010)Wilson, Johansson, Duty, Persson,
Sandberg, and Delsing]{Wilson:2010}
\bibinfo{author}{\bibnamefont{Wilson},
\bibfnamefont{C. M.}}, \bibinfo{author}{\bibfnamefont{G.}~%
\bibnamefont{Johansson}}, \bibinfo{author}{\bibfnamefont{T.}~%
\bibnamefont{Duty}}, \bibinfo{author}{\bibfnamefont{F.}~%
\bibnamefont{Persson}}, \bibinfo{author}{\bibfnamefont{M.}~%
\bibnamefont{Sandberg}}, and \bibinfo{author}{\bibfnamefont{P.}~%
\bibnamefont{Delsing}}, \bibinfo{year}{2010}, \bibinfo{journal}{Phys. Rev. B}
\textbf{\bibinfo{volume}{81}}, \bibinfo{pages}{024520}.

\bibitem[Wittig(2005)Wittig]{Wittig:2005}
\bibinfo{author}{\bibnamefont{Wittig},
\bibfnamefont{C.}}, \bibinfo{year}{2005}, \bibinfo{journal}{J. Phys. Chem. B}
\textbf{\bibinfo{volume}{109}}, \bibinfo{pages}{8428}.

\bibitem[Wubs \emph{et~al.}(2005) Wubs, Saito, Kohler, Kayanuma, and Hanggi]%
{Wubs:2005} \bibinfo{author}{\bibnamefont{Wubs},
\bibfnamefont{M.}}, \bibinfo{author}{\bibfnamefont{K.}~\bibnamefont{Saito}}, %
\bibinfo{author}{\bibfnamefont{S.}~\bibnamefont{Kohler}}, %
\bibinfo{author}{\bibfnamefont{Y.}~\bibnamefont{Kayanuma}}, and %
\bibinfo{author}{\bibfnamefont{P.}~\bibnamefont{H\"{a}nggi}}, %
\bibinfo{year}{2005}, \bibinfo{journal}{New J. Phys.} \textbf{%
\bibinfo{volume}{7}}, \bibinfo{pages}{218}.

\bibitem[Wubs \emph{et~al.}(2006) Wubs, Saito, Kohler, Hanggi, and Kayanuma]%
{Wubs:2006} \bibinfo{author}{\bibnamefont{Wubs},
\bibfnamefont{M.}}, \bibinfo{author}{\bibfnamefont{K.}~\bibnamefont{Saito}}, %
\bibinfo{author}{\bibfnamefont{S.}~\bibnamefont{Kohler}}, %
\bibinfo{author}{\bibfnamefont{P.}~\bibnamefont{H\"{a}nggi}}, and %
\bibinfo{author}{\bibfnamefont{Y.}~\bibnamefont{Kayanuma}}, %
\bibinfo{year}{2006}, \bibinfo{journal}{Phys. Rev. Lett.} \textbf{%
\bibinfo{volume}{97}}, \bibinfo{pages}{200404}.

\bibitem[Wubs(2010)Wubs]{Wubs:2010}
\bibinfo{author}{\bibnamefont{Wubs},
\bibfnamefont{M.}}, \bibinfo{year}{2010}, \bibinfo{journal}{Chemical
Physics, in press: http://dx.doi.org/10.1016/j.chemphys.2010.03.003}
%\textbf{\bibinfo{volume}{}}, \bibinfo{pages}{}.

\bibitem[Yoakum \emph{et~al.}(1992)Yoakum, Sirko, and Koch]{Yoakum:1992} %
\bibinfo{author}{\bibnamefont{Yoakum}, \bibfnamefont{S.}}, %
\bibinfo{author}{\bibfnamefont{L.}~\bibnamefont{Sirko}}, and %
\bibinfo{author}{\bibfnamefont{P. M.}~\bibnamefont{Koch}}, %
\bibinfo{year}{1992}, \bibinfo{journal}{Phys. Rev. Lett.} \textbf{%
\bibinfo{volume}{69}}, \bibinfo{pages}{1919}.

\bibitem[You and Nori(2005)You and Nori]{You:2005} \bibinfo{author}{%
\bibnamefont{You}, \bibfnamefont{J. Q.}} and \bibinfo{author}{%
\bibfnamefont{F.}~\bibnamefont{Nori}}, \bibinfo{year}{2005}, %
\bibinfo{journal}{Physics Today} \textbf{\bibinfo{volume}{58(11)}}, %
\bibinfo{pages}{42}.

\bibitem[Yu \emph{et~al.}(2005)Yu, Oliver, Lee, Berggren, Levitov, and
Orlando]{Yu:2005} \bibinfo{author}{\bibnamefont{Yu}, \bibfnamefont{Y.}}, %
\bibinfo{author}{\bibfnamefont{W. D.}~\bibnamefont{Oliver}}, %
\bibinfo{author}{\bibfnamefont{J. C.}~\bibnamefont{Lee}}, %
\bibinfo{author}{\bibfnamefont{K. K.}~\bibnamefont{Berggren}}, %
\bibinfo{author}{\bibfnamefont{L. S.}~\bibnamefont{Levitov}}, and %
\bibinfo{author}{\bibfnamefont{T. P.}~\bibnamefont{Orlando}}, %
\bibinfo{year}{2005}, \eprint{arXiv:cond-mat/0508587}.

\bibitem[Zagoskin and Blais(2007)Zagoskin and Blais]{Zagoskin:2007} %
\bibinfo{author}{\bibnamefont{Zagoskin}, \bibfnamefont{A.}} and %
\bibinfo{author}{\bibfnamefont{A.}~\bibnamefont{Blais}}, \bibinfo{year}{2007}%
, \bibinfo{journal}{Phys. Canada} \textbf{\bibinfo{volume}{63}}, %
\bibinfo{pages}{215}.

\bibitem[Zenesini \emph{et~al.}(2009)Zenesini, Lignier, Tayebirad,
Radogostowicz, Ciampini, Mannella, Wimberger, Morsch, and Arimondo]%
{Zenesini:2009} \bibinfo{author}{\bibnamefont{Zenesini},
\bibfnamefont{A.}}, \bibinfo{author}{\bibfnamefont{H.}~\bibnamefont{Lignier}}%
, \bibinfo{author}{\bibfnamefont{G.}~\bibnamefont{Tayebirad}}, %
\bibinfo{author}{\bibfnamefont{J.}~\bibnamefont{Radogostowicz}}, %
\bibinfo{author}{\bibfnamefont{D.}~\bibnamefont{Ciampini}}, %
\bibinfo{author}{\bibfnamefont{R.}~\bibnamefont{Mannella}}, %
\bibinfo{author}{\bibfnamefont{S.}~\bibnamefont{Wimberger}}, %
\bibinfo{author}{\bibfnamefont{O.}~\bibnamefont{Morsch}}, and %
\bibinfo{author}{\bibfnamefont{E.}~\bibnamefont{Arimondo}}, %
\bibinfo{year}{2009}, \bibinfo{journal}{Phys. Rev. Lett.} \textbf{%
\bibinfo{volume}{103}}, \bibinfo{pages}{090403}.

\bibitem[Zel'dovich(1973)Zel'dovich]{Zel'dovich:1973} \bibinfo{author}{%
\bibnamefont{Zel'dovich}, \bibfnamefont{Ya. B.}}, \bibinfo{year}{1973}, %
\bibinfo{journal}{Sov. Phys. Usp.} \textbf{\bibinfo{volume}{16}}, %
\bibinfo{pages}{427}.

\bibitem[Zener(1932)Zener]{Zener:1932}
\bibinfo{author}{\bibnamefont{Zener},
\bibfnamefont{C.}}, \bibinfo{year}{1932},
\bibinfo{journal}{Proc. R. Soc.
(Lond.) A} \textbf{\bibinfo{volume}{137}}, \bibinfo{pages}{696}.

\bibitem[Zhang \emph{et~al.}(2008a)Zhang, Hanggi, and Gong]{Zhang:2008a} %
\bibinfo{author}{\bibnamefont{Zhang}, \bibfnamefont{Q.}}, %
\bibinfo{author}{\bibfnamefont{P.}~\bibnamefont{H\"{a}nggi}}, and %
\bibinfo{author}{\bibfnamefont{J. B.}~\bibnamefont{Gong}}, %
\bibinfo{year}{2008}, \bibinfo{journal}{Phys. Rev. A} \textbf{%
\bibinfo{volume}{77}}, \bibinfo{pages}{053607}.

\bibitem[Zhang \emph{et~al.}(2008b)Zhang, Hanggi, and Gong]{Zhang:2008b} %
\bibinfo{author}{\bibnamefont{Zhang}, \bibfnamefont{Q.}}, %
\bibinfo{author}{\bibfnamefont{P.}~\bibnamefont{H\"{a}nggi}}, and %
\bibinfo{author}{\bibfnamefont{J. B.}~\bibnamefont{Gong}}, %
\bibinfo{year}{2008}, \bibinfo{journal}{New J. Phys.} \textbf{%
\bibinfo{volume}{10}}, \bibinfo{pages}{073008}.

\bibitem[Zhou \emph{et~al.}(2008a)Zhou, Gong, Liu, Sun, and Nori]%
{Zhou:2008a} \bibinfo{author}{\bibnamefont{Zhou}, \bibfnamefont{L.}}, %
\bibinfo{author}{\bibfnamefont{Z. R.}~\bibnamefont{Gong}}, %
\bibinfo{author}{\bibfnamefont{Y. X.}~\bibnamefont{Liu}}, %
\bibinfo{author}{\bibfnamefont{C. P.}~\bibnamefont{Sun}}, and %
\bibinfo{author}{\bibfnamefont{F.}~\bibnamefont{Nori}}, \bibinfo{year}{2008}%
, \bibinfo{journal}{Phys. Rev. Lett.} \textbf{\bibinfo{volume}{101}}, %
\bibinfo{pages}{100501}.

\bibitem[Zhou \emph{et~al.}(2008b)Zhou, Dong, Liu, Sun, and Nori]%
{Zhou:2008b} \bibinfo{author}{\bibnamefont{Zhou}, \bibfnamefont{L.}}, %
\bibinfo{author}{\bibfnamefont{H.}~\bibnamefont{Dong}}, \bibinfo{author}{%
\bibfnamefont{Y. X.}~\bibnamefont{Liu}}, \bibinfo{author}{\bibfnamefont{C.
P.}~\bibnamefont{Sun}}, and \bibinfo{author}{\bibfnamefont{F.}~%
\bibnamefont{Nori}}, \bibinfo{year}{2008}, \bibinfo{journal}{Phys. Rev. A}
\textbf{\bibinfo{volume}{78}}, \bibinfo{pages}{063827}.

\bibitem[Zhu \emph{et~al.}(2001)Zhu, Teranishi, and Nakamura]{Zhu:2001} %
\bibinfo{author}{\bibnamefont{Zhu}, \bibfnamefont{C.}}, \bibinfo{author}{%
\bibfnamefont{Y.}~\bibnamefont{Teranishi}}, and \bibinfo{author}{%
\bibfnamefont{H.}~\bibnamefont{Nakamura}}, \bibinfo{year}{2001}, %
\bibinfo{journal}{Adv. Chem. Phys.} \textbf{\bibinfo{volume}{117}}, %
\bibinfo{pages}{127}.

\bibitem[Zueco \emph{et~al.}(2008)Zueco, Hanggi, and Kohler]{Zueco:2008} %
\bibinfo{author}{\bibnamefont{Zueco}, \bibfnamefont{D.}}, %
\bibinfo{author}{\bibfnamefont{P.}~\bibnamefont{H\"{a}nggi}}, and %
\bibinfo{author}{\bibfnamefont{S.}~\bibnamefont{Kohler}}, %
\bibinfo{year}{2008}, \bibinfo{journal}{New J. Phys.} \textbf{%
\bibinfo{volume}{10}}, \bibinfo{pages}{115012}.

\bibitem[Zurek(1996)Zurek]{Zurek:1996}
\bibinfo{author}{\bibnamefont{Zurek},
\bibfnamefont{W. H.}}, \bibinfo{year}{1996}, \bibinfo{journal}{Phys. Rep.}
\textbf{\bibinfo{volume}{276}}, \bibinfo{pages}{177}.

\bibitem[Zwanziger \emph{et~al.}(2003)Zwanziger, Werner-Zwanziger, and Gaitan%
]{Zwanziger:2003}
\bibinfo{author}{\bibnamefont{Zwanziger}, \bibfnamefont{J.
W.}}, \bibinfo{author}{\bibfnamefont{U.}~\bibnamefont{Werner-Zwanziger}},
and \bibinfo{author}{\bibfnamefont{F.}~\bibnamefont{Gaitan}}, %
\bibinfo{year}{2003}, \bibinfo{journal}{Chem. Phys. Lett.} \textbf{%
\bibinfo{volume}{375}}, \bibinfo{pages}{429}.
\end{thebibliography}

\end{document}